\documentclass[usenatbib]{mn2e}
\usepackage{graphicx, amsmath, aas_macros, deluxetable}

\def \Msun {\, \rm M_\odot}

\title[The Magellanic System]{The Role of Dwarf Galaxy Interactions in Shaping the Magellanic System and Implications for Magellanic Irregulars } 
\author[Besla et al.]{
{\parbox{\textwidth}{Gurtina Besla$^{1,2}$\thanks{$\!\!$Hubble Fellow   e-mail:gbesla@astro.columbia.edu}, 
Nitya Kallivayalil$^3$, Lars Hernquist$^2$, Roeland P. van der Marel$^4$,  T.J. Cox$^5$, and Du\v{s}an Kere\v{s}$^6$} \vspace{0.4cm}}\\
$\!\!$Department of Astronomy \& Astrophysics, Columbia University, 550 West 120th Street, New York, NY 10027, USA\\
$\!\!$ Harvard-Smithsonian Center for Astrophysics, 60 Garden Street, Cambridge, MA 02138 \\
$\!\!$ Department of Astronomy, Yale University, New Haven, CT 06520, USA \\
$\!\!$ Space Telescope Science Institute, 3700 San Martin Drive, Baltimore, MD 21218 \\
$\!\!$ Carnegie Observatories, 813 Santa Barbara Street, Pasadena, CA 91101, USA \\
$\!\!$ Department of Astronomy and Theoretical Astrophysics Center, University of California, Berkeley, CA 94720-3411, USA }

\begin{document}

\pagerange{\pageref{firstpage}--\pageref{lastpage}}
\pubyear{2012}

\maketitle

\label{firstpage}

\begin{abstract}
We present a novel pair of numerical models of the interaction history between the Large and Small Magellanic Clouds 
(LMC and SMC, respectively) and our Milky Way (MW) in light of recent high precision proper motions from the Hubble Space 
Telescope \citep{Nitya1, Nitya2}. These proper motions imply that the Magellanic Clouds (MCs)
 are moving $\sim$80 km/s faster than 
previously considered. Given these velocities, cosmological simulations of hierarchical structure formation favor a scenario
 where the MCs are currently on their first infall towards our Galaxy \citep{besla2007, BoylanBesla2011, busha2010}. 
We illustrate here that the observed irregular morphology and internal kinematics of the Magellanic System (in gas and stars) are naturally 
explained by interactions between the LMC and SMC, rather than gravitational interactions with the MW.  These conclusions provide 
further support that the MCs are completing their first infall to our system.
In particular, we demonstrate that the Magellanic Stream, a band of HI gas trailing behind the Clouds 150 degrees 
across the sky, can be accounted for by the action of LMC tides on the SMC before the system 
was accreted by the MW. We further demonstrate that the off-center, warped stellar bar of the LMC and its one-armed spiral, can be 
naturally explained by a recent direct collision with its lower mass companion, the SMC. Such structures are 
key morphological characteristics of a class of galaxies referred to as Magellanic Irregulars \citep{deVaucouleurs}, the majority of which 
are not associated with massive spiral galaxies. We infer that dwarf-dwarf galaxy interactions are important drivers for the 
morphological evolution of Magellanic Irregulars and can dramatically affect the efficiency of baryon removal from dwarf galaxies 
via the formation of extended tidal bridges and tails. 
Such interactions are important not only for the evolution of dwarf galaxies but also have direct consequences for the buildup of baryons in our own MW, 
as LMC-mass systems are believed to be the dominant building blocks of MW-type halos. 
\end{abstract}


\section{Introduction}
\label{sec:intro}

The Large Magellanic Cloud (LMC) is the prototype for a class of dwarf galaxies known as Magellanic Irregulars. 
Like the LMC, these galaxies are characterized 
by being gas rich, one-armed spirals with off-center bars \citep{deVaucouleurs}.
Although there are numerous examples of 
Magellanic Irregulars in our Local Volume, they are rarely found about massive spirals.
This has been confirmed by recent studies of the frequency of LMC analogs about Milky Way (MW) type 
galaxies in the SDSS DR7 catalog \citep{liu2010,tollerud2011}.
Based on similar statistics, \citet{deVaucouleurs} concluded that the LMC has necessarily
 experienced little or no distortion due to interactions with the MW, and so
  its characteristic asymmetric features must owe to some other process.

The idea that the LMC's evolution has not been dictated by interactions with the MW is given further credence by 
the distance morphology relationship exhibited by MW and M31 satellites, whereby gas rich satellites are located at larger
galactocentric radii than gas poor spheroidals. The Magellanic Clouds (MCs), at a mere 50-60 kpc away, are notable exceptions to this relationship, 
 leading \citet{vandenbergh2006} to describe them as interlopers in our system. 
 Along the same lines, recent studies indicate that the LMC is much bluer in color relative to analogs in its magnitude range 
\citep{tollerud2011, james2011}. 
This fact is difficult to reconcile with the expected gas loss and quenching of star
formation the LMC should have incurred if it were indeed a long-term companion of the MW \citep{Grcevich2009}. 
These conclusions are further supported by recent proper motion
measurements \citep{Nitya1, Nitya2}, which indicate that the LMC is moving $\sim$80 km/s faster than previously believed \citep{GN}. 
Given the measured energetics of the LMC's orbit today, backward orbital integration schemes \citep{besla2007} and 
statistics from large scale cosmological simulations \citep{BoylanBesla2011, busha2010} indicate that the LMC is likely on its first infall 
towards the MW.  Consequently, the MW cannot have been the driver of its morphological evolution. 

In this study we ask the following:  if not interactions with the MW, then what is the origin of the asymmetric appearance of the LMC
and what is its connection to Magellanic Irregulars in general?

Notably, the LMC has a nearby companion, the Small Magellanic Cloud (SMC). In fact, many Magellanic Irregulars 
also have companions \citep{odewahn1994}, although the frequency of such configurations is debated \citep{wilcots2009}. 
Particularly striking examples include the Magellanic Irregular galaxies NGC 4027 \citep{phookun1992} and NGC 3664 \citep{wilcots2004}; both 
have a low mass companion to which each is connected by a bridge of gas. 
The LMC and SMC are also connected by a bridge of HI gas, known as the Magellanic Bridge \citep{Kerr}, suggesting
that interactions between dwarf pairs may hold clues to understanding the current morphology of Magellanic Irregular galaxies.

In addition to the Magellanic Bridge, the MCs are associated with both leading and trailing streams of gas, referred to as the 
Leading Arm and Magellanic Stream, respectively. 
The Magellanic Stream extends over 150 degrees across the southern sky \citep{nidever2010, nidever2008, Putman2003, Wannier, Mathewson} 
and has been traditionally modeled as the product of MW tides \citep{MF, Heller, Lin95, GSF, GN, Bekki, Connors, Mastro, ruzicka2009, ruzicka2010} 
and ram pressure stripping ~\citep{Mastro, Moore}. A purely hydrodynamic solution cannot pull material forward to 
explain the Leading Arm Feature, meaning that MW tides must be invoked in some form
in all of these models. However, on a first infall, MW tides are negligible until very recently; it is thus difficult to reconcile the new 
  proper motions and updated orbits with the formation of the Magellanic Stream, Bridge and Leading Arm in the context of the existing scenarios.   

Alternatively, \citet{besla2010} (hereafter B10) introduced a model to explain the observed large scale gas morphology of the Magellanic System
through tidal interactions between the LMC and SMC \cite[see also, ][]{diaz2011}.
  Because MW tides are not responsible for removing material from the system, this picture is consistent 
with a first infall scenario. 
In this model, the Magellanic Bridge, Arm and Stream are hypothesized to be analogs of the classical \citet{TT72} tidal bridge and tail scenario and 
should be commonly found about interacting pairs/groups of dwarf galaxies. 

Here we explore whether interactions between the MCs can also account for the 
internal morphology and kinematics of the LMC and therefore shed light on the dynamical state of Magellanic Irregulars more generally.  

In particular, the nature of the LMC's off-centered stellar bar has been a long standing puzzle, as it is not present in any other
tracer of the 
interstellar medium (ISM); it is neither apparent in the HI gas disk nor a site of active star formation as traced by H$\alpha$ emission. 
Strong bars in more massive galaxies serve to funnel gas towards the center; streaming motions and characteristic ``S-shaped" isovelocity 
contours are thus evident in their gas velocity fields.  
While weak large scale streaming motions along the bar may be evident in the LMC HI velocity field \citep{kim1998}, the expected ``S-shaped" isovelocity 
contours are not present.\footnote{\citet{kim1998} comment on the existence of a distorted S-shaped isovelocity contour across the LMC's 
minor axis. However,  \citet{Olsen} find that this feature straightens out when the higher proper motions are accounted for.}
Interestingly, this is also true of many other 
Magellanic Irregulars \citep{wilcots2009}: bars in these systems do not appear to strongly affect the underlying gas distribution. 
 There is also evidence that the bar may be warped relative to the LMC disk plane \citep{sub2003, lah2005, koerwer2009}.
Using relative distance measurements to Cepheids,  \citet{nikolaev2004} concluded that the bar is in fact located $\sim$0.5 kpc in front of the main disk. 
For this reason it has been described as a ``levitating" bar. \citet{zaritsky2004b} suggests that this may be a result of viewing a triaxial stellar bulge that is 
embedded in a highly obscuring thick disk.   Along the same lines,  \citet{zhao2000} postulate that the off-centered bar is an unvirialized
structure, inclined relative to the plane of the LMC disk by as much as 25 degrees in order to explain the microlensing optical depth observed
towards the LMC. Clearly, the nature of the LMC's bar is an ongoing subject of debate.

We posit here that a recent direct collision between the LMC and SMC has left the LMC with a warped, off-centered stellar
bar and pronounced one-armed spiral. We further claim that such asymmetric structures are characteristic of
Magellanic type galaxies undergoing minor mergers.  In this study we
 illustrate that such a scenario is consistent with a first infall towards our MW and 
can simultaneously explain both the morphology and kinematics of the LMC as well as the 
 large scale gas morphology of the Magellanic System.  Thus, Magellanic Irregulars with nearby companions
should also be associated with faint extended gaseous tails and bridges.  As in the Magellanic System, such features 
 could hold  $\sim$50\% of the baryonic mass of the original system, indicating that dwarf-dwarf tidal interactions are an important 
 mechanism for the loss of baryons in low mass systems  \citep[see also,][]{DOnghia2009}, as
a consequence of resonant interactions between spinning disks \citep{DOnghia2010}.

We stress that the goal of our study here is not to reproduce every detail of the Magellanic System, as we have not conducted a 
complete parameter search of all the possible orbital configurations, mass ratios and gas fractions, which
influence the 
final outcome. The aim of this investigation is rather
to determine which of the observed peculiarities of the Magellanic System 
can be directly linked to interactions between the MCs.    

Moreover, our work has broader implications for understanding the properties of accreted satellites.  Minor mergers
are frequent events that shape galaxies and their halos; however, little attention has been given to the accretion of binary pairs 
or groups of smaller galaxies \citep[but see,][]{DOnghia2008, sales2007}.
This study represents a first step towards understanding the morphological evolution and gas loss 
rates of such galaxies immediately after their capture by a massive host.  LMC mass objects are expected to be the primary 
building blocks of MW type galaxies \citep{stewart2008}, making this study of direct relevance to our understanding of the evolution 
of the MW.

In this paper we begin by outlining our methodology and introducing two possible models for the interaction history of the MCs, one of which invokes a recent
direct collision between the MCs.  In the subsequent sections we discuss the resulting large scale gas structure and internal structure and kinematics of the
LMC.  The results for the SMC and the expected stellar counterpart to the Magellanic Stream will be presented in future work.

\section{Methodology}
\label{sec:Methods}

We follow the general method outlined in B10 to set up the initial galaxy models and orbits in order to 
reproduce the observed large scale gaseous structure of the Magellanic System. Details about our 
numerical methods, initial conditions and chosen orbital parameters are described below.

\subsection{Numerical Methods}

All of the numerical simulations performed in this work use the N-body smoothed-particle hydrodynamics (SPH) 
code, Gadget3 \citep{Gadget2}.	
The Gadget3 code incorporates a subresolution multiphase model of the ISM that includes radiative cooling 
\citep{springel2003}, and incorporates a fully conservative approach to
integrating the equations of motion \citep{springel2002}.
 Star formation from the cold phase (i.e. all cold gas - no distinction is made between atomic and molecular components) 
 follows a Schmidt volume
density law  $\rho_{\rm SFR} \propto \rho_{\rm gas}^N$ (with $N=1.5$) that is normalized to approximate the star formation rate
of the MW. A local threshold volume density cutoff of 0.004 $\Msun pc^{-3}$ is adopted, below which stars do not form. 
As pointed out by many authors \citep[e.g., ][]{kuhlen2011, Hop2011, gnedin2010b, robertson2008}, such a star formation prescription is likely inappropriate for 
dwarf galaxies. We discuss the implications of our adopted prescriptions to our results in $\S$~\ref{sec:SF}.

Stellar feedback in the form of galactic winds
is not employed in our simulations; however, we comment on the relative importance of 
outflows to the formation of the Magellanic Stream in Appendix~\ref{sec:Feedback}.  

We note that the reliability of SPH for cosmological simulations has recently been
called into question by \citet{Vogelsberger2011,Sijacki2011,Keres2011,Torrey2011,
Bauer2011}.  However, comparisons between SPH and calculations done with the
moving mesh code Arepo \citep{Springel2010}, show good agreement for applications
involving galaxy collisions, at least when the subresolution model mentioned
above is used to represent star-forming gas (Hayward et al. 2011, in prep).  The tests done
by e.g. \citet{Sijacki2011} indicate that SPH can fail when applied to 
situations in which gas
in very different phases are in motion relative to one another.  The use
of an effective equation of state to describe the ISM effectively circumvents this
issue because the different phases of the gas are not modeled explicitly.

\subsection{Initial Conditions}

The initial conditions for the construction of the LMC and SMC galaxies used for all models are outlined in 
Table~\ref{ch4table1}. As in B10, the total initial mass of the LMC is determined using  
current halo occupation models to relate the observed stellar mass of the LMC to its original
halo mass before infall into the MW halo \citep{guo2010}. 
Reflecting their stellar mass ratio, the SMC is then chosen to be 10 times less massive than the LMC.  
Consequently, the MCs are modeled here to have infall masses an order of magnitude larger than employed 
in previous models.
The number of particles of each component (gas, stars, dark matter) are chosen such 
that the mass resolution per particle of a given type is roughly the same in both galaxies. 

The SMC is modeled with an extended gaseous disk with a scale length 3 times that of the stellar component. Much larger
ratios are common for isolated dwarfs found in voids \citep{kreckel2011}, and
neutral hydrogen observations of SMC-like dwarfs with the Westerbork Synthesis Radio Telescope by \citet{swaters2002} find 
HI disk scale lengths ranging from 1.4-4.5 kpc \citep{Connors}. 
Our adopted scale length of 3.3 kpc is consistent with the upper end of the observed range. 

The LMC is modeled with gas and stellar disks with the same scale length, rather than with an extended gaseous disk. 
In reality, the interaction with the ambient hot gaseous halo of the MW would serve to truncate the LMC's extended
 gas disk.  The scale height of the stellar disk
 is taken as 0.2 of the disk scale length. The modeled scale height of the LMC's stellar disk is thus initially
 $z_0$= 0.34 kpc  (the observed value today is R$_{\rm disk}$ = 1.4 kpc and $z_0$= 0.27 kpc; van der Marel et al. 2002). 
  The gaseous disk height is determined by self-gravity and the pressurization of the ISM, as prescribed by the chosen 
  effective equation of state \citep{spring2005}.    
  
  The dark matter halos of the LMC and SMC follow Hernquist potentials \citep{Hernquist}. The scale radius for the Hernquist potential 
  ($r_H$) is related to the scale radius of the corresponding NFW halo ($r_S = R_{200}/C$) \citep{NFW} as described in \citet{spring2005}:
  
  \begin{equation}
  r_H = r_S \, \sqrt[]{ 2 \left ( {\rm ln}(1+C) - \frac{C}{1+C}  \right)}, 
  \end{equation}
  where $C$ is the concentration parameter.  Values for $C$, $r_S$ and $r_H$ are listed in Table ~\ref{ch4table1}. 
  
 The MW is modeled as a static NFW potential with a total mass of $1.5 \times 10^{12} \Msun$, $C=12$, 
  virial radius of $R_{vir}$ = 300 kpc, and $R_{200}$ = 220 kpc (radius where the average density is 200 times the critical
  density of the Universe).   As in B10, dynamical friction from the MW halo is not explicitly accounted for, but is expected to have little 
  impact on the orbit in a first passage (see Besla et al. 2007, Figure 4).  Dynamical friction between the MCs, on the other hand, plays a much more 
  important role in their orbital evolution and is captured explicitly by modeling these two galaxies with live dark matter halos.

 \begin{table*}\footnotesize
 \centering
 \begin{minipage}{210mm}
 \caption{L/SMC Initial Conditions \label{ch4table1}}
 \begin{tabular}{@{}lcc@{}}
 \hline
  Property & LMC & SMC \\
  \hline
  \hline
  M$_\ast$ ($\Msun$) &  $2.5 \times 10^{9}$ & $2.6\times 10^8$\tablenotemark{a} \\  
 M$_{gas}$ ($\Msun$)  & $1.1 \times 10^{9} $ & $7.9\times 10^8$ \\ 
 f$_{gas}$ \tablenotemark{b} & 0.3 &  0.75\\
 M$_{total}$ ($\Msun$) \tablenotemark{c}& $1.8\times 10^{11} $ & $2.1 \times 10^{10}$ \\ 
 $R_{200}$ (kpc)\tablenotemark{d} &  117.1  &  57.1  \\ 
 $C$  &  9 & 15 \\
 $r_S$ (kpc)\tablenotemark{e} & 13.0 & 3.8 \\
 $r_H$ (kpc)\tablenotemark{e} &  21.4   &   7.3  \\
 Stellar Disk scale length (kpc) &  1.7 & 1.1\tablenotemark{a} \\
 Gas Disk scale length (kpc) &  1.7 & 3.3  \\
 Gravitational Softening Gas/Stars (kpc)  & 0.1 & 0.1 \\ 
 Gravitational Softening Halo (kpc) &  0.29 & 0.29 \\ 
 Nstars  & 10$^6$ & 10$^5$\\
 Ngas & $3 \times 10^5$ & $3 \times 10^5$\\
 Nhalo & 10$^5$ & 10$^4$ \\
 q\tablenotemark{f} &  0.3  & 0.3 \\ 
  \hline 
  \tablenotetext{a} {Note that the initial stellar mass and stellar disk scale length chosen for the SMC deviate from the values adopted in B10. Here the disk is 
  chosen to be more extended in order to increase the number of stars removed by LMC tides. Other changes in parameter values are minor and 
  reflect attempts to match various observed mass constraints for the MCs (see Table~\ref{ch4table:Model}).} 
   \tablenotetext{b}{The gas fraction relative to the total disk mass (stars + gas). The gas fractions of isolated dwarf galaxies are known to be large, e.g. \citet{geha2006} }
\tablenotetext{c}{The total mass of the LMC/SMC at infall is determined using the observed stellar mass of the LMC(SMC) 
M$_\ast$ =  $3 \times 10^{9} \Msun$ ($3 \times 10^{8}\Msun$) 
\citep{vanderMarel, Stan2004} and the relations from \citet{guo2010}. The total halo mass used to define the Hernquist dark matter profile
 is then M$_{halo}$ = M$_{total}$ - M$_\ast$ - M$_{gas}$.}
\tablenotetext{d}{The radius where the average enclosed density is 200 times the critical density of the universe}
\tablenotetext{e}{The scale radius for the NFW profile ($r_S$), which is used to define the scale radius of the Hernquist profile ($r_H$), following \citet{spring2005}.}
\tablenotetext{f}{The effective equation of state parameter, q, defines the pressurization of the ISM following \citep{spring2005}.}
\end{tabular}
 \end{minipage}
  \end{table*}

\begin{figure*}
\begin{center}
\mbox{
\includegraphics[width=2.5in]{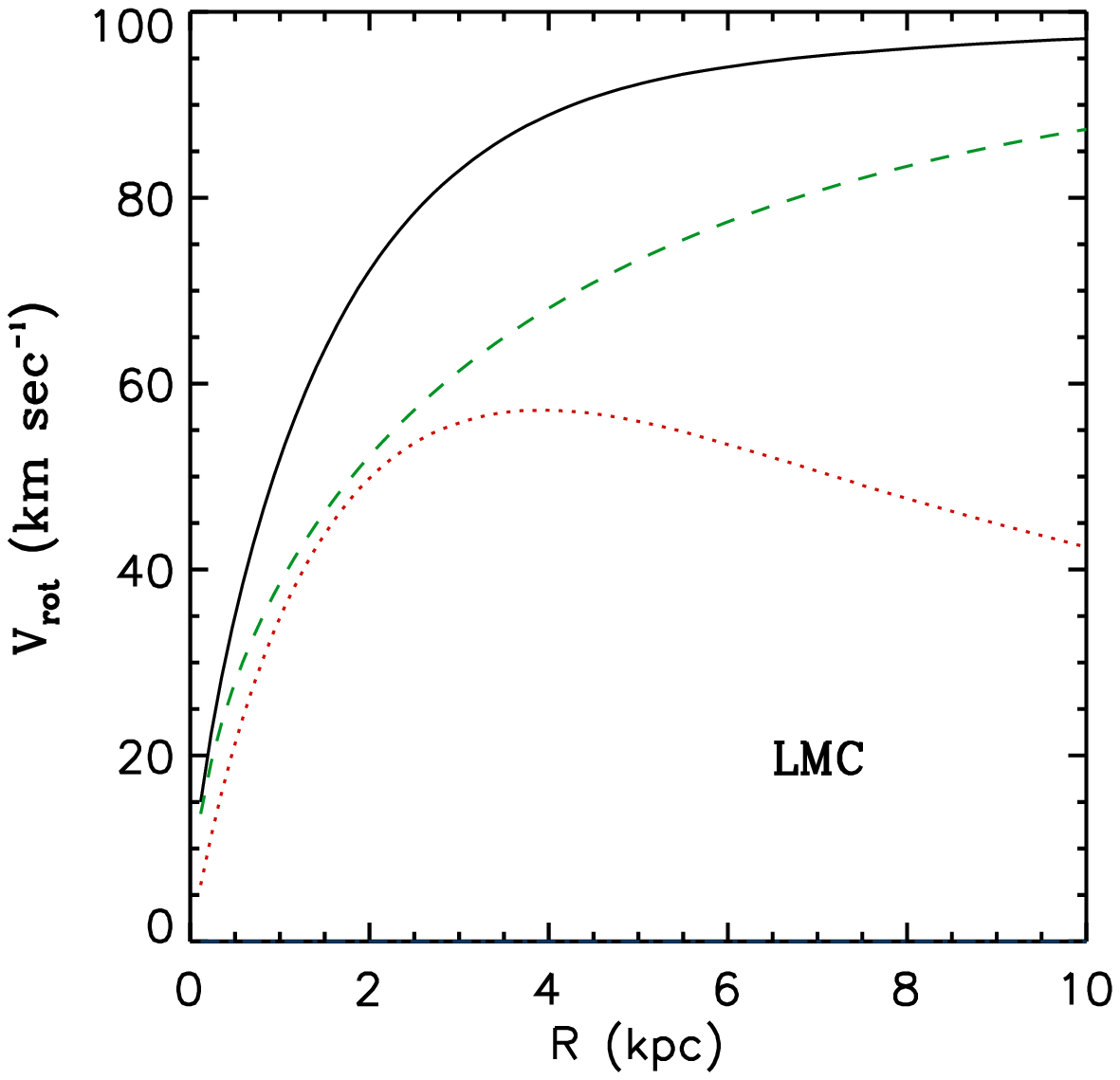}
 \includegraphics[width=2.5in]{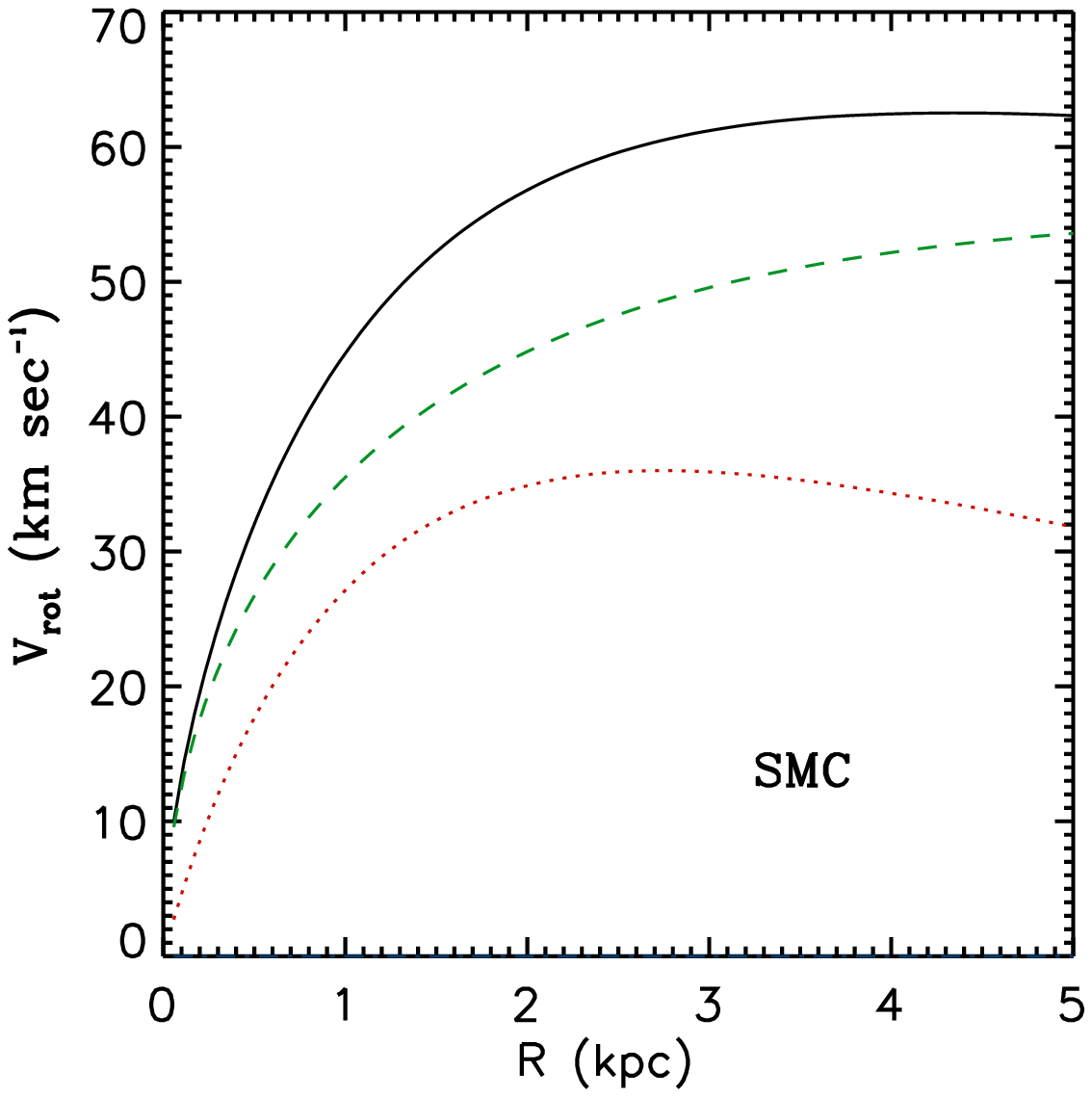}}
 \end{center}
 \caption{\label{ch4fig:RotCurve} The initial rotation curve is plotted for the LMC (left) and SMC (right). The different 
 lines indicate the contribution from the dark matter halo (green, dashed) and the disk (red, dotted).  The solid 
 black line indicates the total rotation curve. }
 \end{figure*}

The resulting rotation curves for the MCs are plotted in Figure ~\ref{ch4fig:RotCurve}. 
The initial SMC rotation curve peaks at $V_{rot} = 60$ km/s at 3 kpc from the center, as expected from HI kinematics \citep{Stan2004}; 
the SMC is initially a well-behaved disk galaxy.   
The initial simulated LMC rotation curve peaks at $V_{rot} = 95$ km/s, which is within the observed range \citep{Staveley, Olsen, vanderMarel}. 

\subsection{Orbit of the SMC about the LMC and Definition of Models}

Following the method outlined in B10, the MCs are evolved as an isolated
interacting binary pair over a period of $\sim$7 Gyr since the SMC first crossed within $R_{200} = 117$ kpc of the LMC. 

The SMC is placed on an eccentric orbit about the LMC (ecc = 0.7).
Higher orbital eccentricities for the SMC result in fly-by encounters between the MCs, while lower values cause the SMC's
orbit to decay too quickly.   

The simulation is stopped at characteristic points in time, defining two models for the orbital history of the 
SMC about the LMC, referred to as Model 1 and Model 2.  Model 1 is stopped after 5.1 Gyr and Model 2 after 5.9 Gyr.
 Thus, Model 1 and 2 differ based on the number of passages the SMC has completed about the LMC. 
 In Model 1, the SMC has completed 2 passages about the LMC, 
  whereas in Model 2 it has completed 3. 
 The stopping times are chosen such that 1 Gyr after this time, the LMC
 will have travelled from a distance of 220 kpc ($R_{200}$ for the MW) to its current location and the SMC will have completed
 the desired number of orbits about the LMC. 
 
  The choice of these two models is motivated by the overarching goal of this study to assess the role of 
  interactions between the MCs to their evolution.  Given the chosen L/SMC mass ratio of 1:10, it is unlikely that the SMC could have survived 
  more than 3 passages about the LMC, making Model 2 a maximal interaction scenario.  Model 1 is 
  very similar to the solution presented in B10 - the analysis of such a model is a direct extension of the B10 work.  
  Tidal forces between the MCs have been acknowledged as playing an important role in the formation of the Magellanic Stream in many previous studies. 
In fact, in \citet{Connors2006}, the tidal force from the LMC on the SMC dominates over MW tides for most of the SMC's orbit. But to explain the Leading Arm Feature and extent 
of the Magellanic Stream, MW tides have been invoked in all of these studies. Instead, here and in B10 we illustrate how such extended structures can form without requiring the MCs to 
complete an orbit about the MW.
  
\begin{figure*}
\begin{center}
\mbox{
\includegraphics[width=2.5in]{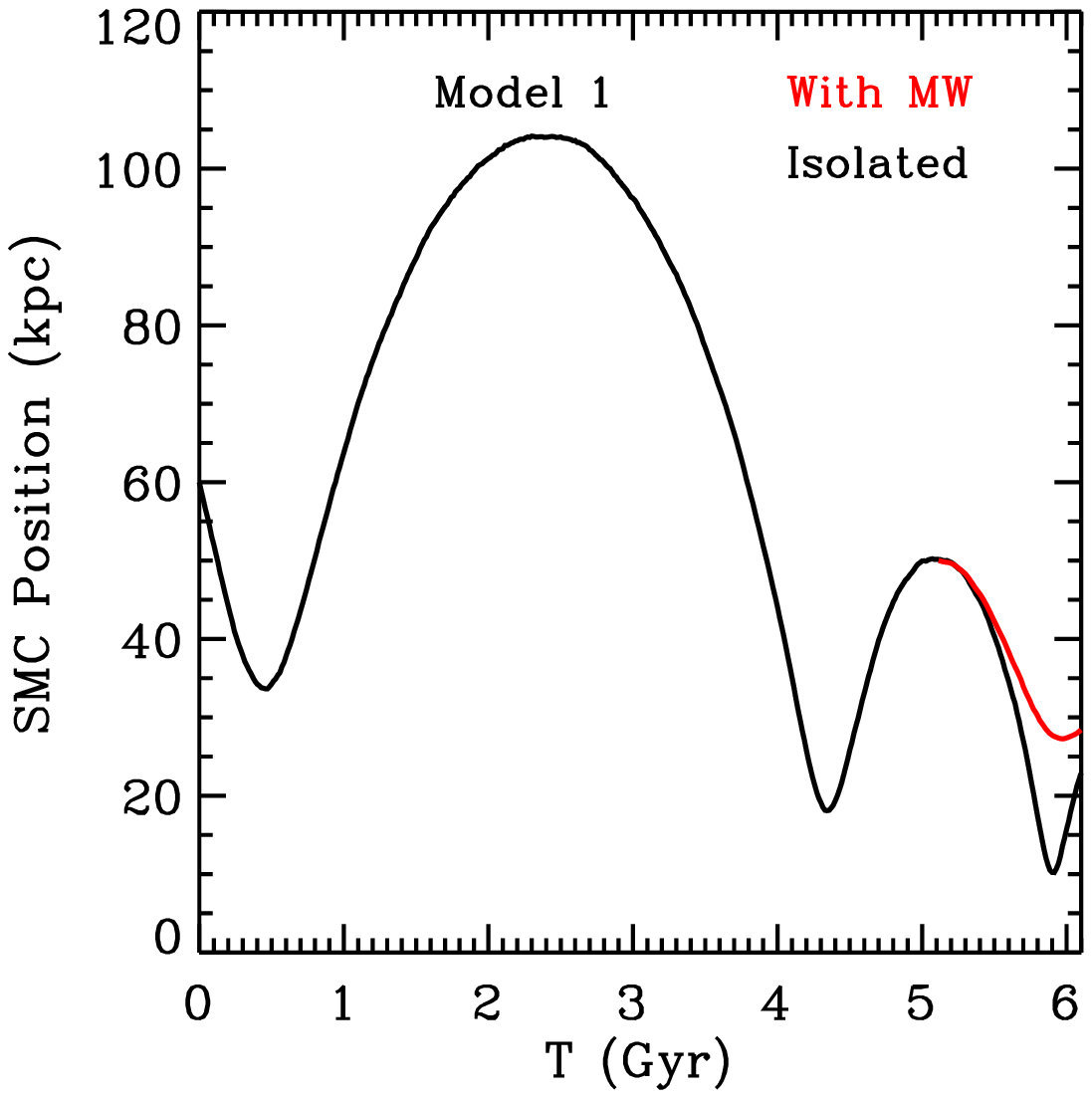}
 \includegraphics[width=2.5in]{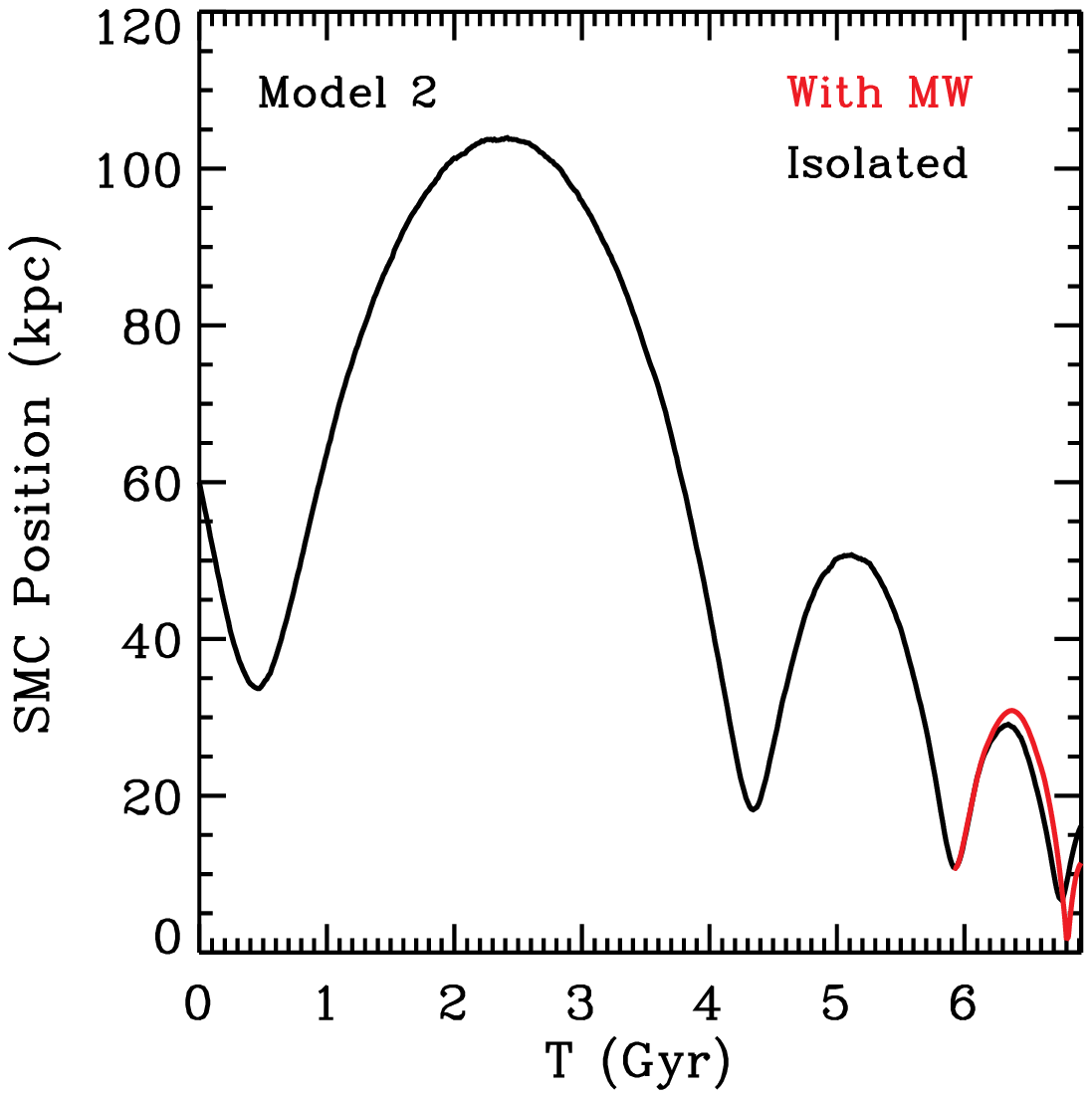}} \\ 
 \mbox{
 \includegraphics[width=3in, clip=true, trim=0 3in 0 2.5in]{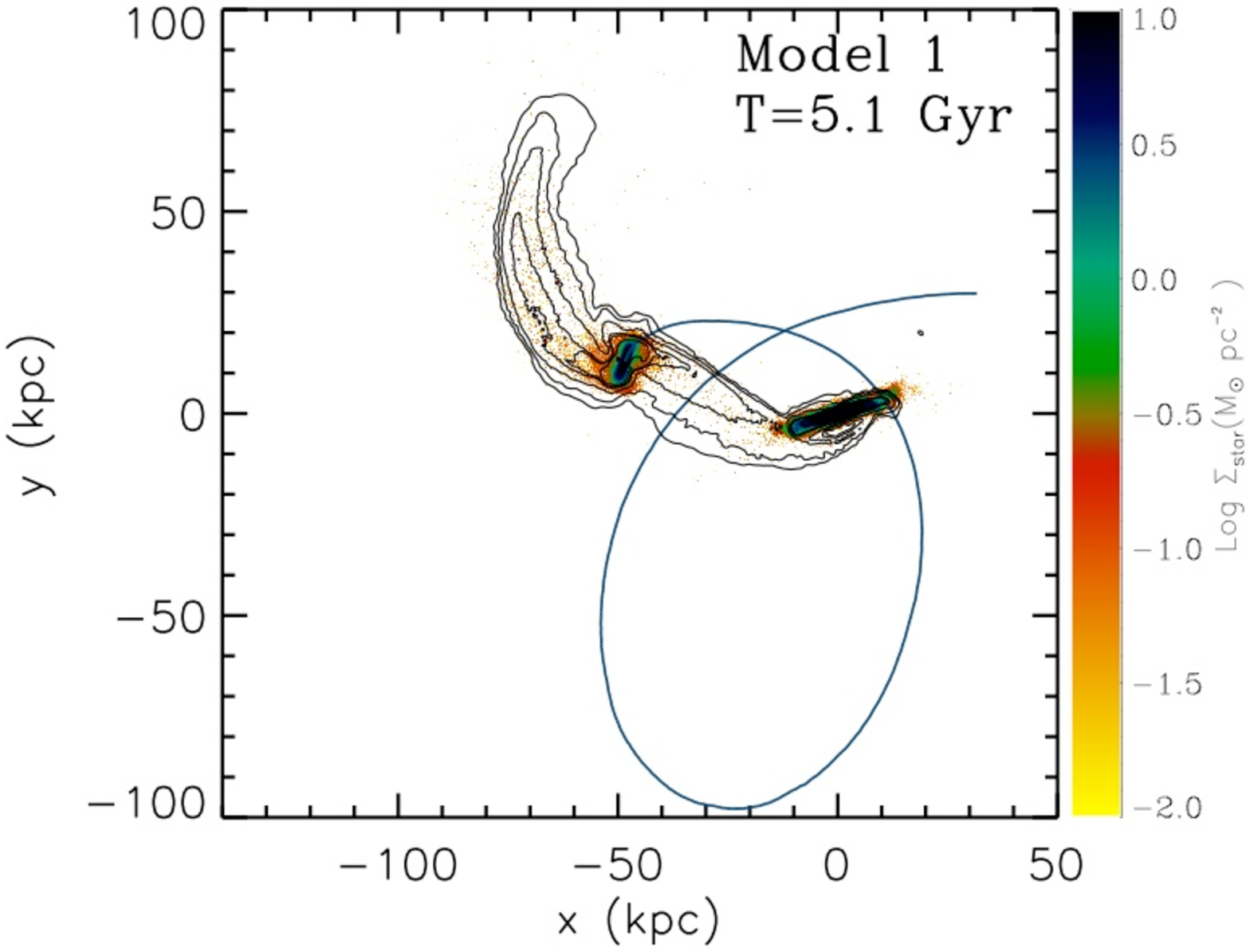}
 \includegraphics[width=3in, clip=true, trim=0 3in 0 2.5in]{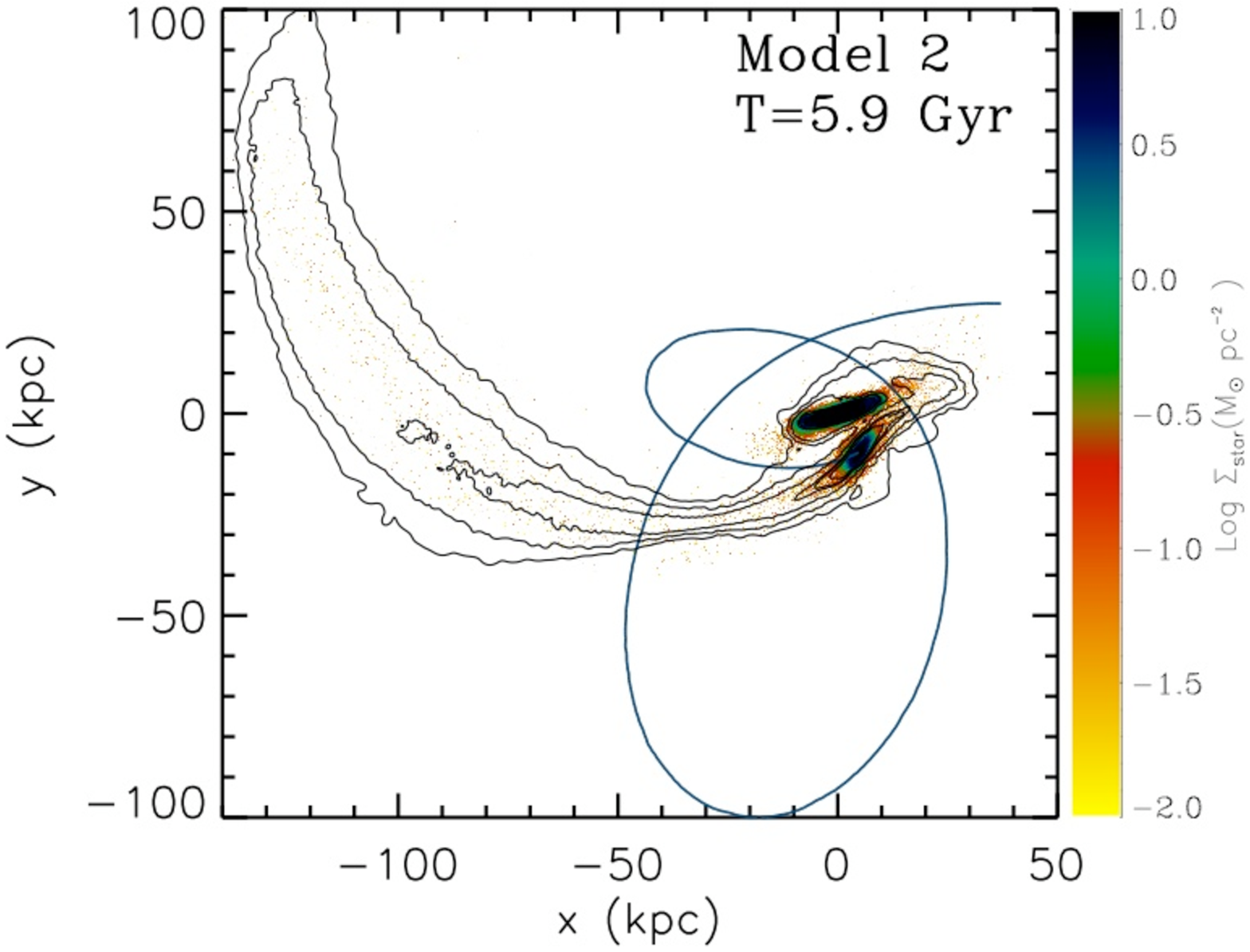}}
 \end{center}
 \caption{\label{ch4fig:SLOrbit} {\it Top Panel:} The separation of the SMC from the LMC is plotted as a function of time for Model 1 (left) 
 and Model 2 (right). The black line denotes the orbital history when the 
 LMC and SMC are modeled as an isolated binary pair: the MW's
 potential is not included.  The red line indicates how the separation between the MCs is modified after the pair first cross within $R_{200} = 220$ kpc
 of the MW. This occurs after 5.1 Gyr for Model 1 and after 5.9 Gyr for Model 2;  these times are referred to as the stopping times for 
 each model. 
 {\it Bottom Panel:}  The gas column density of the isolated LMC-SMC system is projected in the binary orbital plane as contours over the 
 stellar distribution at the stopping time for Model 1 
 (left) and Model 2 (right). Gas contours span a range of $10^{18}-10^{20}$ cm$^{-2}$, where each contour represents an increase
 in column density by a factor of 1.5. 
  These images thus depict the gas distribution of the simulated system {\it before} it is influenced by the MW's gravitational potential. 
 An extended tail of gas is stripped from the SMC and a bridge of gas connects it to the LMC in both models (see also Figure 1 of B10).  }
 
 \end{figure*}

The orbit of the SMC about the LMC in Model 1 and Model 2 is plotted in the top panel of Figure~\ref{ch4fig:SLOrbit}. 
The black line indicates the evolution of the system in isolation (no MW potential) and is continued 1 Gyr past the respective
stopping point for each model.  
The red line shows how the orbit of the SMC is modified if instead the binary pair is captured by the MW after the stopping point. 

In Model 1, MW tides work to keep the LMC and SMC further apart than they would have been in isolation. In Model 2 the 
opposite occurs; the MW's gravitational pull forces the SMC to collide directly with the LMC.  The distinct outcomes occur  
because of differences in the SMC's separation from the MW relative to its separation from the LMC at the pericenter of
 its orbit about the MW.  At pericenter,  the SMC is closer to the LMC in Model 2 than in Model 1
 (see Figure~\ref{ch4fig:LSMWOrbit}).

The bottom panels of Figure~\ref{ch4fig:SLOrbit} show the gas distribution of the L/SMC at the stopping points for each model
as contours plotted over the stellar distribution. 
After 5.1 Gyr (stopping point for Model 1) the SMC is at the apocenter of its second orbit about the LMC. 
After 5.9 Gyr (stopping point for Model 2) the SMC is just completing its second orbit.  As outlined in B10, a tidal bridge 
and extended tail forms as a result of the action of LMC tides on the SMC; these features form before the system is captured 
by the MW.

\subsection{Orbit of the MCs about the MW}

\begin{figure*}
\centering
\mbox{{\includegraphics[width=3in, clip=true, trim=0.5in 2in 0.5in 2in]{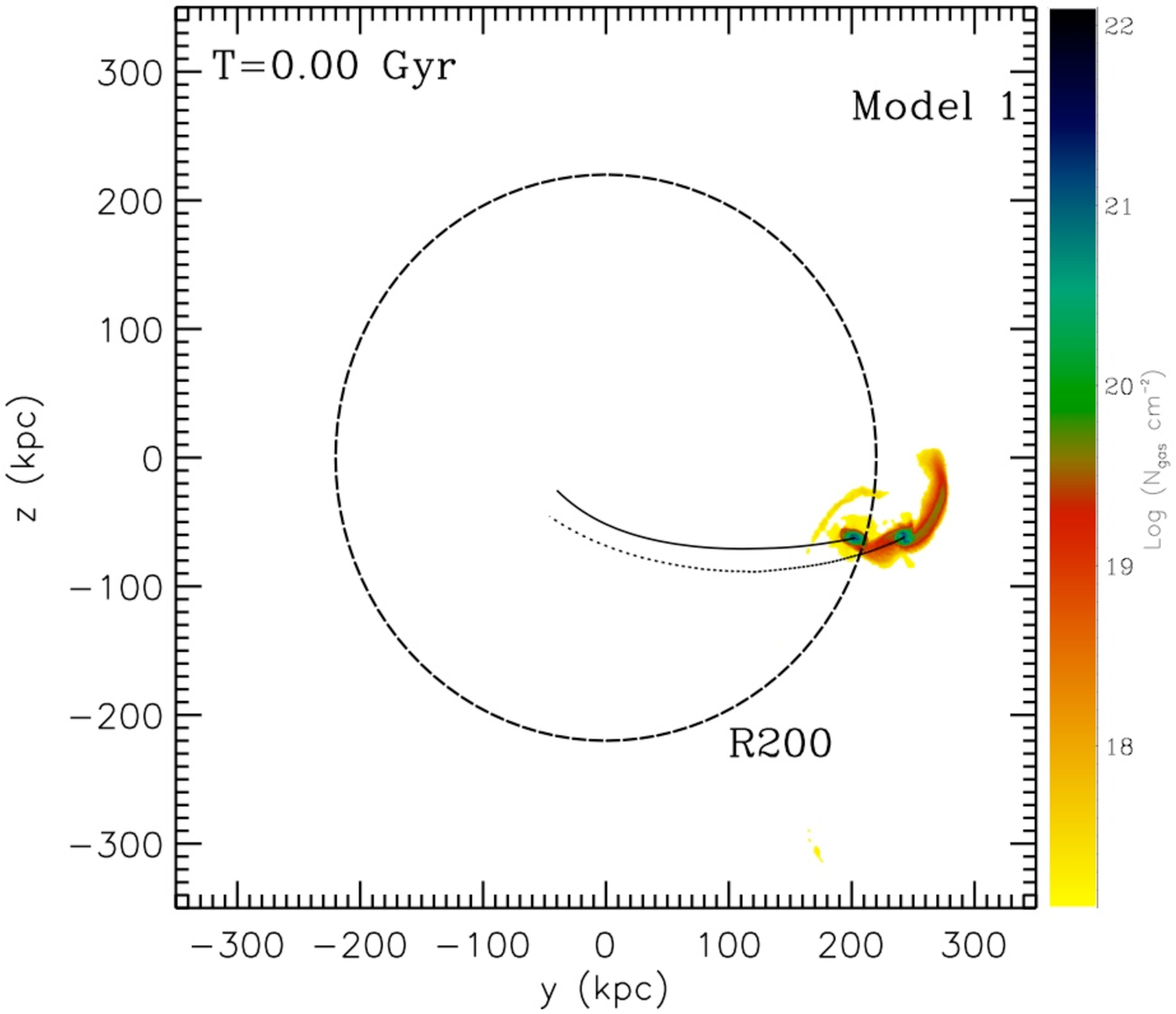}}
{\includegraphics[width=3in, clip=true, trim=0.5in 2in 0.5in 2in]{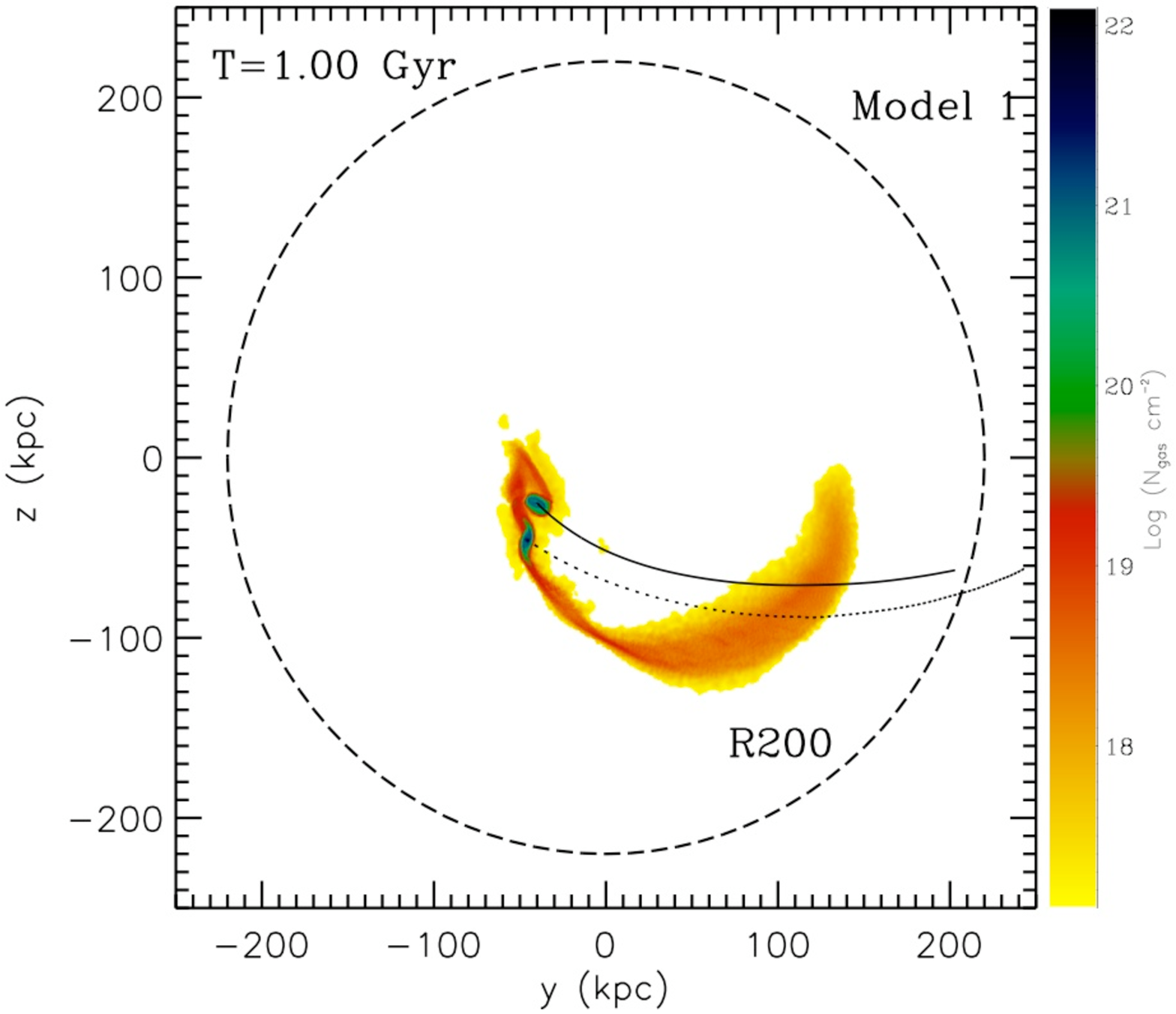}}}\\
 \mbox{
 {\includegraphics[width=3in, clip=true, trim=0.5in 2in 0.5in 2in]{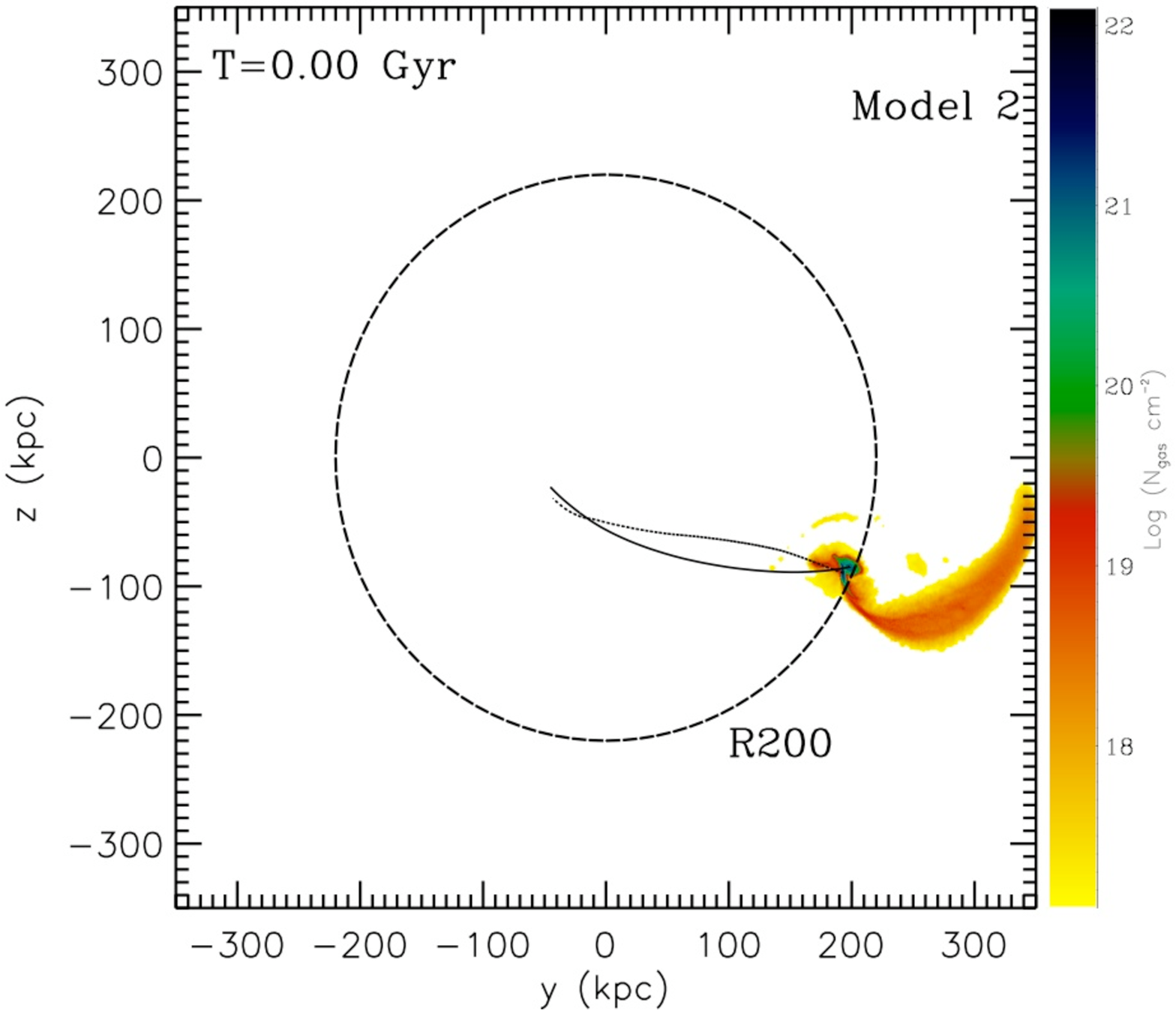}}
 {\includegraphics[width=3in, clip=true, trim=0.5in 2in 0.5in 2in]{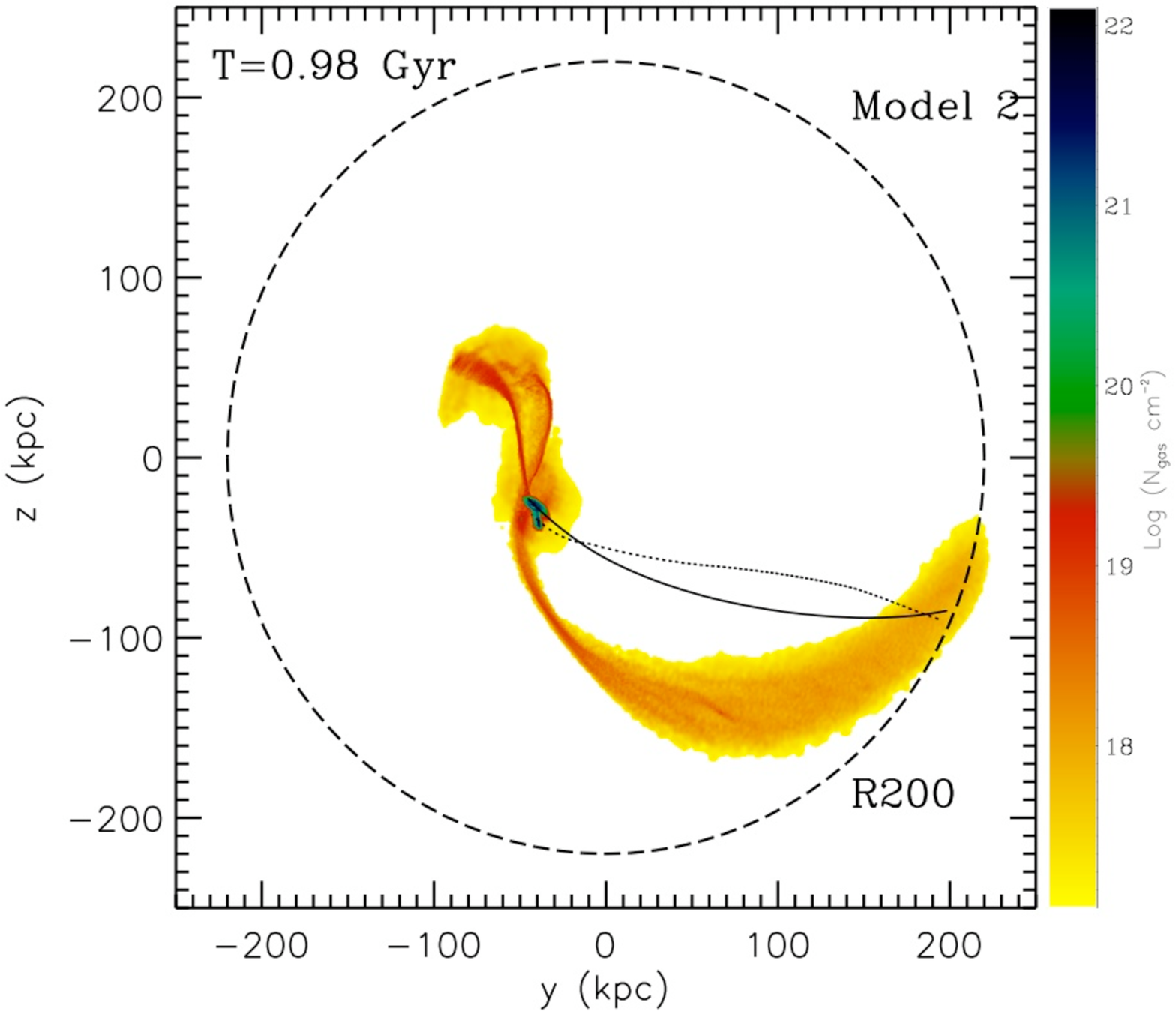}}}
 \caption{\label{ch4fig:YZplane}  Projected gas column densities in the YZ galactocentric plane for the simulated Magellanic System when the LMC 
 first crosses within $R_{200}$ of the MW (time=0, left) and today (time$\sim$1.0 Gyr, right).  The circle indicates the location of $R_{200} = 220$ kpc.
  The results for Model 1 are plotted in the top row and 
 Model 2 is on the bottom. The LMC's(SMC's) orbital path is denoted by the solid(dashed) line.}  
 \end{figure*}

At the stopping time (5.1 Gyr for Model 1 and 5.9 Gyr for Model 2), the isolated MC pair is placed at  
$R_{200} = 220$ kpc from MW's galactic center, as illustrated in Figure ~\ref{ch4fig:YZplane} (left panels). 
$R_{200}$ is chosen as the starting radius because the MW tidal field does not distort the orbit of the SMC relative to the isolated orbit 
until well within that radius; it takes 500 Myr for the red and black lines in the top panels of Figure~\ref{ch4fig:SLOrbit} to deviate
 after the stopping point (i.e. 500 Myr 
after they cross $R_{200}$). As such, the overall interaction history of the MCs is well-described by the isolated system 
before this point.

The galaxies travel to their current locations on orbits consistent with the HST proper motions for the LMC, 
as indicated in Figure ~\ref{ch4fig:YZplane} (right panels). This takes 1 Gyr in both models since the 
LMC's orbit about the MW is roughly the same in both cases - it is the SMC's orbit that differs.   

\begin{figure*}
\begin{center}
\mbox{{\includegraphics[width=2.5in, clip=true, trim= 0.0 0.6in 0 0.0in]{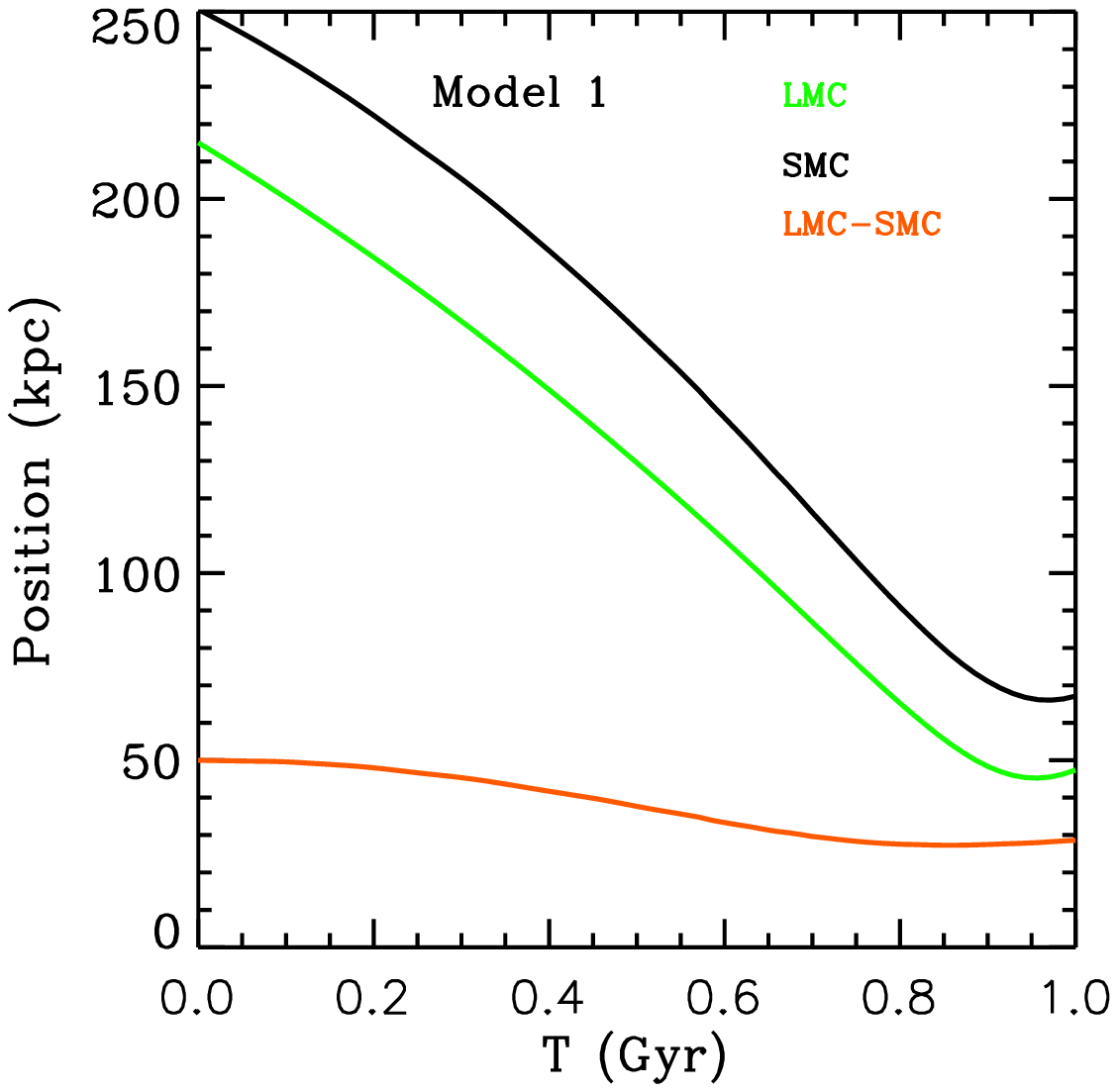}}
{\includegraphics[width=2.03in, clip=true, trim= 0.9in 0.6in 0 0.2in]{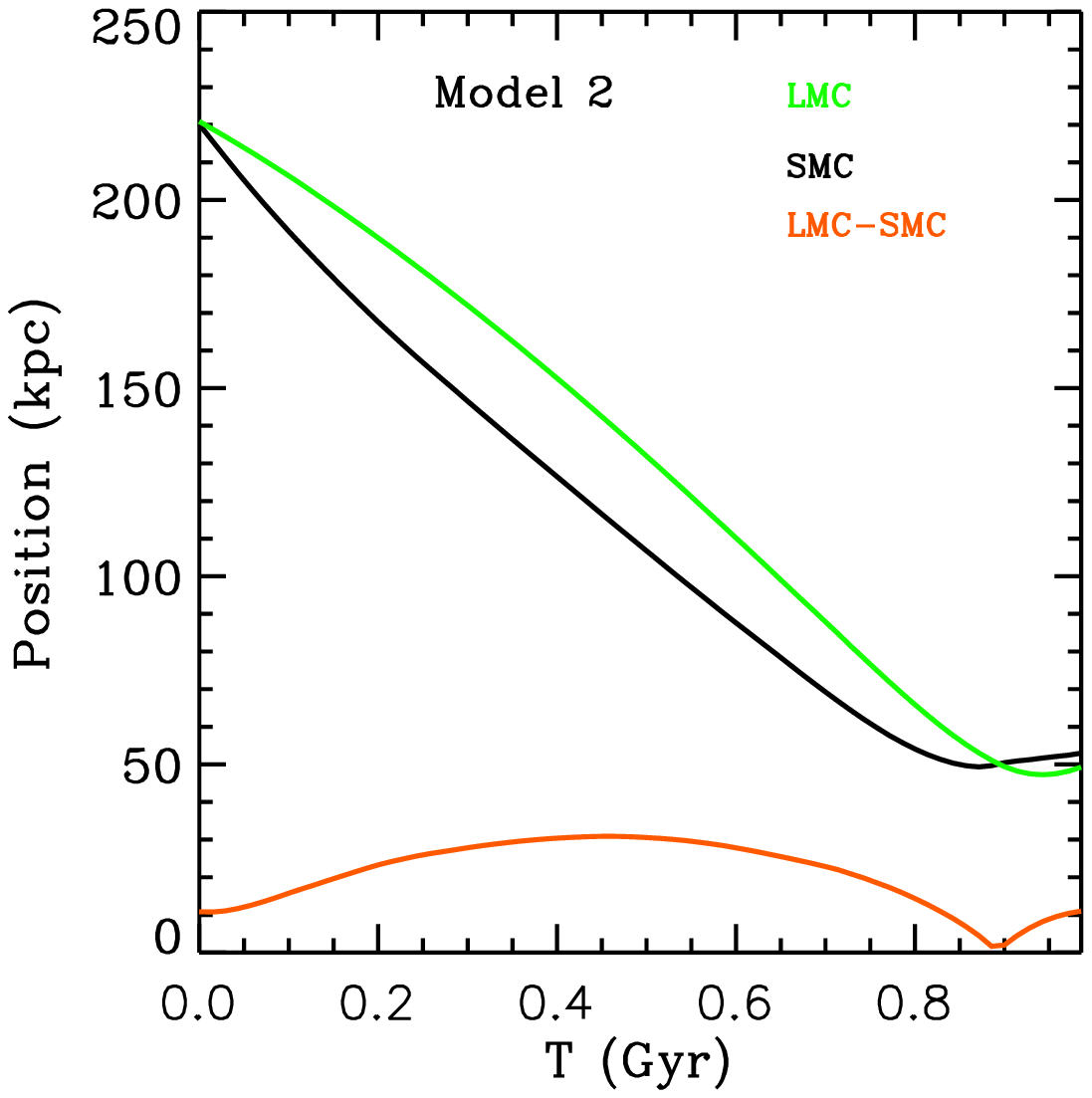}}}\\
 \mbox{
 {\includegraphics[width=2.5in, clip=true, trim= 0.5 0.1in 0.0 0.0in]{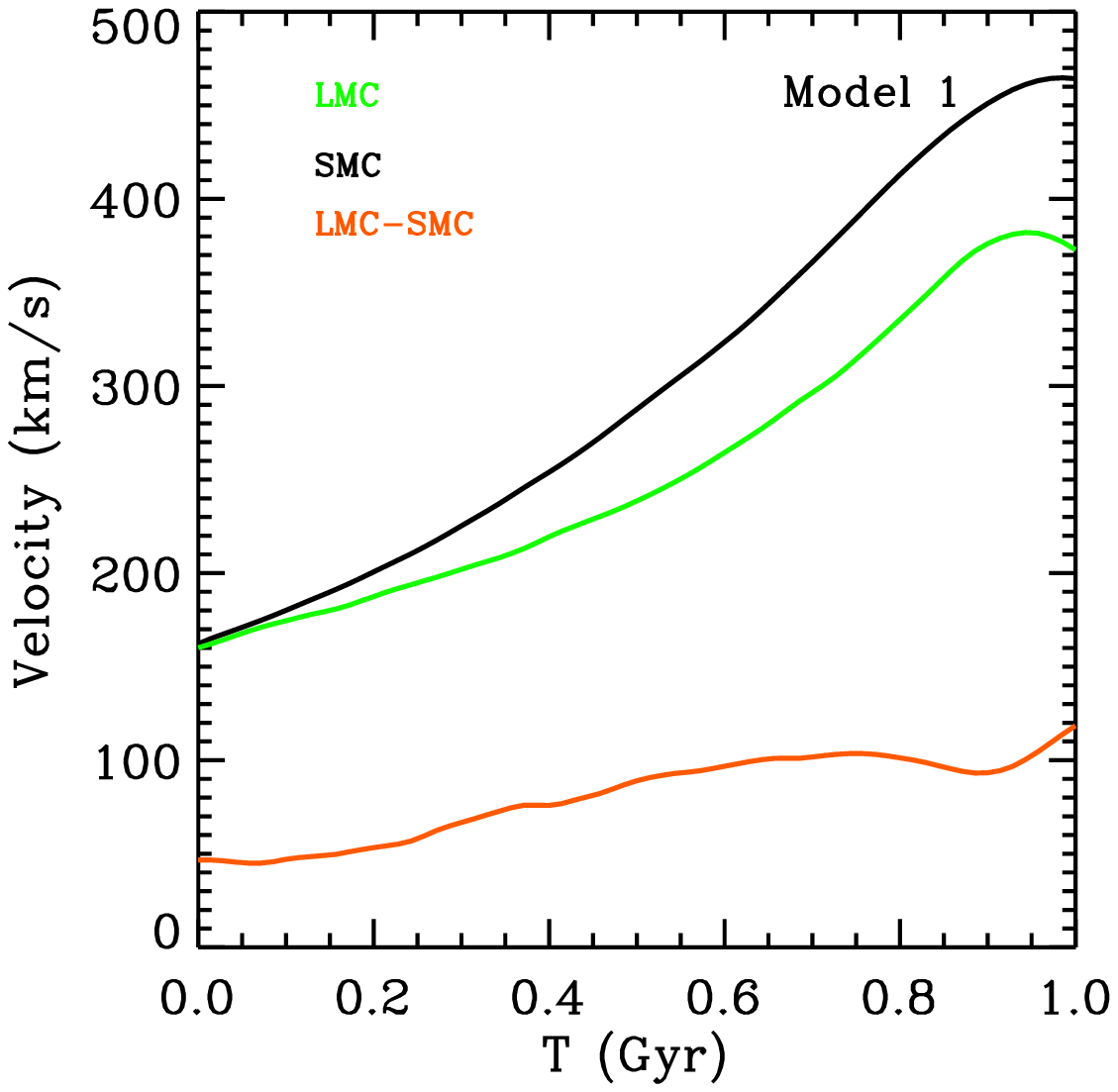}}
 {\includegraphics[width=2.03in, clip=true, trim = 0.94in 0.1in 0 0.2in]{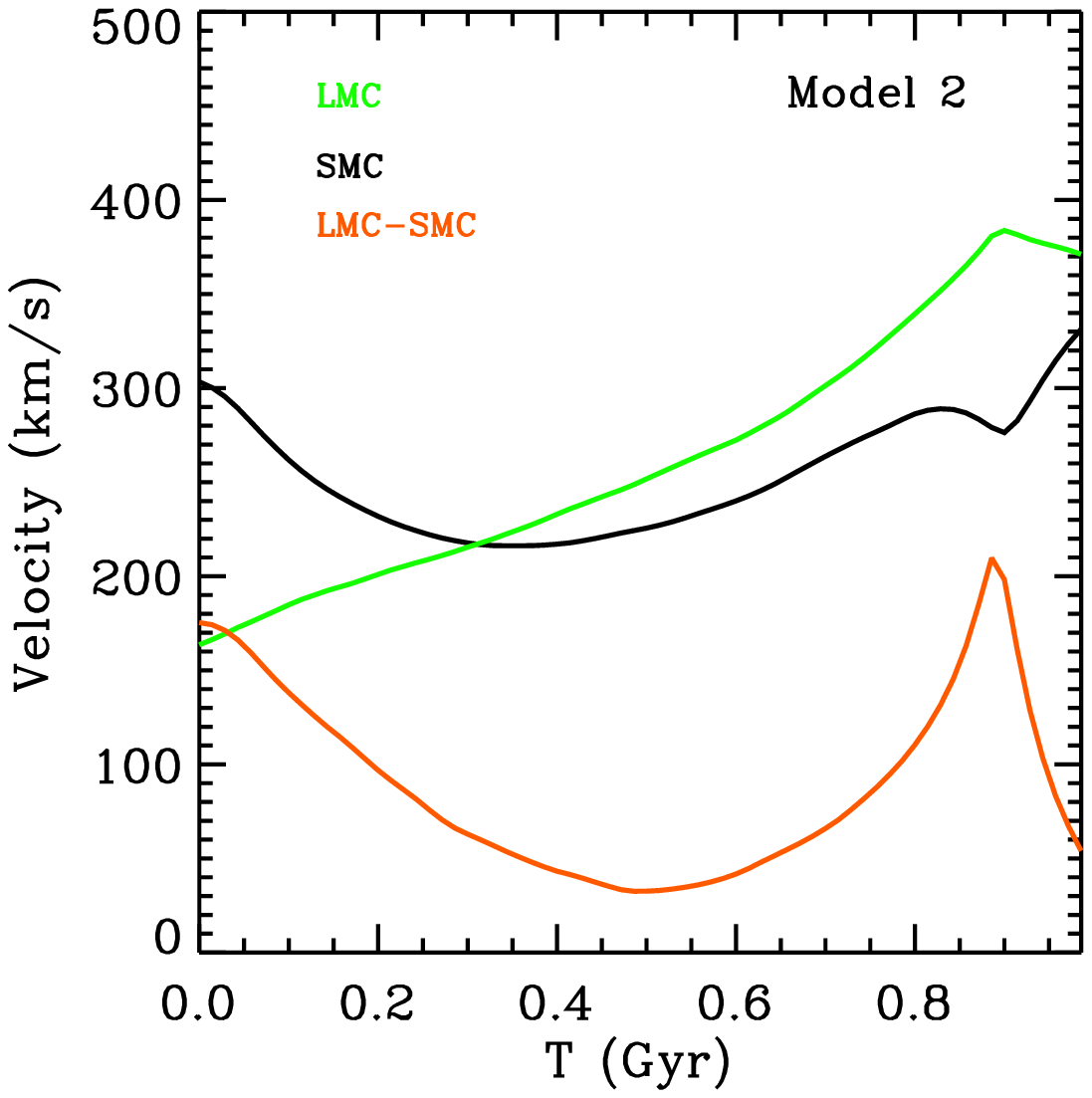}}}
 \caption{\label{ch4fig:LSMWOrbit}  The Galactocentric position (top) and velocities (bottom) of the LMC (green) and SMC (black)  are
  plotted as a function of time since the MCs first crossed within R$_{200}$ = 220 kpc of the MW. 
 The relative separation and velocity between the MCs is plotted in orange. The results of Model 1 are plotted in the left column
 and those of Model 2 are in the right column.  The Galactocentric velocities determined by 
 \citep[hereafter K1 and K2]{Nitya1, Nitya2} are 378 $\pm$ 18 km/s for the LMC and 
 302 $\pm$ 52 km/s for the SMC (errors quoted are 1 $\sigma$).  
  In Model 1, the velocity of the SMC is higher than observed. In Model 2, the galactocentric position of the SMC is too small (53 kpc vs. 60 kpc); consequently the relative separation between the MCs is also too small (11 kpc vs. 23 kpc).  Note that in Model 2, the separation between the LMC and SMC approaches zero $\sim$ 100 Myr ago, indicative of
  a direct collision. }
 \end{center}
 \end{figure*}

The Galactocentric position and velocities of the MCs are plotted in Figure~\ref{ch4fig:LSMWOrbit}
 as a function of time since they first crossed within $R_{200}$ of the MW.  
 Also plotted are the relative positions and velocities between the MCs (orange line). 
 
 In Model 2 the SMC completes an 
 additional passage about the LMC since entering the virial radius, versus in Model 1. This additional passage
 results in a direct collision between the MCs and the formation of a new bridge. Tidal bridges and tails are formed at each 
 pericentric passage of the SMC about the LMC \citep{TT72}. 
 Thus, in Model 2, the bridge connecting the MCs will have formed $\sim$100 Myr 
 ago, during this direct collision (separation approaching zero).

\subsection{Comparison of Modeled Orbits with Data} 

\begin{figure*}
\begin{center}
\mbox{
\includegraphics[width=3in]{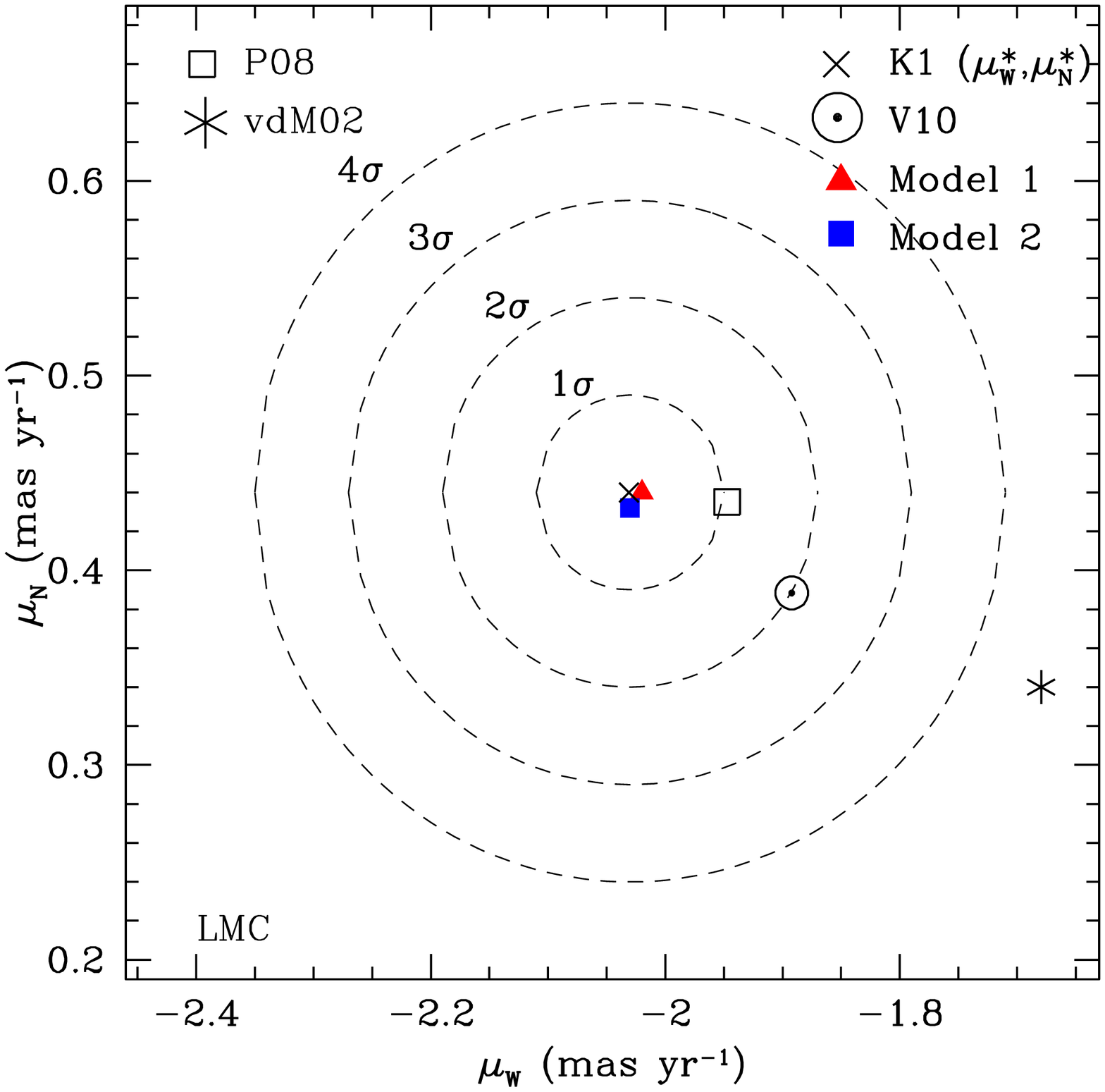}}
 \end{center}
 \caption{\label{ch4fig:PM} Concentric circles indicate the 4$\sigma$ error space for the K1 proper motion error space
 for the LMC (where the mean value is indicated by the X). 
 The asterisk indicates the mean of all proper motion estimates for the LMC prior to 2002 \citep{vanderMarel}. 
 The circled dot indicates the recent proper motion estimate by \citet{vieira2010}  
 and the open square shows the reanalysis of the K1 proper motion data by \citet{piatek2008}.  The red triangle and blue square 
  indicate the proper motion of the LMC today in Model 1 and 2, respectively. These values were chosen to closely match the K1 data.    }
 \end{figure*}

For each model, the initial velocities and positions of the LMC and SMC at $R_{200}$ = 220 kpc from the MW and their final values today 
are summarized and compared to data from \citet{Nitya1} (hereafter K1)  and \citet{Nitya2} (hereafter K2)  in Table ~\ref{ch4table:Model}. 
The proper motion error space for the K1 HST proper motion measurements of the LMC is indicated in Figure ~\ref{ch4fig:PM}. 
Over-plotted are various other measurements for the LMC's proper motion and the simulated Model 1 and 2 results.  
 The final LMC velocities and positions are designed to be within 1$\sigma$ of the observations in both models.  The differences between the two models 
 are the orbital parameters for the SMC. 
 
In Model 1 the SMC velocity is significantly larger than that indicated by the HST proper motions of K2.  In Model 2, the SMC velocities 
are in better agreement, however the separation between the LMC and SMC is smaller than observed (by about 10 kpc).  As such, the line-of-sight 
velocity and proper motions for the SMC are also different than observed. 

 While the SMC velocities and
positions are not perfect matches to the observations, it is unlikely that significantly new insight would be gained as to the physical 
processes at work if an exact solution were found. 
Slightly different choices of orbital parameters and timing in the orbit can change the SMC's final position and velocity, but not the physical picture. 
This is practically illustrated by comparing the resulting large-scale gas distribution in Model 1 and 2 (see $\S$ ~\ref{sec:Stream}); despite 
differences in the SMC orbital properties, the same overall scenario has produced similar global features (i.e. a Leading Arm, Bridge and Stream).
 To match the exact properties of the Magellanic System, a more detailed study, varying orbital parameters, L/SMC mass ratios, MW mass, etc, is required;
 this is beyond the scope of the present study.

We have obtained another epoch of data with WFC3, resulting in an average time baseline of 7 years (Kallivayalil et al. in prep).
 These new data are expected to reduce the errors on the proper motions by a factor of 3, potentially narrowing parameter space the SMC's error space. 
  We note that, within this error space, the exact choice of LMC and SMC velocity today will not alter the 
 physical picture presented in this work, which is that 
 tidal interactions between the two Clouds are the main driver for their morphological and kinematic evolution. 

\begin{table*}
\centering
 \begin{minipage}{200mm}\footnotesize
\caption{Initial and Final Orbital Parameters: Model 1 and 2  \label{ch4table:Model} }
\begin{tabular}{@{}lcccccc@{}}
\hline
				& 							&    \multicolumn{2}{c}{ Model 1} 			 				&  \multicolumn{2}{c}{Model 2 }			& Observed Today \\
\hline
Galaxy 	&  Parameter  				& At R200   &  Today       & At R200  &    Today    &     K1 and K2 \\
\hline
\hline
LMC				&  (x,y,z) (kpc) 						&  (35, 203, -63)                    &  (-1, -40, -25 )              &  (48, 198,-85)                    &  (-1, -42, -26)                &  (-0.8, -41.5, -26.9)   \\
				&  (vx,vy,vz) (kpc)					&  (-14, -157, -29)                 &  (-72, -267, 250)          & (-17, -160, -29)                & (-82, -263, 249)           & (-87 $\pm$ 12, -268 $\pm$ 11, 252 $\pm$ 16) \\
				& V$_{\rm los}$ (km/s) 				&                                              &   262                              &                                            & 259                                & 262  $\pm$ 3.4 \\
				& PM (W, N) (mas/yr) 				&                                              &   (-2.02, 0.44)                &                                           &  (-2.03, 0.43)                &  (-2.03 $\pm$ 0.08, 0.44 $\pm$ 0.05) \\
SMC	\footnote{The SMC proper motions are not included in this table: since the line of sight velocities are not well-matched to the observations the proper motions can't be meaningfully compared to the data.}			& position (x,y,z) (kpc)				&  (5, 243, -62)                        & (18, -46, -46)               & (56, 193, -90)                  & (6, -39, -35)                  &  (15.3, -36.9, -43.3) \\
				& velocity  (vx,vy,vz) (kpc)				& (6, -146, -70)                        &  (-88, -384, 246)         & (-51, -289, 88)               & (-66, -258, 198)           &  (-87 $\pm$48, -247 $\pm$42, 149 $\pm$ 37) \\
				&  V$_{\rm los}$ (km/s) 				&                                                &  215			         &                                        &  201                                 & 146 $\pm$ 0.6 \\
\hline
\end{tabular}
\end{minipage}
\end{table*}

\section{Large Scale Gas Morphology}
\label{sec:Stream}

\begin{figure*}
\begin{center}
\mbox{\, {\includegraphics[width=2.5in, height=6.55in, angle=-90, clip=true, trim=1.6in 0.39in 2.8in 0.4in]{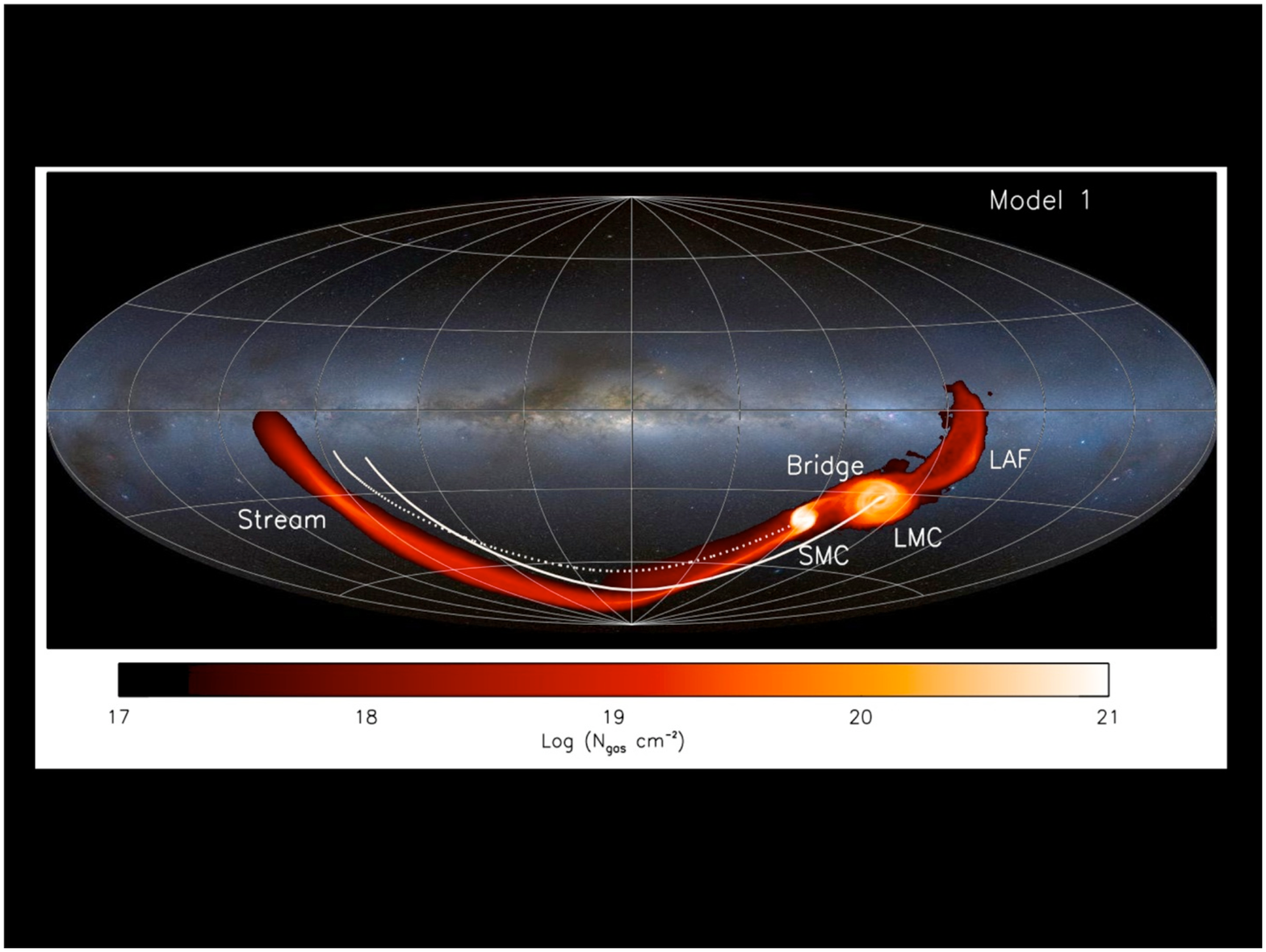}}}\\ 
 \mbox{ {\includegraphics[width=3.2in, height=6.55in, angle=-90, clip=true, trim=1.6in 0.39in 1.68in 0.4in]{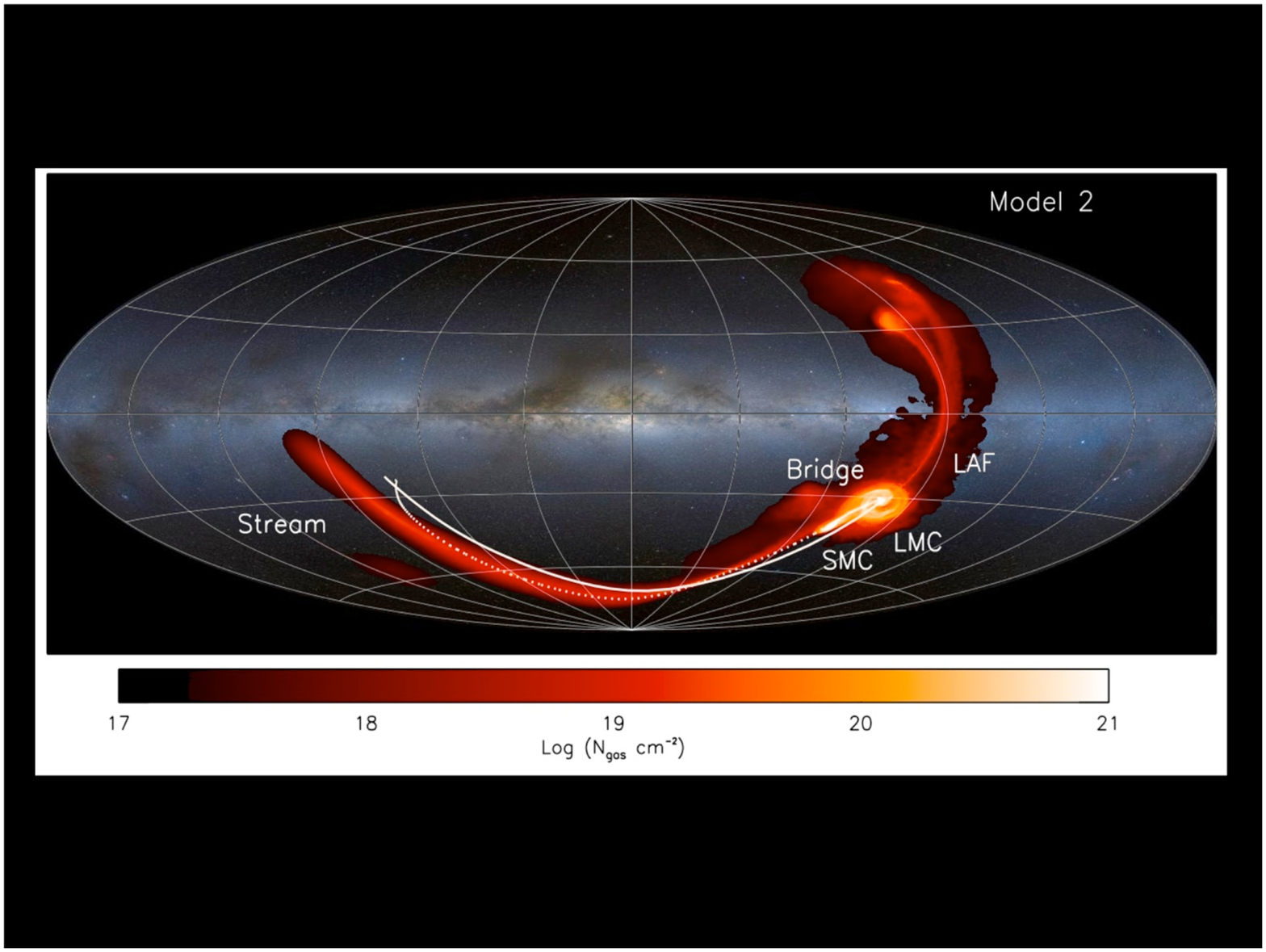}}}
 \end{center}
 \caption{\label{ch4fig:Aitoff}  Hammer-Aitoff projection of the total gas distribution of the simulated Magellanic System (red scale) for Model 1 (top) and Model 2 (bottom) 
 is plotted over an image of the MW  \citep[blue, white and brown colors; ][]{Mellinger}. 
 The orbital trajectory of the LMC(SMC) is indicated by the solid(dotted) white line.  Various components of the Magellanic System are labelled, where 
  LAF stands for Leading Arm Feature. }
 \end{figure*}

The resulting large scale gas distributions in Models 1 and 2 are shown in a Hammer-Aitoff projection in Figure ~\ref{ch4fig:Aitoff}.  
In both models, the final gas distribution can be described as an
 extended tail, a leading component and a bridge of gas connecting
 the two galaxies. As such, the main components of the Magellanic System are reproduced by both models.
 Moreover, in both cases the simulated stream stretches $\sim$150 degrees across the sky, 
as observed \citep{nidever2008}.

\begin{figure*}
\begin{center}
\mbox{{\includegraphics[width=6.5in, clip=true, trim= 0 4.65in 0 4.2in]{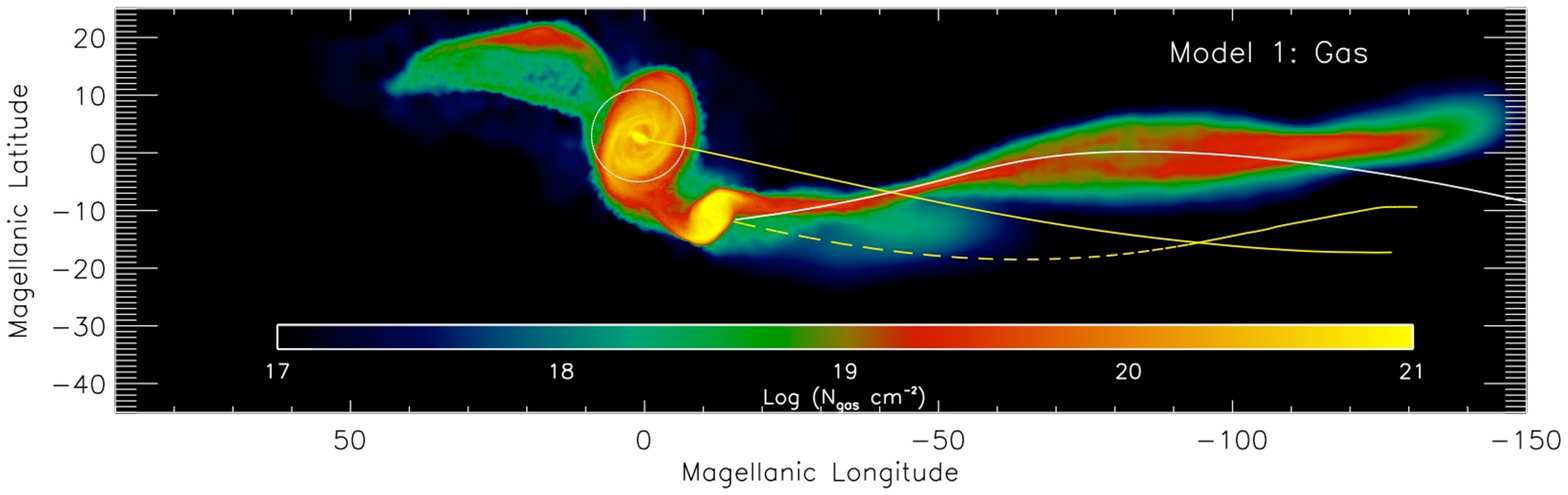}}}\\
 \mbox{ {\includegraphics[width=6.5in , clip=true, trim= 0 4.2in  0 4.2in]{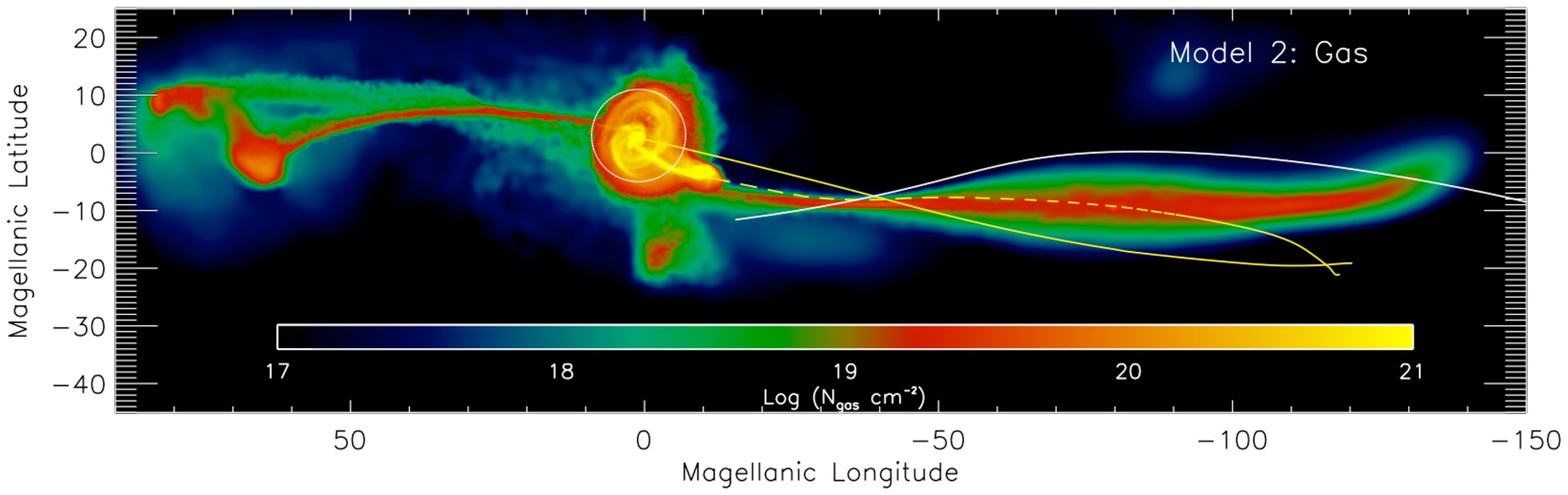}}}
 \end{center}
 \caption{\label{ch4fig:MS}  The total gas distribution of the simulated Magellanic System for Model 1 (top) and Model 2 (bottom) is plotted in 
 Magellanic Coordinates. 
 The orbital trajectory of the LMC(SMC) is indicated by the solid(dashed) yellow line. The actual location of the Magellanic Stream is
 roughly traced by the solid white line. The white circle represents the observed radius of the LMC.    }
 \end{figure*}

In Figure~\ref{ch4fig:MS} the simulated stream is plotted in Magellanic Coordinates, a variation of the galactic coordinate system where 
the Stream is straight \citep{nidever2008}.  
In both models, the simulated stream deviates away from the projected location of the past orbits on the plane of the sky, 
as expected according to the recent proper motions \citep[see e.g., Figure 8 in][]{besla2007}.
The deviation is a natural result of the proposed stream formation mechanism. It occurs largely because the Stream 
is removed in the binary LMC-SMC orbital plane by LMC tides. This binary plane is not parallel to the LMC-SMC-MW 
orbital plane, thus the Stream is not coincident with the orbit of the MCs about the MW.  
A second factor is the orientation of the SMC's disk; the location of the simulated stream can be modified by changing this angle. In both of these 
models the SMC disk is initially oriented 90 degrees with respect to the SMC-LMC orbital plane.
The deviation between the simulated stream and the orbits is more pronounced in Model 1 than in Model 2. However, this could be 
altered if the SMC disk were oriented differently initially and is not a physical distinction between the models; the magnitude of the offset is a tunable parameter. 
Note that this offset is not expected in a ram pressure solution for the Stream, as the material should be removed along the 
direction of motion (see Appendix~\ref{sec:Ram}). 

The structure of the Leading Arm Feature (LAF) is distinct in each model. 
 In Model 1 the LAF represents material that was stripped from the SMC on earlier passages 
and captured by the LMC. Since this material is bound to the LMC it does not extend further than 50 degrees (see Figure ~\ref{ch4fig:MS}). 
 In Model 2, the LAF is better described as a tidal tail or loop, resulting 
from tidal stripping of the SMC on its most recent orbit about the LMC. The tidal tail gains energy and deviates to larger 
distances away from the SMC's orbit  \citep[e.g., ][]{choi2007}. The resulting simulated LAF in Model 2 spans 80-90 degrees across the sky, which 
is larger than observed. 
 Unlike the Stream, this material is leading to the MCs and so will experience a significant ram pressure headwind. 
 Consequently, its final appearance, position and angular extent on the plane of the sky
 cannot be well captured without including hydrodynamic effects ~\citep[see Appendix~\ref{sec:Ram}, ][]{binney2011}. 
Model 2 does illustrate, however, that the observed 70 degree span of the LAF \citep{nidever2010} can be reproduced
without invoking a previous orbit about the MW. 

\begin{figure*}
\begin{center}
\mbox{{\includegraphics[width=6.5in]{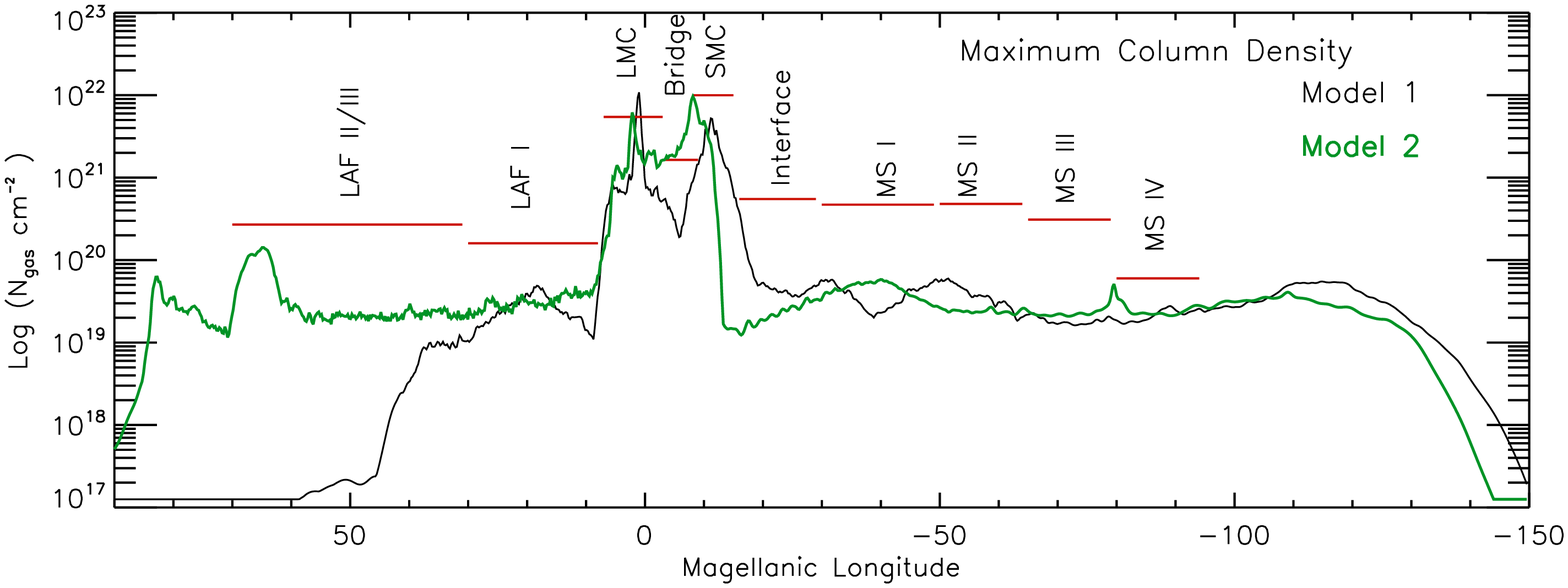}}}
 \end{center}
 \caption{\label{ch4fig:Column} The maximum total gas column density of the simulated Magellanic System is plotted as a function of Magellanic Longitude
 for Model 1 (black line) and Model 2 (green thick line). 
 The red lines indicate the observed maximum HI column density from the \citet{Bruens2005} data set for each marked region: 
  MS stands for Magellanic Stream and LAF for Leading Arm Feature. The various roman numerals refer to specific sections of the MS/LAF as
 defined in \citet{Putman2003} and \citet{Bruens2005}.  Neither model reproduces the observed HI column density gradient in the MS.}
 \end{figure*}

The Magellanic Stream is observed to have a pronounced HI column density gradient along its extent \citep{nidever2010, Putman2003, Bruens2005}. 
 The maximum gas column density along the simulated Magellanic System is determined from Figure~\ref{ch4fig:MS} and plotted as a function of 
Magellanic Longitude in Figure~\ref{ch4fig:Column}.  Both models underestimate the 
observed values, which are indicated by the solid red line \citep{Bruens2005}. 
There are a number of possible explanations for this discrepancy.  This problem could be alleviated if the gas reservoir in the SMC were depleted less 
efficiently at earlier times; for example, if the SMC's gas disk were initially less extended or if star formation was not quite so efficient. 
Ram pressure stripping has falso not been modeled and could also increase the amount of gas removed as the MCs get closer 
to the MW.  Furthermore, hydrodynamic instabilities are not well modeled with SPH \citep{Agertz2007,Sijacki2011}, 
and so clumping of the gas is not captured in these simulations. This is a process that will clearly influence the resulting gas column density in the Stream
 \citep{bland2007, nigra2010}. 

At the same time, there are notable consistencies between the models and the data. 
\citet{nidever2010} find that the column density along the LAF is fairly constant along its $\sim$70 degree span ($\sim4\times10^{19}$ cm$^{-1}$); this is true of Model 2. 
The column density in the bridge in Model 2 also matches the observations: the maximum column density in the Bridge is 
$1.64 \times 10^{21}$ cm$^{-2}$ \citep{Bruens2005};  the simulated bridge column density for Model 1 is too low. 
Also in Model 2 the simulated column density of the SMC is higher than that of the LMC, as observed \citep[the maximum column density is 
$5.45 \times 10^{21}$ cm$^{-2}$ and  $9.98 \times 10^{21}$ cm$^{-2}$ for the LMC and SMC, respectively]{Bruens2005}.

\begin{figure*}
\begin{center}
\mbox{{\includegraphics[width=6.5in, clip=true, trim = 0 0.6in 0 0]{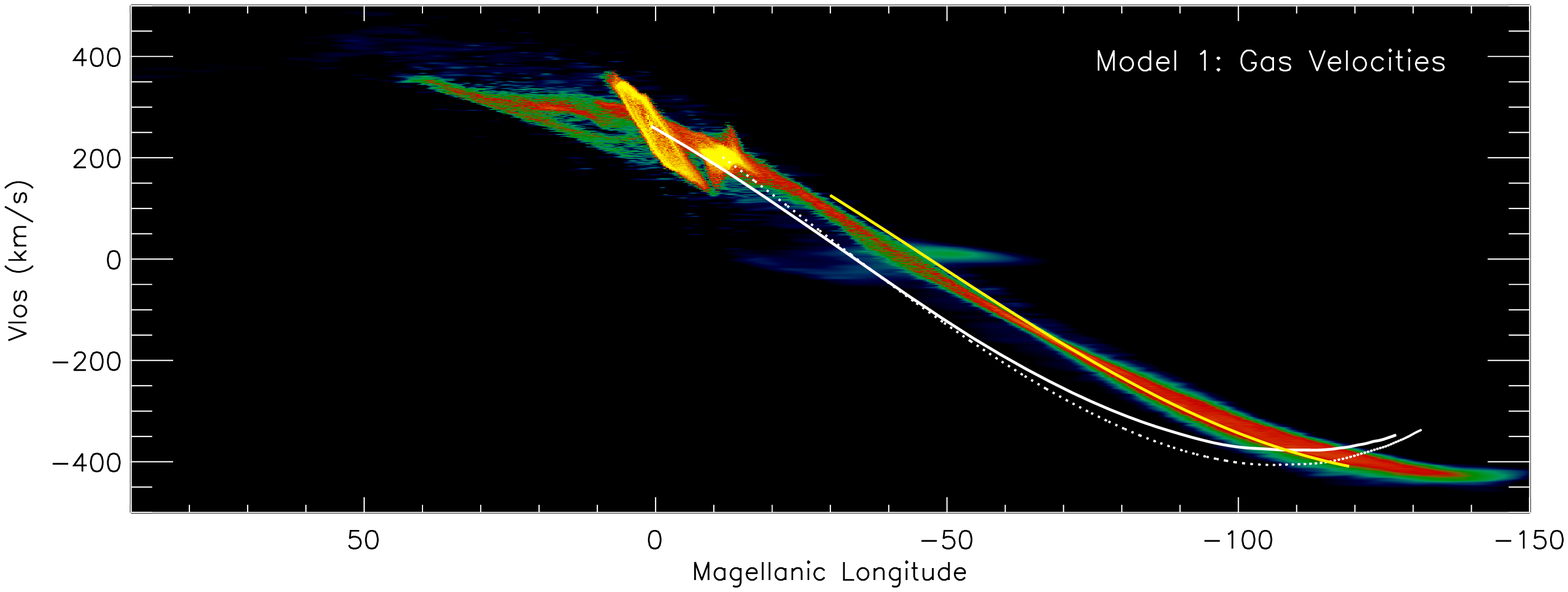}}}\\
 \mbox{ {\includegraphics[width=6.45in, clip=true, trim = 0.1in 0 0 0 ]{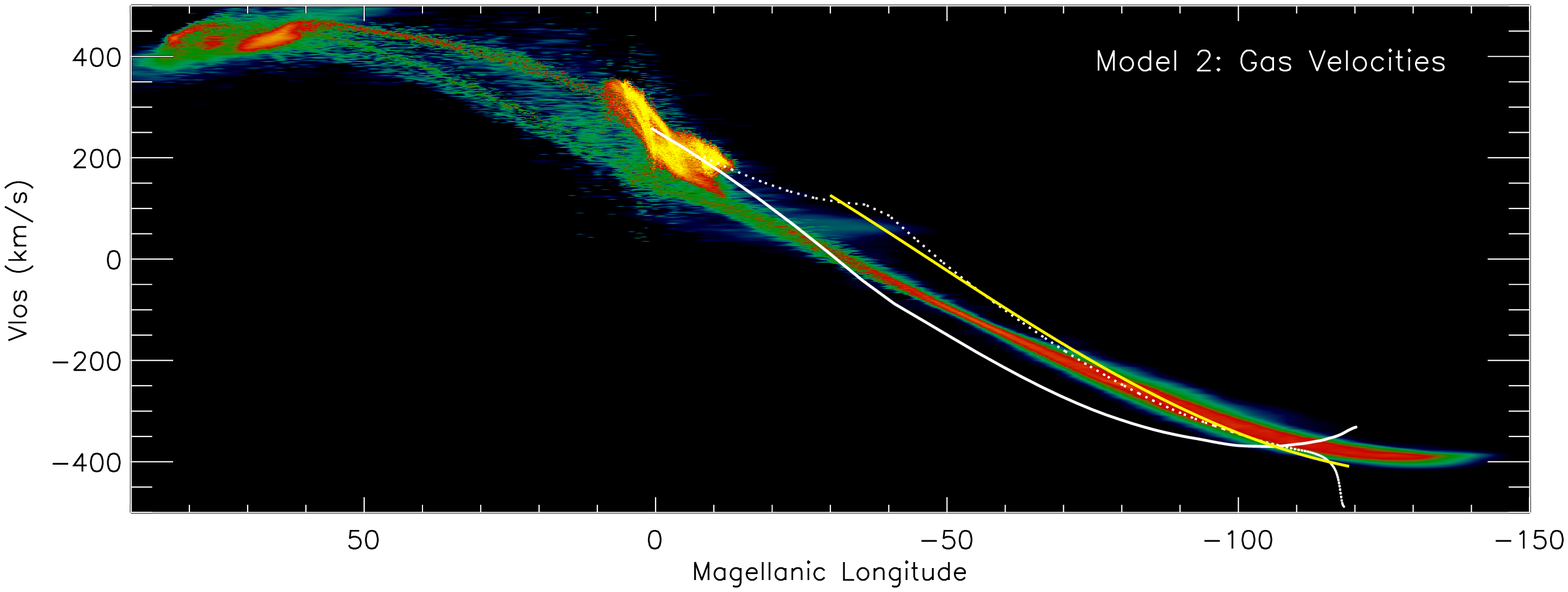}}}
 \end{center}
 \caption{\label{ch4fig:Vel} The line-of-sight velocities for the total gas distribution of the simulated Magellanic System are plotted as a function of Magellanic Longitude
 for Model 1 (top) and Model 2 (bottom). 
 The yellow line is a fit to the data of \citet{nidever2010}. The modeled line-of-sight velocities along the past orbit of the LMC(SMC) are plotted as the solid (dotted) 
 white line.  }
 \end{figure*}

There is a well-defined velocity gradient along the length of the Stream \citep{Putman2003, nidever2008}, ranging from 200 km/s near the 
Clouds to -400 km/s.  The simulated results are plotted in Figure~\ref{ch4fig:Vel} for both models.
The observed line-of-sight velocities along the MS (yellow line) and the rest of the system are well-traced by Model 1.  
Model 2 also reproduces the observed range of velocities, 
 but the slope of the velocity gradient along the MS is not well-matched to the data.  
 
 Given that the only difference between Model 1 and Model 2 
 is the SMC's orbital parameters, it is doubtful that this discrepancy in predicted velocities owes to missing physics.  Rather, a detailed search of the 
 SMC's orbital parameter space will likely yield better matches for Model 2.  It is possible that gas drag effects (not modeled here) may also 
 modify the velocity profile - particularly in the LAF, where the velocities are currently too high in Model 2.

\begin{figure*}
\begin{center}
\mbox{{\includegraphics[width=3in, clip=true, trim = 0 1.3in 1.6in 1in]{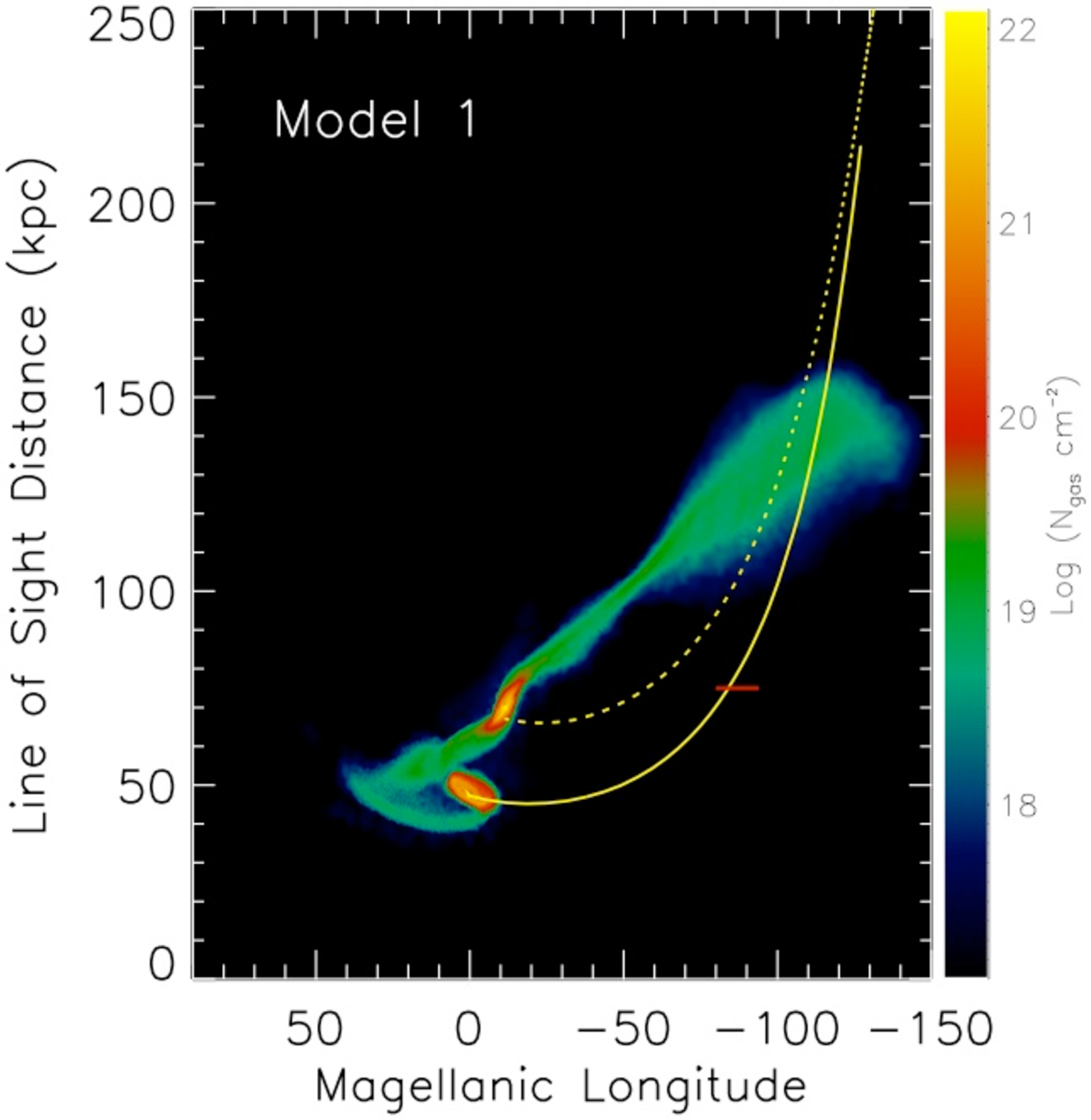}} {\includegraphics[width=3in, , clip=true, trim = 1.6in 1.3in 0 1in]{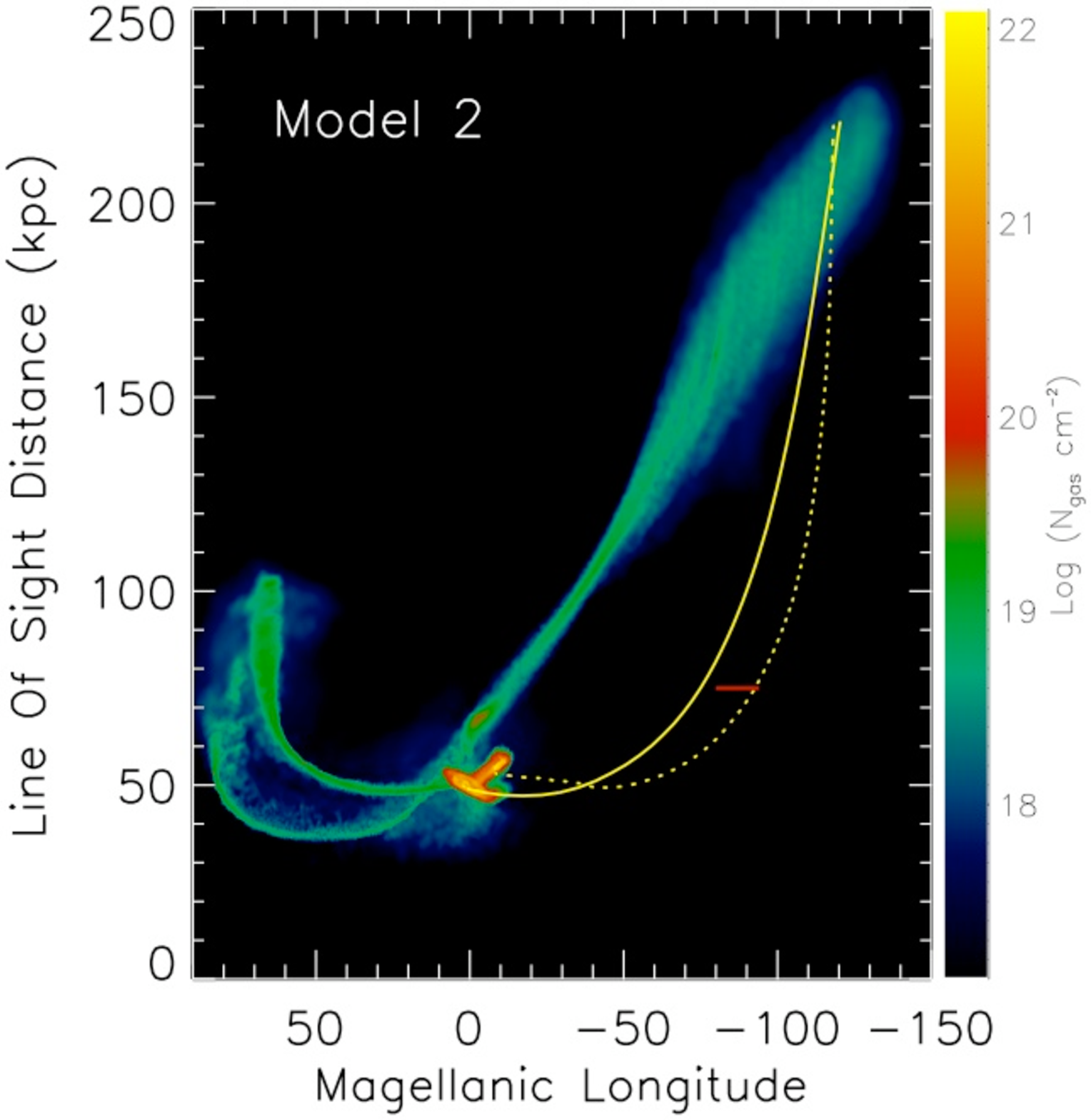}}}
 \end{center}
 \caption{\label{ch4fig:Dist} The line-of-sight distances for the gas distribution of the simulated system are plotted as a function of Magellanic Longitude
 for Model 1 (left) and Model 2 (right). 
 The modeled line-of-sight distance along the orbit of the LMC(SMC) is plotted as the solid (dashed) yellow line. The solid red line indicates the distance 
 estimate from ~\citet{jin2008}. }
 \end{figure*}

No direct distance measures exist for the Magellanic Stream, as no stellar counterpart has yet been identified  \citep{NoStars}. 
\citet{jin2008} present a geometrical method to determine distances along streams with well defined velocity gradients. Using this method they 
find the tip of the 100 degree long Stream defined in \citet{Putman2003} (i.e. not including the extension recently described by Nidever et al. 2010) to be located
at 75 kpc from the Galactic center.
The line-of-sight distances of the gas in the simulated Magellanic System are plotted in Figure~\ref{ch4fig:Dist}.  The stream produced by Model 1 is generally closer 
(80-150 kpc) than that of Model 2 (80-230 kpc).  Both simulated streams are further away than predicted by the \citet{jin2008} method;
however, gas drag and changes in the model parameters (such as increasing the MW mass) can alter the distance to the simulated stream.

\section{LMC Morphology}
\label{sec:Bar}

In this section
 we study in detail the resulting structure of the simulated LMC stellar and gaseous disks
 in our two models of the large scale gas distribution of the 
Magellanic System.  

\begin{figure*}
\begin{center}
\mbox{{\includegraphics[width=2.5in, clip=true, trim=1.5in 2.8in 1.55in 2.9in]{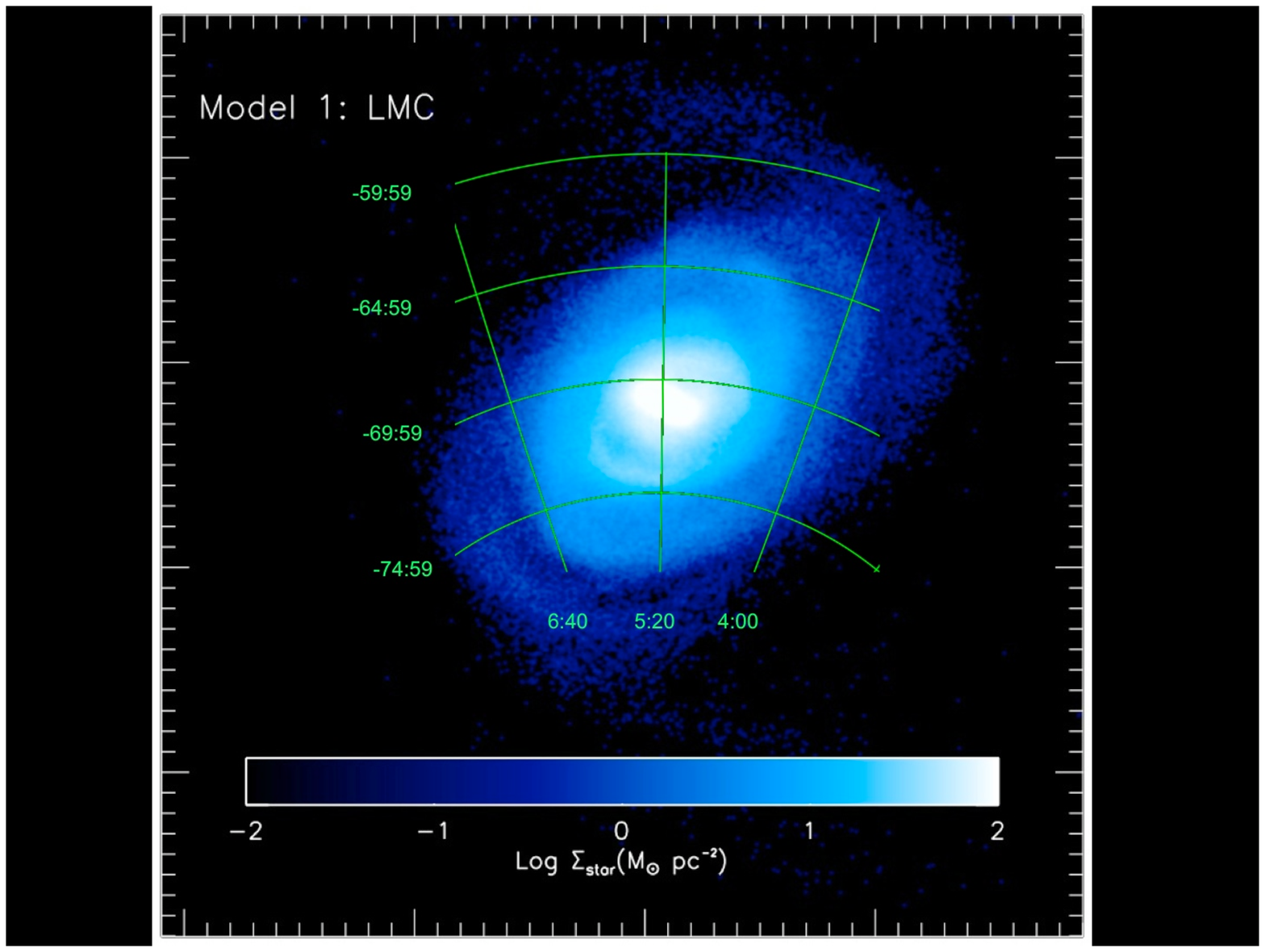}}}
\mbox{\includegraphics[width=2.5in,  clip=true, trim=1.5in 2.8in 1.55in 2.9in]{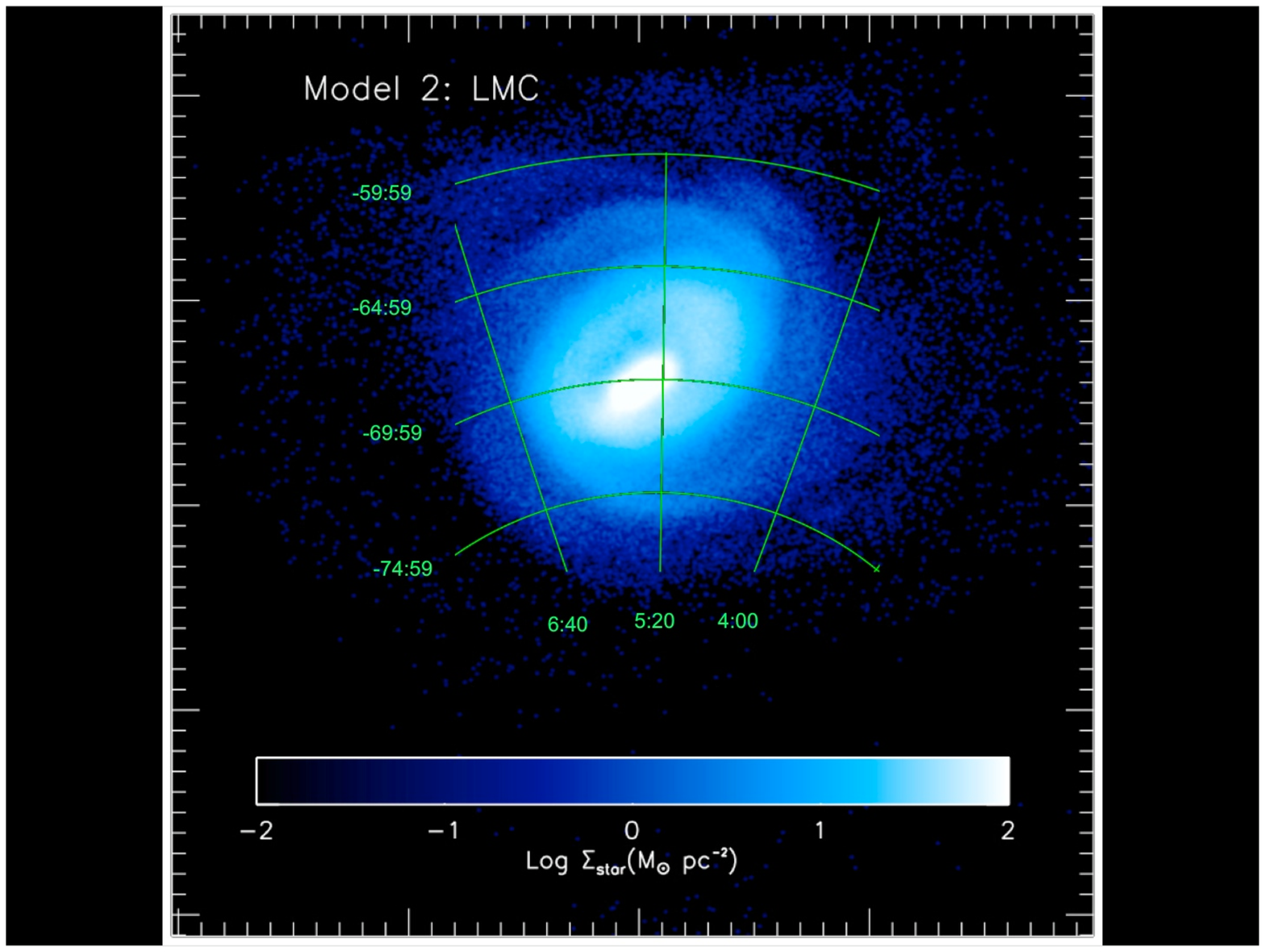}}
 \end{center}
 \caption{\label{ch6fig:LMCLOS}  The stellar surface density of the LMC disk is plotted in the line-of-sight frame for Model 1 (top) and 
 Model 2 (bottom).  RA and DEC grids are overplotted in green across the face of the disk.
 The stellar distribution in Model 2 is significantly more disturbed than in Model 1. In particular, the bar in Model 2 is off-center relative 
 to the stellar disk. The Model 2 LMC disk is also significantly disturbed in the upper left (North-East). }
 \end{figure*}

Figure ~\ref{ch6fig:LMCLOS} shows the LMC's stellar disk in Model 1 (left) and Model 2 (right) in our line-of-sight view. 
The RA and DEC coordinate grid is overplotted in green across the face of the disk. 
In both models the LMC disk is inclined $\sim$35 degrees with respect to the plane of the sky, as observed 
(i.e. despite the recent collision of the SMC in Model 2, the inclination of the LMC's disk remains unchanged). 

The Model 1 disk is fairly uniform and symmetric. 
In Model 2, however, there are perturbations induced in the LMC's stellar disk by the recent encounter with the 
SMC. In particular, there are significant distortions in the North-East. 
Only LMC stellar particles are plotted in these images, and so these structures are in the plane of the 
LMC disk and do not represent tidal debris from the SMC.  Such structures are observed in deep 
observations of the periphery of the LMC's disk  \citep[][ Martinez-Delgado et al. in prep]{deVaucouleurs}.

The initial LMC disk was bar unstable, and so 
the bar feature in both models was present since the beginning of the simulation - it was not induced by external tidal perturbations from the SMC 
or MW.  
Interestingly, in Model 2, the bar of the LMC is now off-centered relative to the disk,
 as observed.  No such perturbations are observed in Model 1: without a close encounter the LMC looks like a symmetric 
 spiral disk galaxy and it is doubtful that such a galaxy would be classified as a Magellanic Irregular galaxy.

\begin{figure*}
\begin{center}
\mbox{{\includegraphics[width=3in, clip=true, trim = 0 2.2in 0 2.2in]{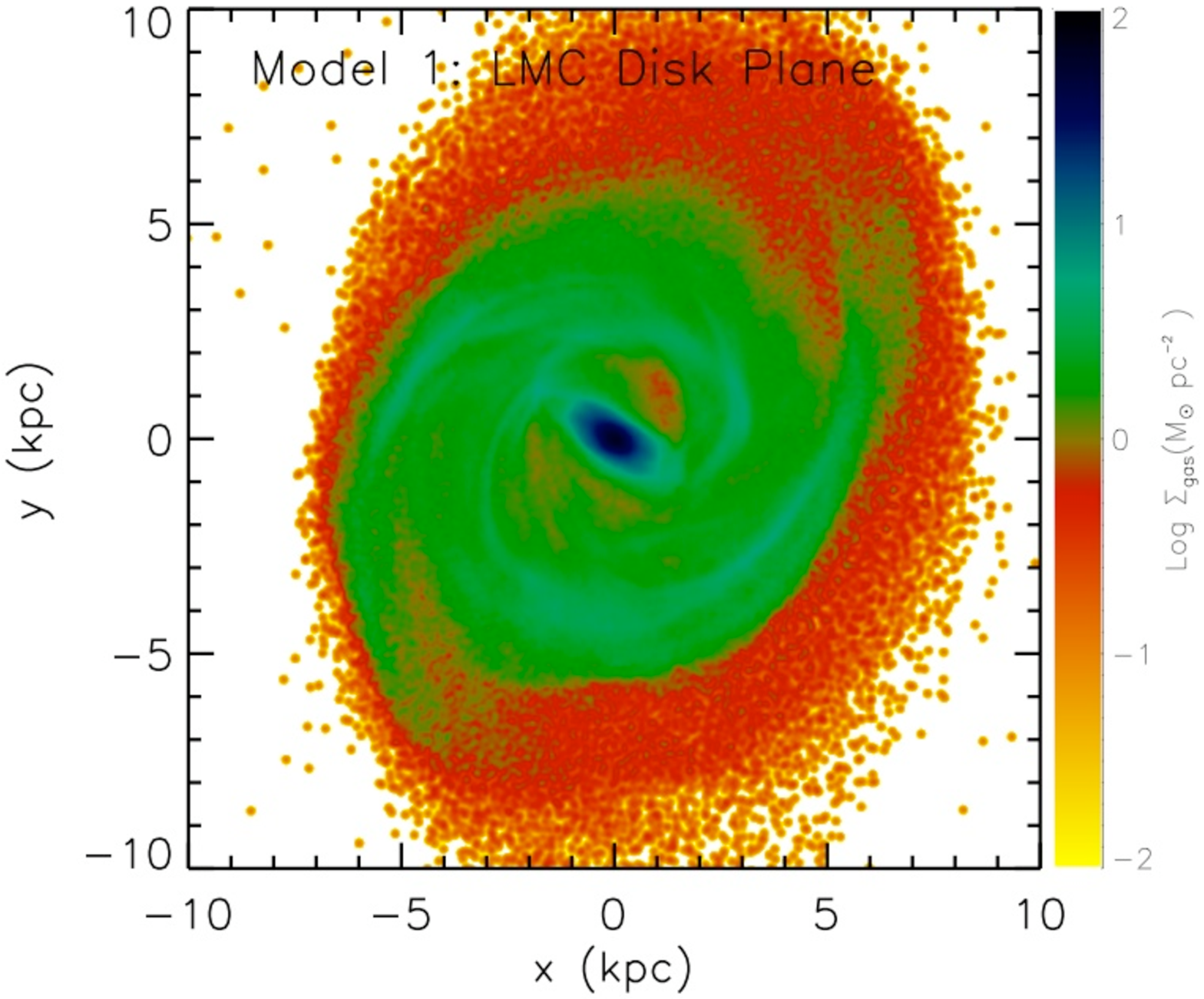}}
{\includegraphics[width=3in, clip=true, trim = 0 2.2in 0 2.2in]{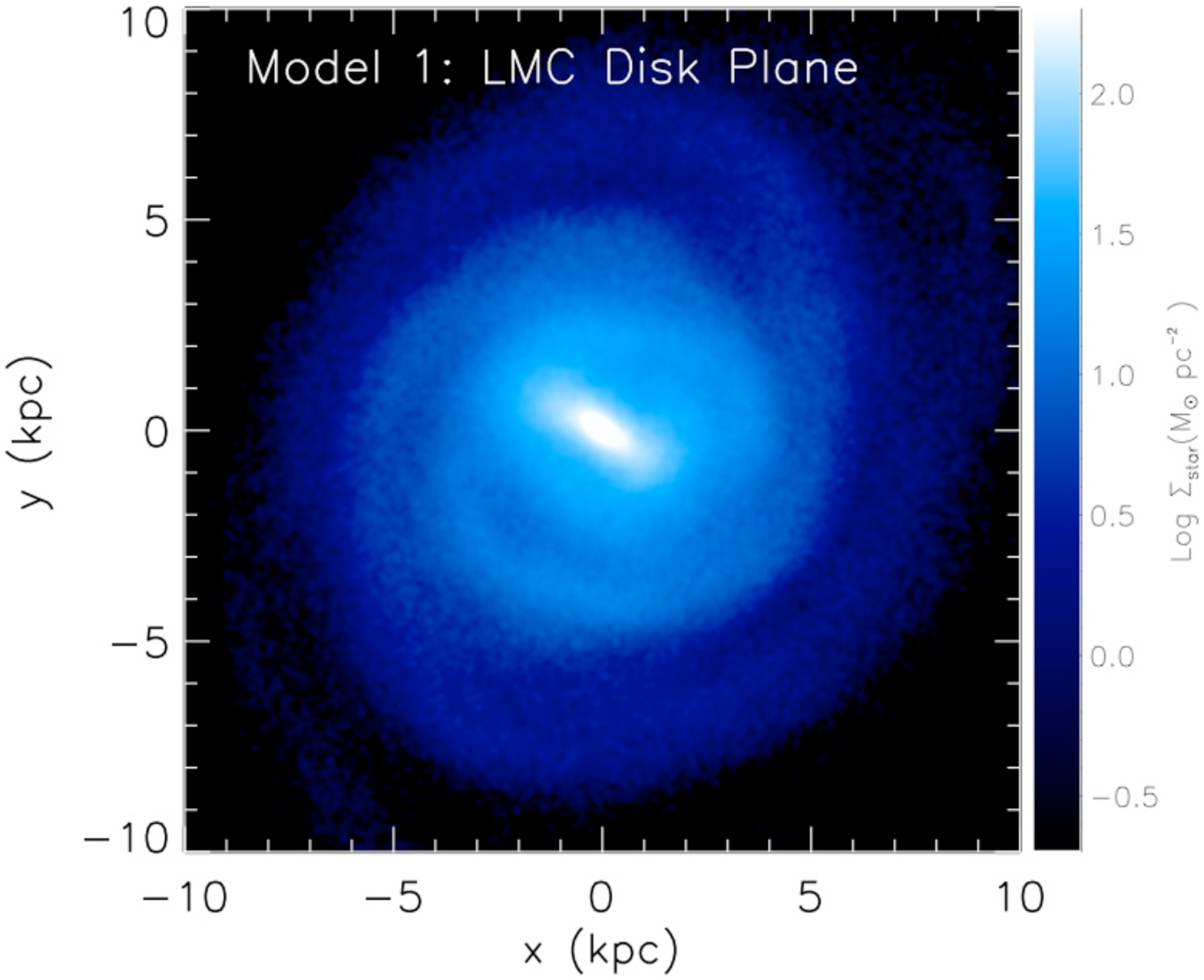}}}\\
\mbox{{\includegraphics[width=3in, clip=true, trim = 0 2.2in 0 2.2in]{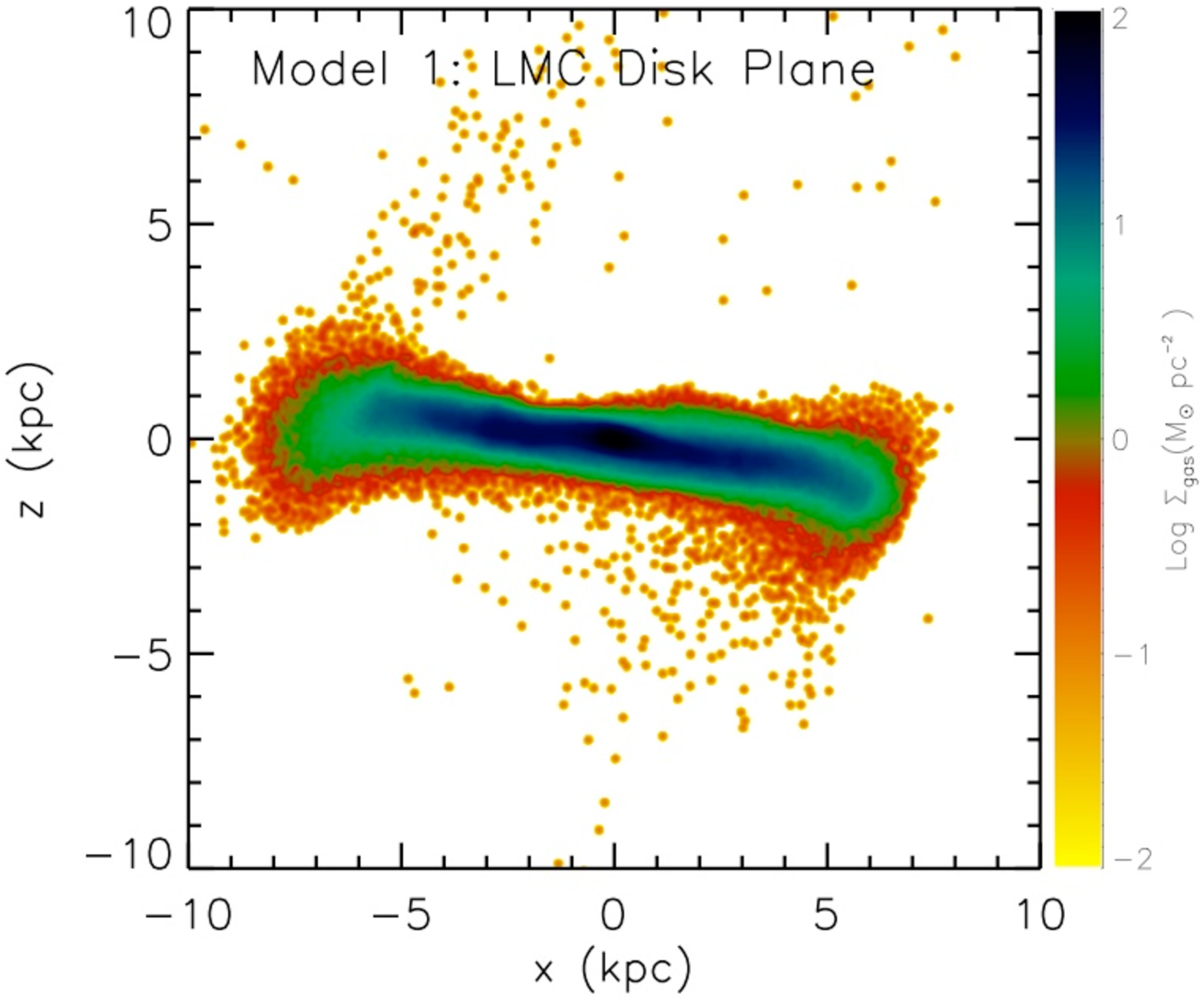}}
{\includegraphics[width=3in, clip=true, trim = 0 2.2in 0 2.2in]{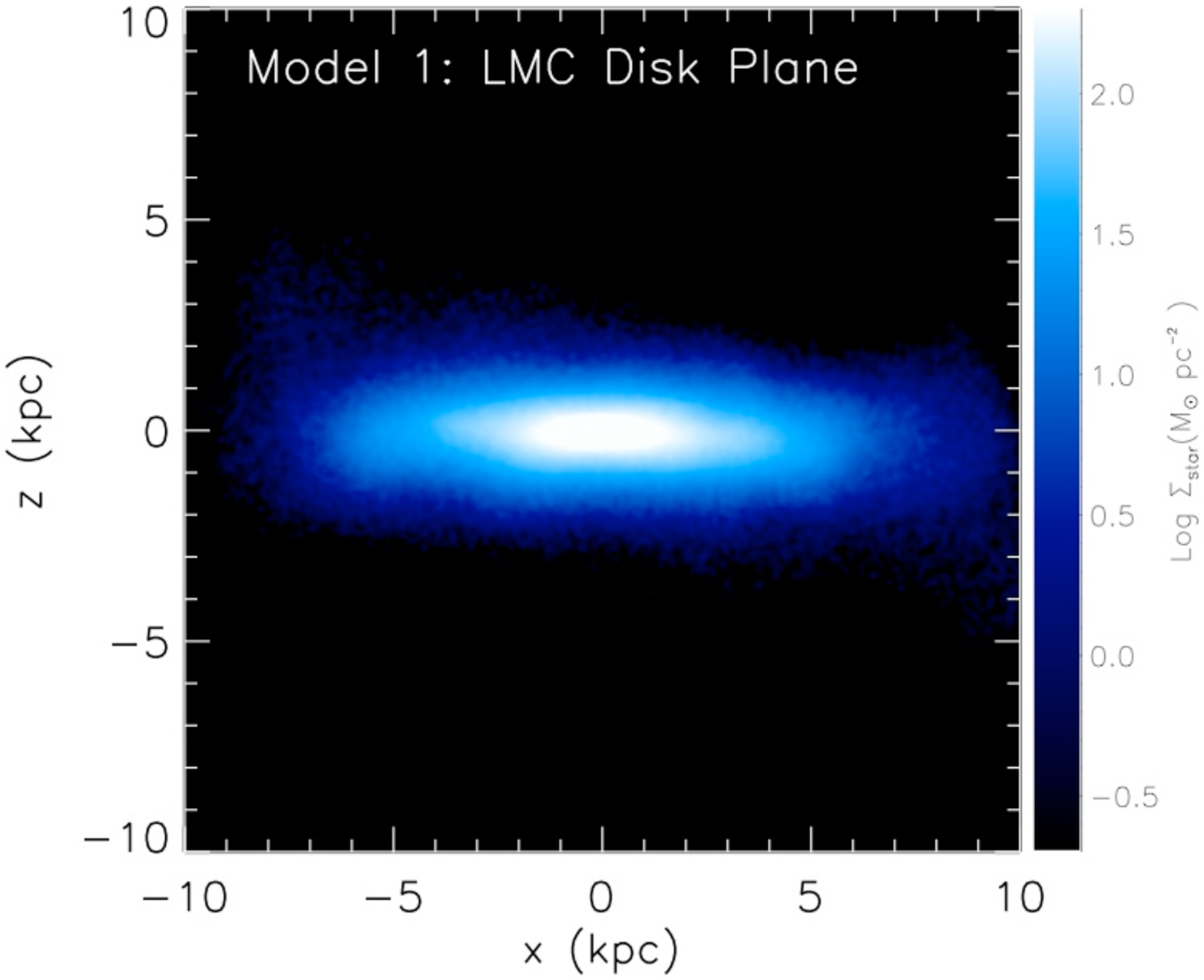}}}\\
 \mbox{
 {\includegraphics[width=3in, clip=true, trim = 0 2.2in 0 2.2in]{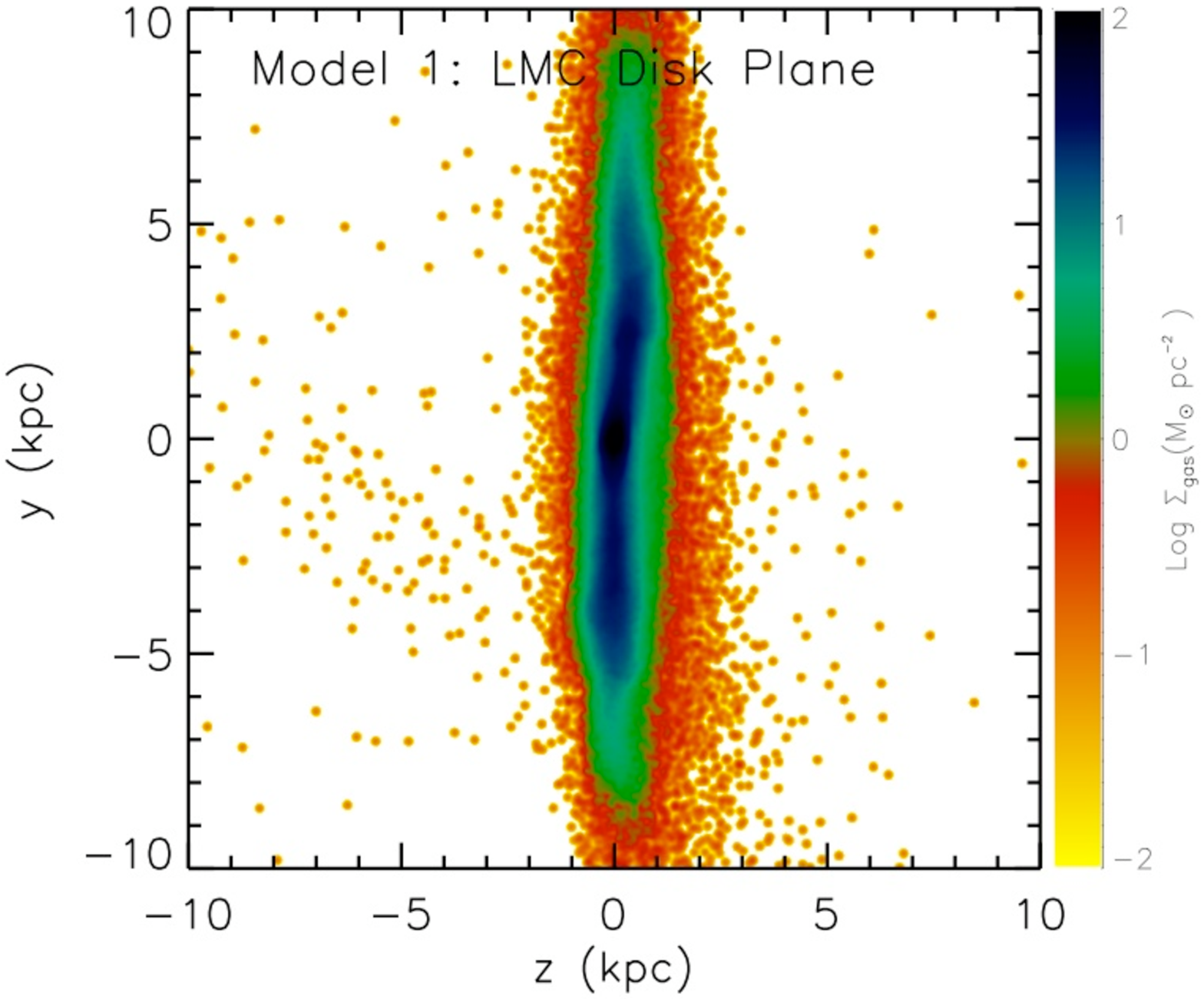}}
 {\includegraphics[width=3in, clip=true, trim = 0 2.2in 0 2.2in]{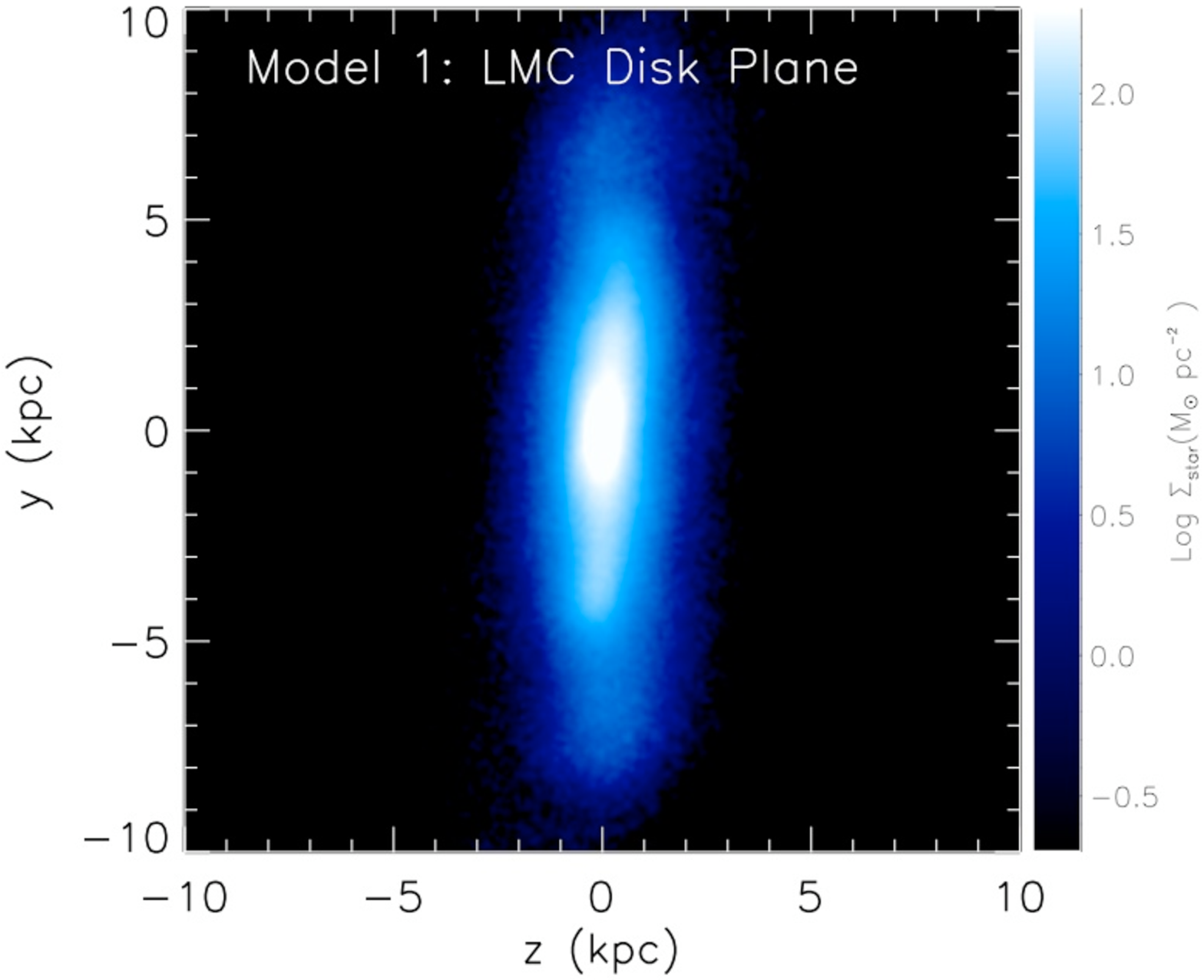}} \hspace{0.01in}}
 \end{center}
 \caption{\label{ch6fig:LMCModel1}  The gas (left) and stellar (right) density of the LMC disk for Model 1. 
 {\it Top}: face-on view (x,y plane).  {\it Middle}: edge-on view along the x axis. {\it Bottom}: along the
  y axis.   Both the gas and stellar projections have a centered, in-plane bar.}
 \end{figure*}

\begin{figure*}
\begin{center}
\mbox{{\includegraphics[width=3in, clip=true, trim = 0 2.2in 0 2.2in]{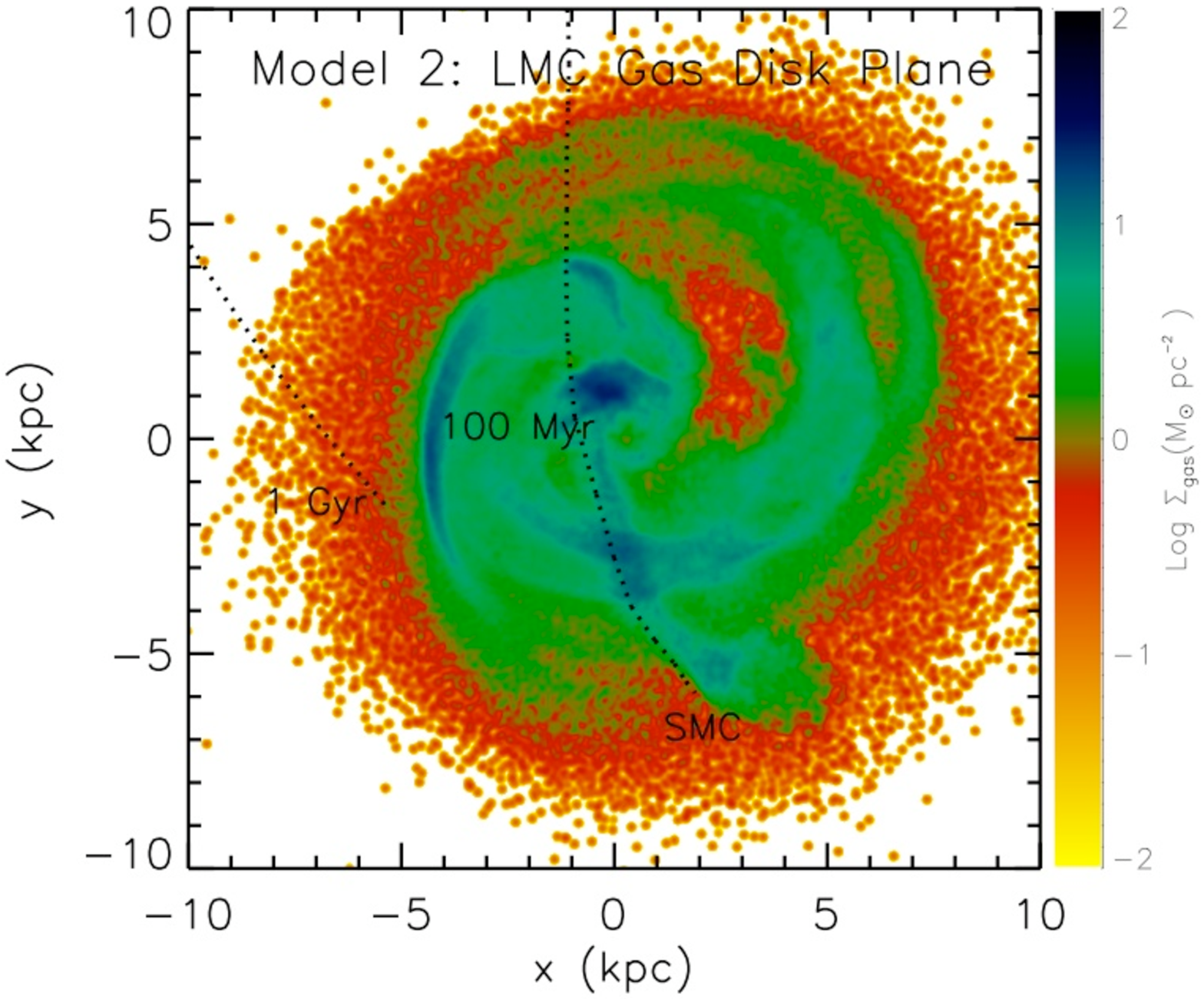}}
{\includegraphics[width=3in, clip=true, trim = 0 2.2in 0 2.2in]{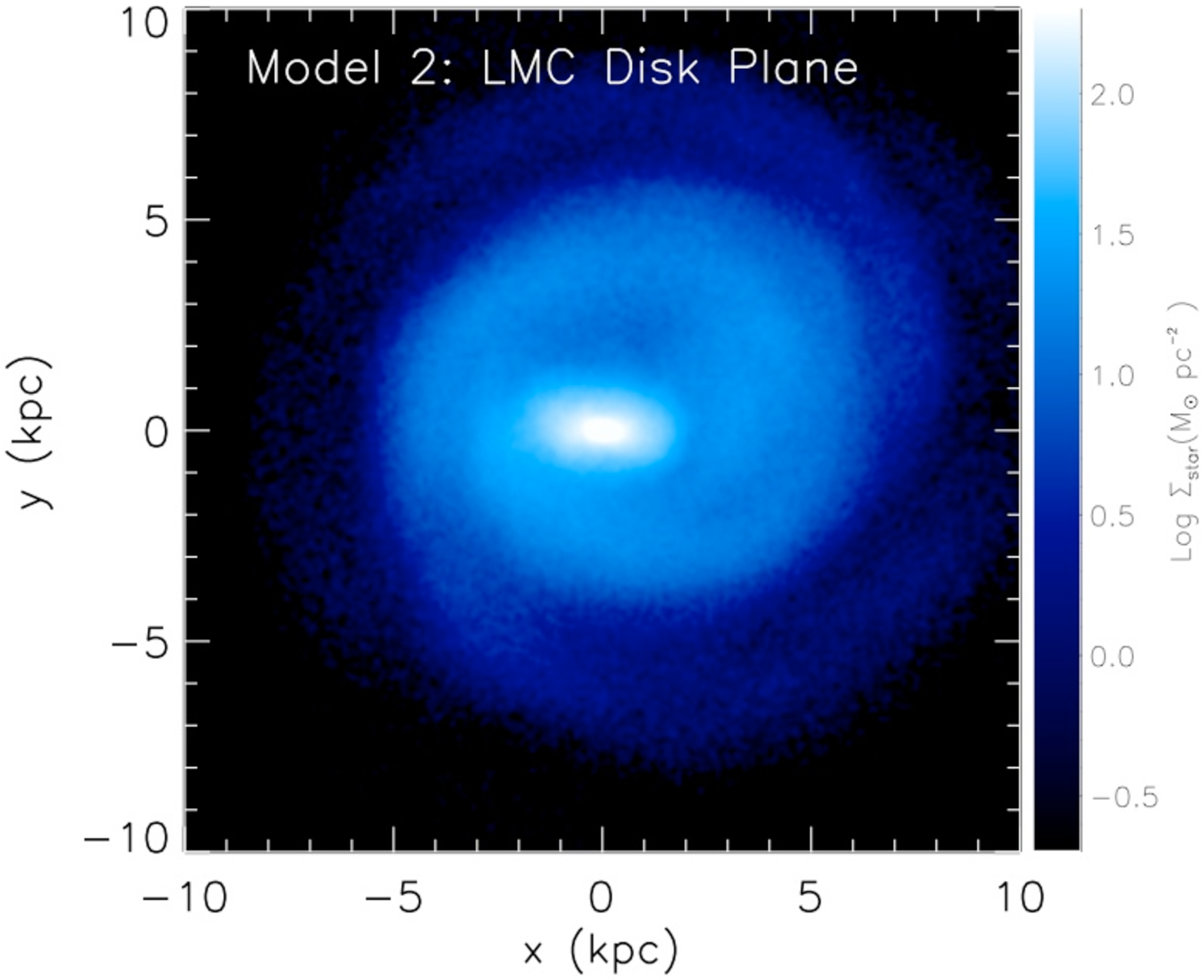}}}\\
\mbox{{\includegraphics[width=3in, clip=true, trim = 0 2.2in 0 2.2in]{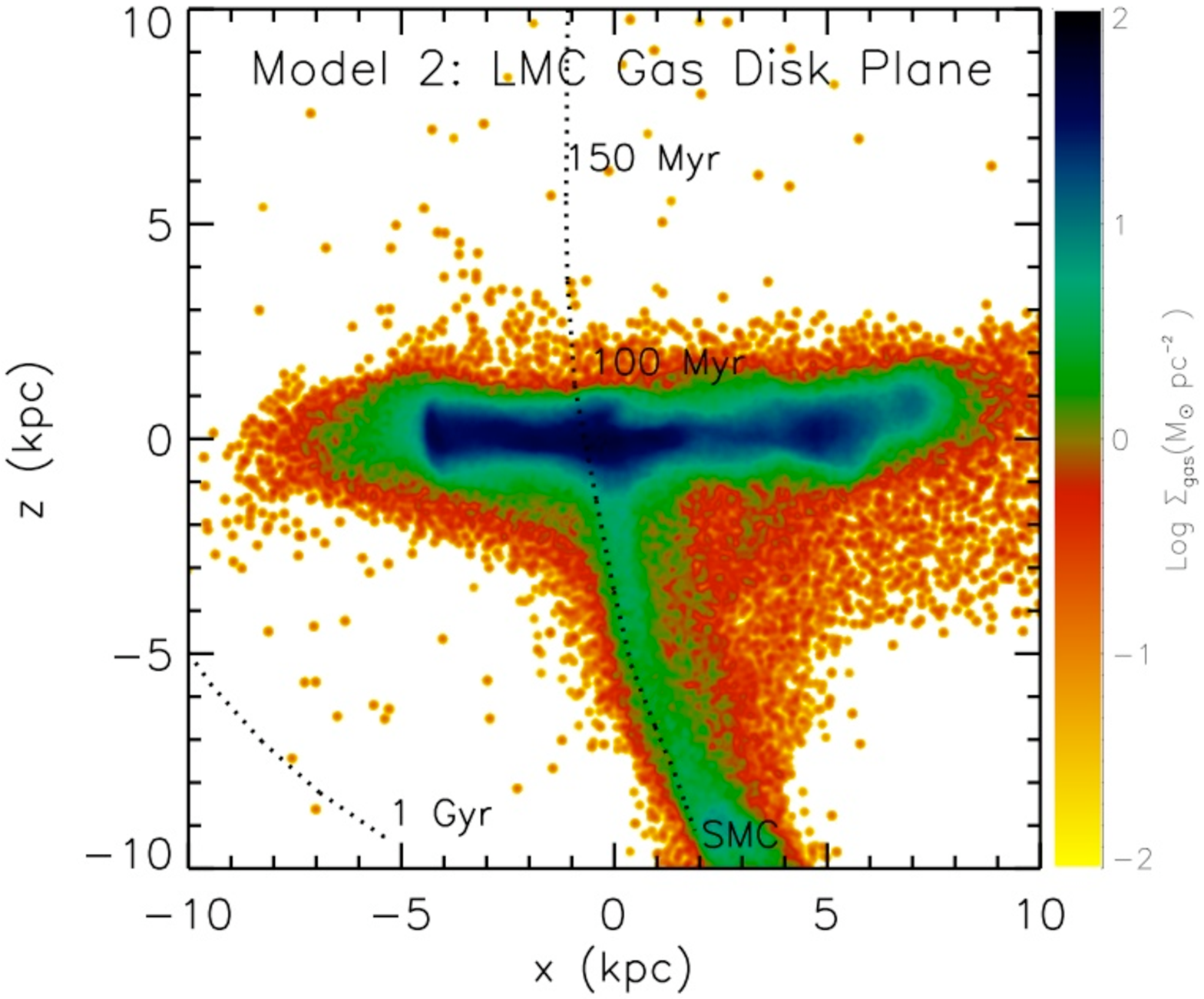}}
{\includegraphics[width=3in, clip=true, trim = 0 2.2in 0 2.2in]{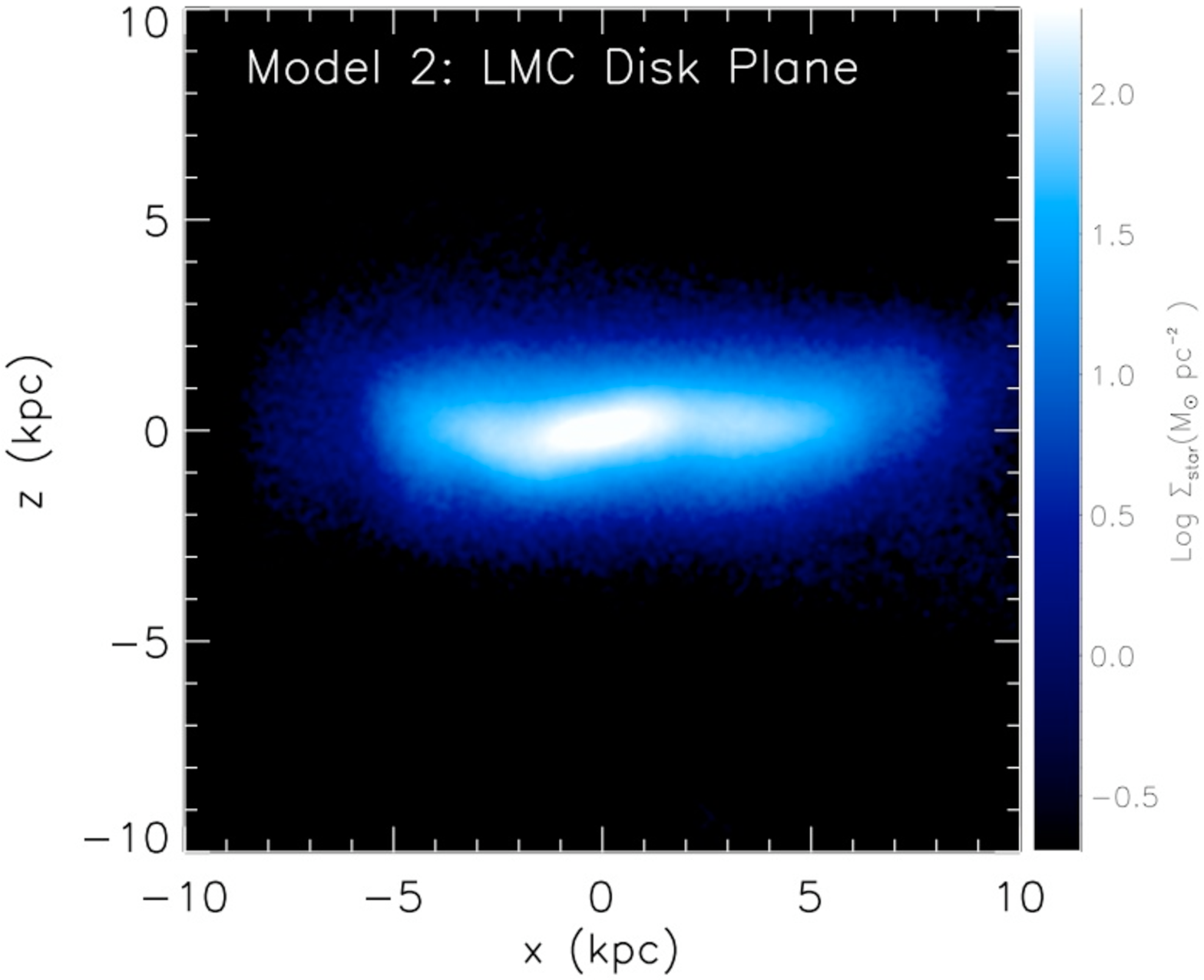}}}\\
 \mbox{
 {\includegraphics[width=3in, clip=true, trim = 0 2.2in 0 2.2in]{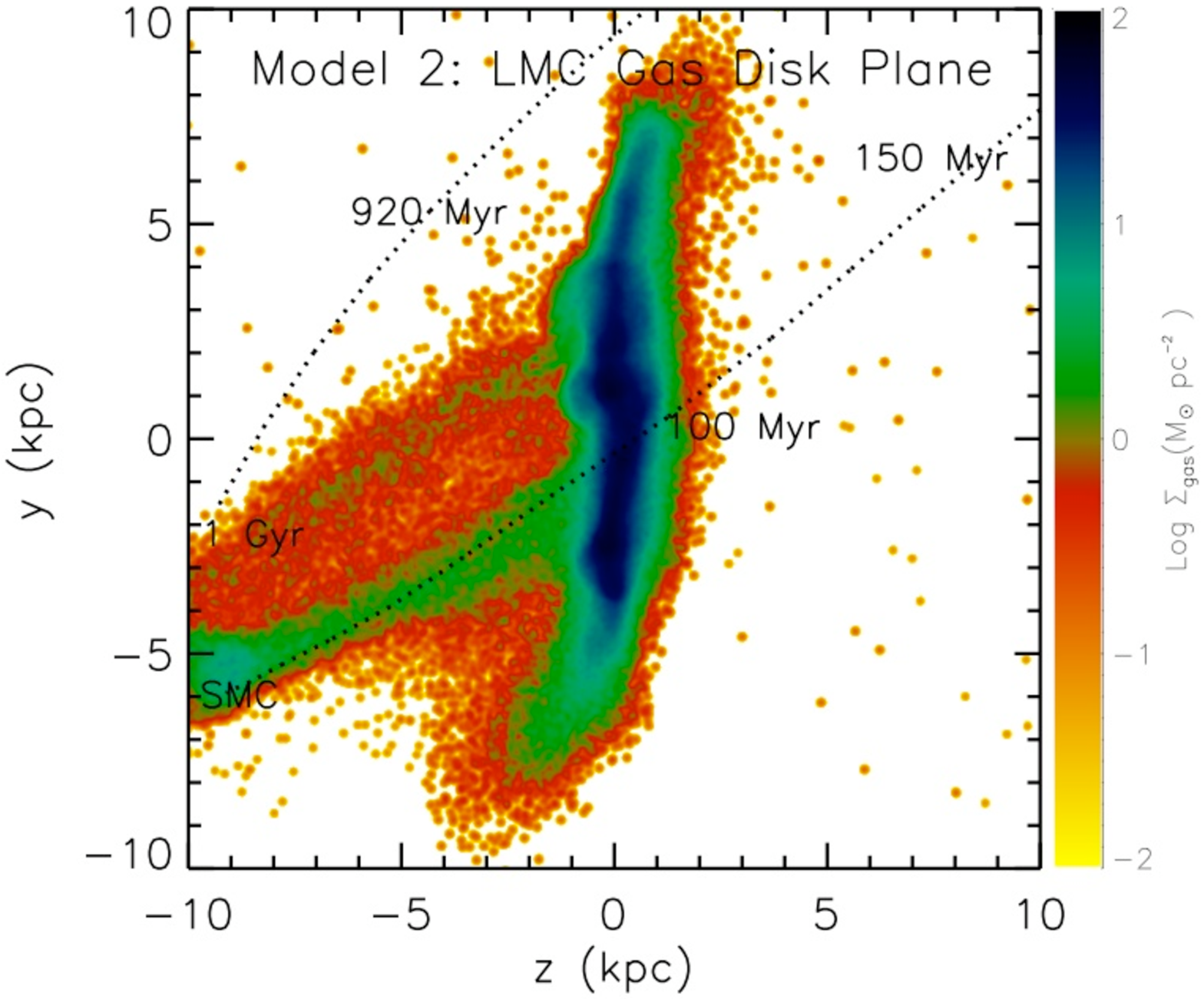}}
 {\includegraphics[width=3in, clip=true, trim = 0 2.2in 0 2.2in]{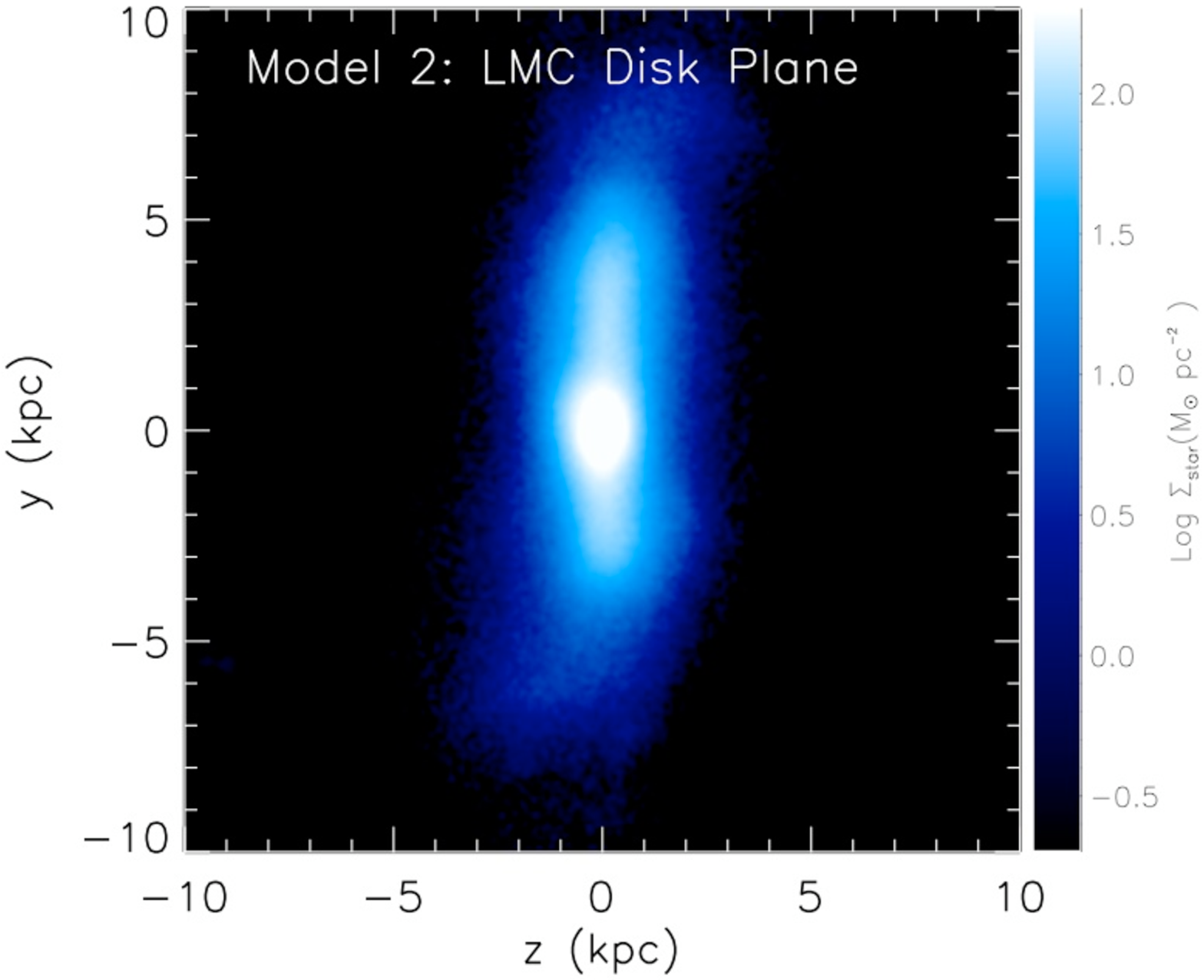}} \hspace{0.01in}}
 \end{center}
 \caption{\label{ch6fig:LMCModel2}  Same as Figure~\ref{ch6fig:LMCModel1} but for Model 2. The stellar bar is clearly off-centered
 and warped relative to the stellar disk; it is also absent in the gas. The SMC's orbit in an LMC-centric frame is overplotted (dashed line) in 
 the middle and bottom panels of the LMC's gas distribution. Various times are also marked along the SMC's past orbit. The SMC collides with the 
 LMC 100 Myr ago and the SMC's current position is marked.  An extra-planar stream of gas is pulled out from the LMC by the passage of the SMC. }
 \end{figure*}

In Figures~\ref{ch6fig:LMCModel1} and ~\ref{ch6fig:LMCModel2} we take a closer look at the LMC's gas and stellar disk by deprojecting
the disk from the line-of-sight frame into a Cartesian coordinate system centered on the LMC disk plane for 
both Model 1 and Model 2.  
Only particles associated with the LMC are plotted. The images are centered on the peak of the stellar density distribution (i.e. the photometric center).

In Model 1, the bar of the LMC is clearly visible in both the stellar and gaseous disks.  This is in disagreement with the observed  
HI maps of \citet{kim1998}: there is no distinguishable bar structure in the observed global HI emission that is comparable
to the optical bar. 
 The results for Model 2 illustrate the consequences of a recent (100-300 Myr ago) direct 
collision between the LMC and SMC. In this scenario, Model 1 represents the state of the LMC disk 
before the collision occurred, where the LMC is a symmetric barred spiral. 
After the collision, the bar has become off-centered with respect to the underlying disk and it has almost 
disappeared from the gas disk of the LMC. 
The reason becomes clear when the disk is viewed edge-on along the x-axis.  The bar has become warped 
by $\sim$10-15 degrees relative to the LMC disk 
plane and is therefore inefficient at  funneling gas in the way it could in Model 1.   This warp of the bar is less
extreme than that required by models such as 
\citet{zhao2000} for the microlensing optical depth and may be consistent with the observations of  \citet{sub2009b}.  
The simulated offset bar is also consistent with the structure of the observed bar,
 which is described as a cigar shaped structure with dimensions of 1 $\times$ 3 kpc.

 From the edge-on view of the Model 2 gas disk, 
it is clear that LMC gas particles have been pulled out of the disk by the passage of the SMC through the LMC. 
This causes the appearance of a gaseous ``arm" in the face-on view of the disk.  Such ``arms" are seen in the LMC gaseous disk \citep{kim1998} and are 
 believed to be related to the Magellanic Bridge and Leading Arm Feature \citep{nidever2008}.   In our interpretation, at least one of these ``arms"
 is extra-planar and located behind the LMC disk. The fact that LMC gas is removed from the disk towards the Bridge indicates that the formation of the 
 Magellanic Bridge has been aided by hydrodynamic gas drag.  It is not purely a tidal feature.  
 The Bridge is known to be quite metal poor along two sightlines towards early type stars ~\citep[see Table~\ref{metalTable} and][]{lehner2008}.
 However, a full census of the metallicity across the Bridge does not yet exist. Regardless, it is clear that LMC gas cannot have contaminated the entirety of the Bridge.
   The Model 2 scenario predicts that the majority of the Bridge material originated from the SMC, but there should be some contribution from LMC gas that increases
   in importance with proximity to the LMC.  Model 2 thus predicts that Bridge material towards the LMC should be increasingly metal enriched.
This should not be true in Model 1.  
 
 In the edge-on view of the Model 1 disk,  the gas disk appears to be tilted relative to the stellar distribution. This is likely because of the infall of gas 
from the SMC that forms the Magellanic Bridge.  
The outer stellar and gas disks in Model 2 are significantly warped and distorted in the edge-on view. The true disk is also observed 
to be both flared \citep{alves2000} and warped \citep{vanCioni2001, olsen2002, nikolaev2004}.
   Such results are in keeping with a study of Magellanic-type spirals by \citet{wilcots1996}, who also suggest that the observed 
 lopsidedness in their HI disks may be a result of minor mergers.

In Model 2 the gas disk has also formed a pronounced arc in the upper right.  Since our star formation 
prescriptions depend sensitively on the gas density, this arc of gas will also be actively forming stars (see $\S$~\ref{sec:SF}),  
giving the LMC the appearance of a one-armed spiral.  
 A number of numerical studies have been conducted on the resulting structure of a large galaxy after a direct collision with 
 a smaller companion in the context of explaining the origin of ring galaxies  \citep{lynds1976, weil1993, struck1997}.
 In particular, \citet{struck1997} finds that, in some cases, a one-armed spiral structure can be excited in the larger galaxy. Also, 
 \citet{Bekki2009bar} explored a scenario where the LMC bar becomes off-center as a result of a recent encounter with a dark 10$^{8} \Msun$ companion.
  The specific asymmetries induced depend sensitively on the mass ratio, inclination and the location of the smaller companion's passage
   through the larger galaxy.  A future study will explore these parameters in depth in the context of the LMC-SMC encounter and 
   assess the longevity of the resulting asymmetric structures.   For example, \citet{levine1998} have illustrated that disk-lopsidedness can be long-lived 
   if the disk is displaced from the center of the dark matter potential and spinning in a sense that is retrograde to its orbit about that center.

Finally we note that, while we have not discussed the results for the SMC in detail, the simulated SMC morphology in Model 2
 is consistent with the observations of a ``bar"-like main body with a stellar wing leading towards the LMC and a 
 significant line-of-sight depth.  We will discuss these results in depth in a future paper (Besla et al. 2012 in prep.).

\section{LMC Kinematics}
\label{sec:Kin}

The internal kinematics of the LMC has been quantified by many tracers. 
The LMC's rotation curve rises roughly linearly to a radius of $\sim$ 4 kpc, after which it stays flat at a value of $V_{rot}$. 
The observed rotation curve has been noted to peak at different 
values depending on the kinematic tracer being studied. 
The HI kinematics yield $V_{rot}$ = 80 km/s \citep{Staveley}, data from 
red supergiants gives $V_{rot}$ = 107 km/s \citep{Olsen} and carbon stars yields $V_{rot}$=61 km/s \citep{vanderMarel}.  
However, recently \citet{olsen2011} examined the kinematics of a combined
 population of massive red supergiants, oxygen-rich and carbon-rich AGB stars in the LMC.  After correcting for the 
 LMC's space motion and the asymmetric drift in the AGB population, they find a consistent rotation curve between all 
 kinematic tracers with $V_{rot}$ = 87 $\pm 5$ km/s.  This is in accord with the HI rotation curve and the initial value adopted
 in this study  ($\sim$ 95 km/s).

The LMC disk initial conditions (gas fraction, equation of state) are chosen such that the 
LMC kinematics are representative of a symmetric, bar-unstable disk galaxy. It remains to be seen whether 
MW tides will introduce kinematic anomalies in the disk in a first infall scenario, or, perhaps more significantly,
 whether the LMC can retain a kinematically stable disk after a direct collision with the SMC (i.e. in Model 2).

\begin{figure*}
\begin{center}
\mbox{{\includegraphics[width=6.5in, clip=true, trim = 0 4.5in 0 4.1in ]{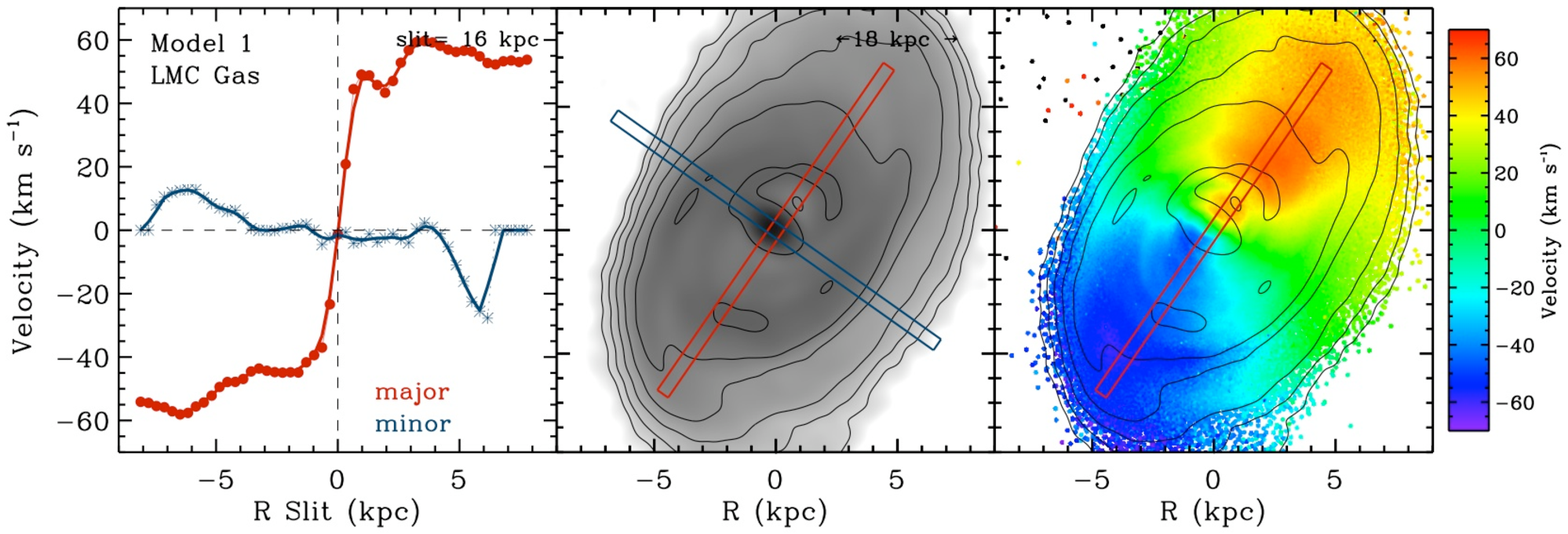}}}\\
\mbox{{\includegraphics[width=6.5in, clip=true, trim = 0 4.5in 0 4.1in]{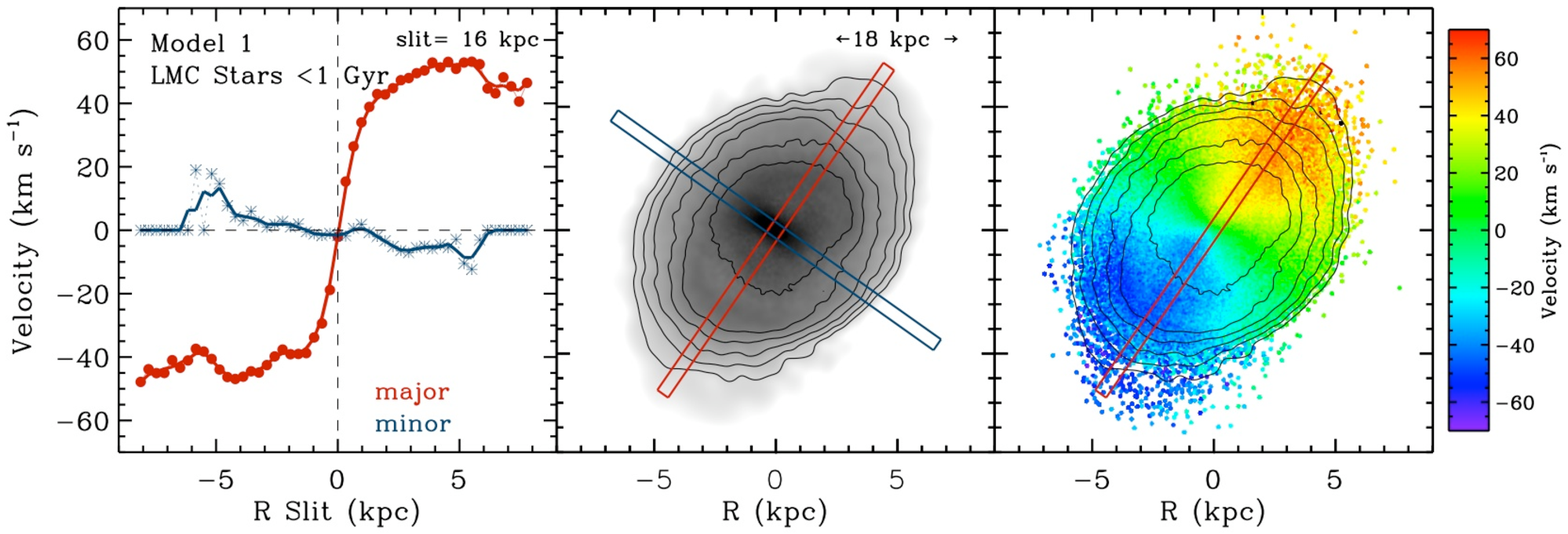}}}\\
\mbox{{\includegraphics[width=6.5in,clip=true, trim = 0 4in 0 4.1in]{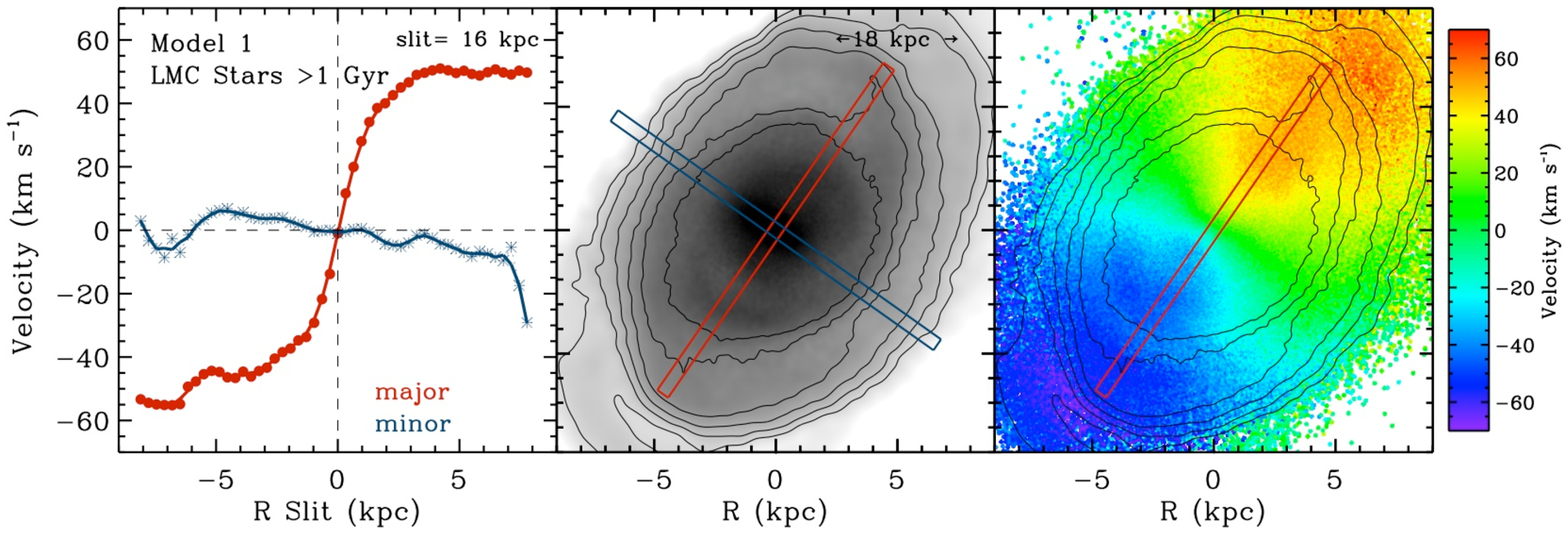}}}\\
 \caption{\label{ch6fig:LMCkinModel1} The kinematics of the simulated LMC's gaseous disk (top), young stellar disk ($< 1$ Gyr; middle) 
 and old stellar disk ($>$ 1 Gyr; bottom) in Model 1 are illustrated in the line-of-sight frame (North to the top, East to the left; 
 the SMC is located to the South-West). 
 The {\it right panels} show the line-of-sight velocity field, where the center-of-mass velocity of the respective kinematic tracer has been 
 subtracted. Density contours and major axis slit location are also indicated. The color gradient denotes material moving
 towards (blue) and away (red) from the observer.
  The {\it central panels} show the surface density of the tracer, with density contours overplotted. The central and right panel boxes spans 18 kpc
  a side. Each box is centered on the stellar density peak.  Slits are placed along the major (red) and 
 minor (blue) stellar kinematic axes as indicated. The major kinematic axis slit is inclined 55 degrees counter clock wise from the x-axis in all panels. 
 The minor kinematic axis is placed 90 degrees with respect to the major axis, although in practice this does not exactly trace the 
 zero velocity regions.  
 Note that even if the gas kinematics are being examined, the red slits still denote the locations of the {\it stellar} kinematic axes.
In the {\it left panel}, the line-of-sight velocities are plotted along the slit. The dashed vertical line indicates the location of the 
peak stellar density (photometric center); in Model 1 it is coincident with the kinematic centers of all tracers.}
\end{center}
 \end{figure*}

\begin{figure*}
\begin{center}
\mbox{{\includegraphics[width=6.5in, clip=true, trim = 0 4.5in 0 4.1in ]{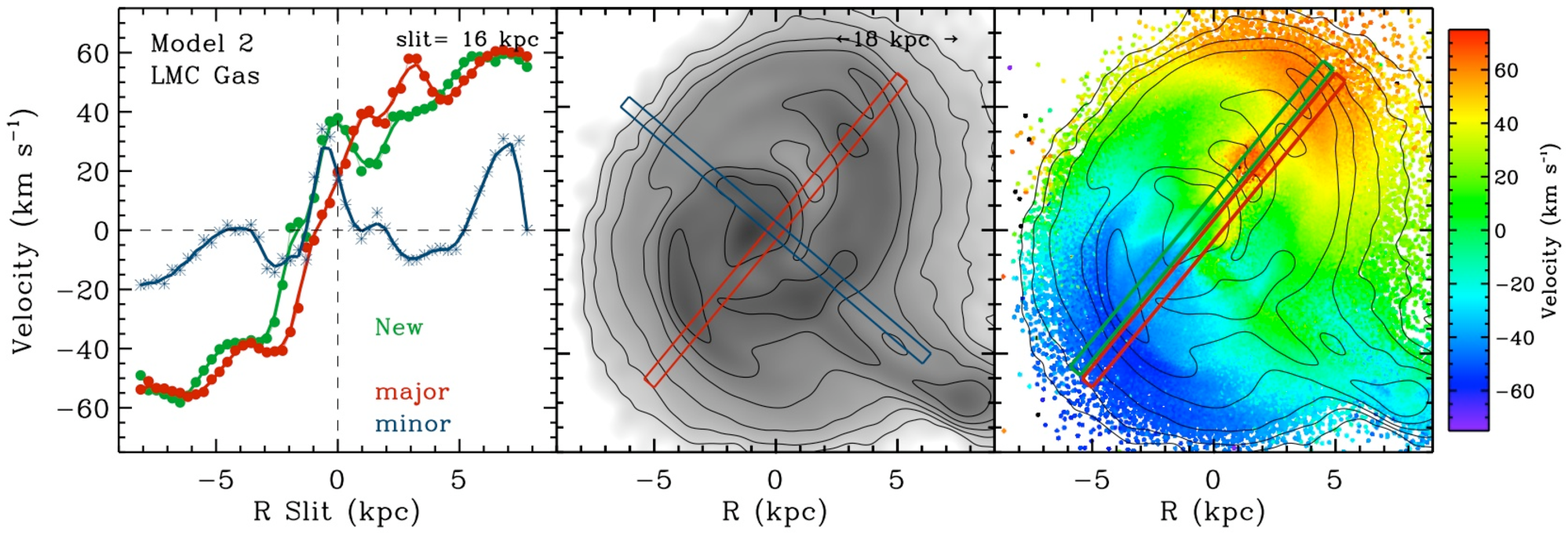}}}\\
\mbox{{\includegraphics[width=6.5in, clip=true, trim = 0 4.5in 0 4.1in ]{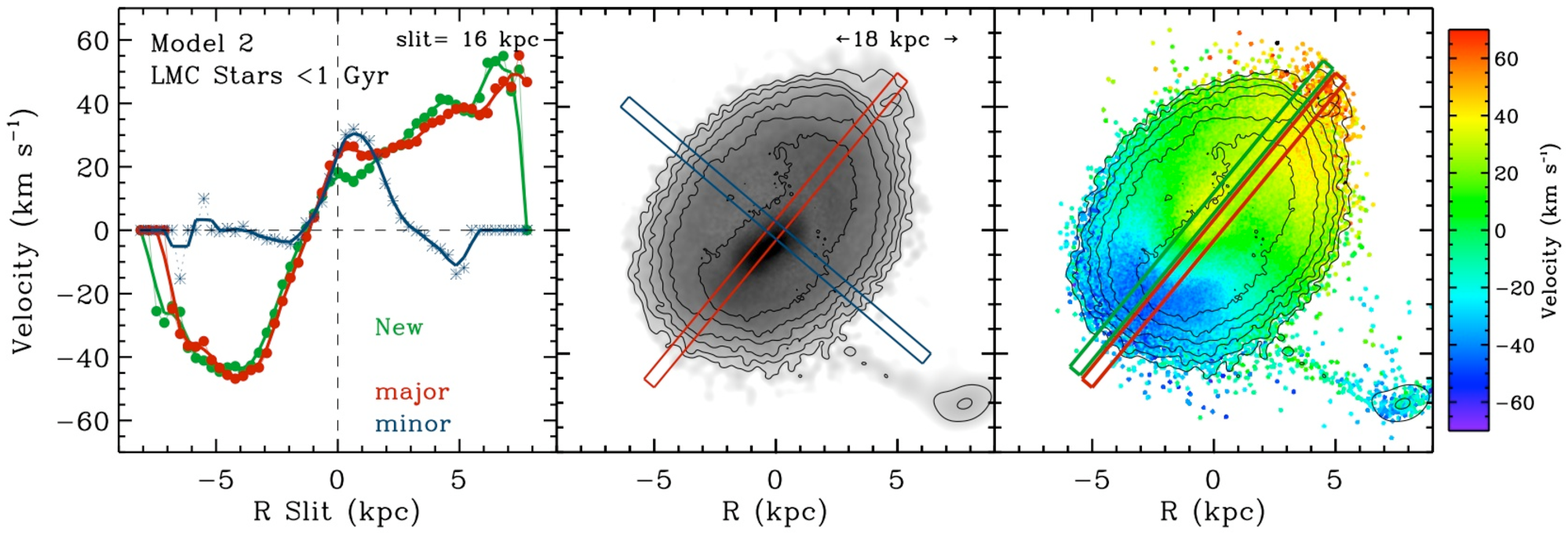}}}\\
\mbox{{\includegraphics[width=6.5in, clip=true, trim = 0 4in 0 4.1in]{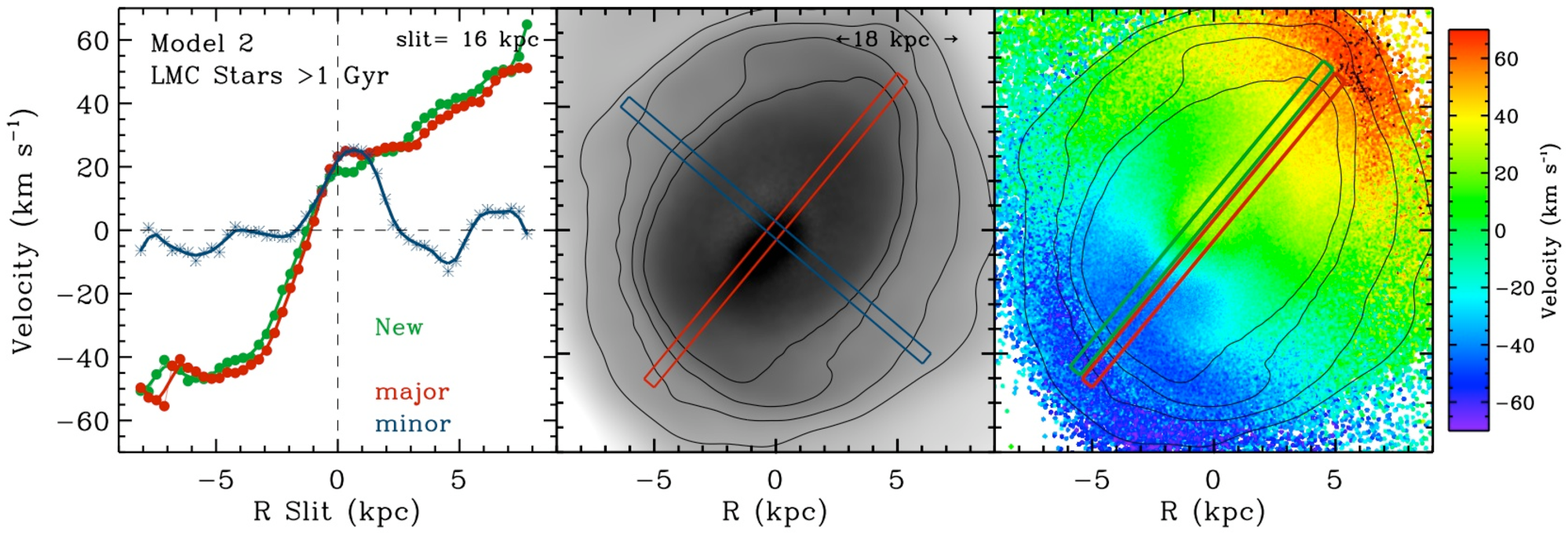}}}\\
 \end{center}
 \caption{\label{ch6fig:LMCkinModel2} The same as in Figure ~\ref{ch6fig:LMCkinModel1}, except for Model 2. The kinematic major 
 axis (red) is inclined 50 degrees counter clock wise from the x-axis in all panels. In the left panel it is clear that the kinematic center 
 of all tracer populations are coincident, but not with the photometric center (dashed vertical line). Furthermore, the shape of the velocity field 
 as traced by gas vs. stars is different 
 across the face of the disk.  This is illustrated by the inclusion of a new slit (in green), placed 1 kpc above the major axis.  The line-of-sight 
 velocities of the gas along the 'New' slit deviate from their stellar counterpart. This offset is likely a result of the warped stellar bar.  }
 \end{figure*}

In Figures ~\ref{ch6fig:LMCkinModel1} and ~\ref{ch6fig:LMCkinModel2} the kinematics of the LMC disk in Models 1 and 2, respectively, 
are broken down 
for various kinematic tracers: gas (top panel), young stars (middle) and old stars (bottom), in the line-of-sight frame. In each panel the disk 
is centered on the stellar center of mass. The center panels show the surface density of the tracer population. 
A slit is placed along the largest 
velocity gradient of the LMC's older stellar distribution and is defined as the major kinematic axis (red). A second slit is placed 90 degrees with 
respect to the major axis and referred to as the ``minor" kinematic axis (blue).
The position angle of the kinematic major axis of the simulated disk for Model 1 is 55 degrees counter-clockwise from the x-axis in all panels 
and 50 degrees for Model 2.
The line-of-sight velocities along the slits are plotted in the left panel.  
The right panel shows the full line-of-sight velocity field.  The middle and right boxes are 18 kpc a side whereas the 
left box is scaled to the length of the slit (16 kpc along the x-axis).


In both models the LMC disk retains a pronounced velocity gradient along the same major axis; perhaps surprisingly, the LMC disk retains a 
well defined rotation curve despite a direct collision with the SMC.  Indeed, \citet{hopkins2008, hopkins2009} showed that disks can survive 
even a 1:1 mass ratio major merger.   However, the disk kinematics in Model 2 are more distorted than in Model 1, particularly the 0 velocity field.  

There are observed asymmetries in the LMC's gas and stellar kinematics. 
It has long been noted that the HI kinematic center is offset by $\sim$1 kpc from {\it both} the stellar kinematic and photometric center, which is roughly
centered on the stellar bar \citep[as illustrated in][]{cole2005}.  However, upcoming work by Kallivayalil et al. (in prep) using a 3rd epoch of HST 
data provides proper motions of high enough accuracy to independently constrain all parameters of the LMC rotation field and geometry, including the dynamical center. 
The best-fit stellar dynamical center from the proper motions {\it agrees} with the HI dynamical center determined by Kim et al. (1998), but remains offset from the 
photometric center (van der Marel \& Kallivayalil, in prep.).  

Each panel in Figures ~\ref{ch6fig:LMCkinModel1} and ~\ref{ch6fig:LMCkinModel2} is centered on the peak of the stellar density of the simulated LMC disk; 
i.e. the photometric center.
In Model 1, the stellar density peak is coincident with the kinematic centers of all tracers (vertical dashed line in left panel crosses zero where the 
major slit velocities do). Thus, contrary to observations, the stellar and gas kinematic centers are coincident with the photometric center and the center of the stellar bar.  

In Model 2, the zero velocity field of the stars and gas is twisted such that the velocity gradient does not cross zero at the location of the stellar density peak. 
The kinematic centers of all tracers are offset by about 1 kpc from the photometric centers, as observed. 
 However, the gas and stellar kinematics are somewhat discrepant from each other.  This is illustrated by placing a third slit
1 kpc above the major axis (illustrated in the right hand panel of Figure~\ref{ch6fig:LMCkinModel2}, in green). The old and young stellar line-of-sight velocities along this slit
are similar to those along the original major axis. The gas line-of-sight velocities, on the other hand, have changed, crossing the zero axis roughly 1 kpc further away than 
seen for the stars. This implies that the shape of the zero velocity field across the face of the gaseous and stellar disks are different and is likely a result of the warped stellar bar.
 Note that neither model predicts strong differences in the rotation curves traced by the young ($<$ 1 Gyr) and older stellar populations, as expected from \citet{olsen2011}. 


 Again, the simulated SMC kinematics will be presented in a forthcoming paper. However, we mention here that the resulting kinematics are much more consistent with 
 Model 2 than Model 1.  It appears that a direct collision with the LMC is required to erase the initial velocity gradient in the older stellar population.  \citet{zaritsky2000} 
 was unable to find a pronounced velocity gradient in the RGB population within a radius of 2 kpc of the center of mass, despite the existence of a 60 km/s 
 velocity gradient in the gas \citep{Stan2004}.  In our simulations, the gas
is able to cool, since it is dissipative, and the original gas disk survives the tidal shocks resulting 
 from the direct impact with the LMC, whereas the older stellar population does not. 
In Model 1 (no direct impact with the LMC), there is always a pronounced gradient in both the stellar and gaseous components of the SMC.

\section{Discussion}
\label{sec:Discuss}

In this study we have shown that it is possible to explain the nature of the LMC's off-center stellar bar ($\S$~\ref{sec:Bar}) and
gas and stellar kinematics ($\S$~\ref{sec:Kin}) in a model that self-consistently reproduces the general large scale gas morphology 
of the Magellanic System ($\S$~\ref{sec:Stream}) in a first infall scenario.  
To do this, we invoked a recent direct collision between the LMC and SMC (Model 2). 
Here we discuss some of the testable consequences of such a scenario.

Note that the resulting simulated SMC kinematics and structure and an expected
 stellar counterpart to the Magellanic Stream will be discussed in upcoming papers.

\subsection{The Recent Star Formation History of MCs}
\label{sec:SF}

\begin{figure*}
\includegraphics[width=2.5in]{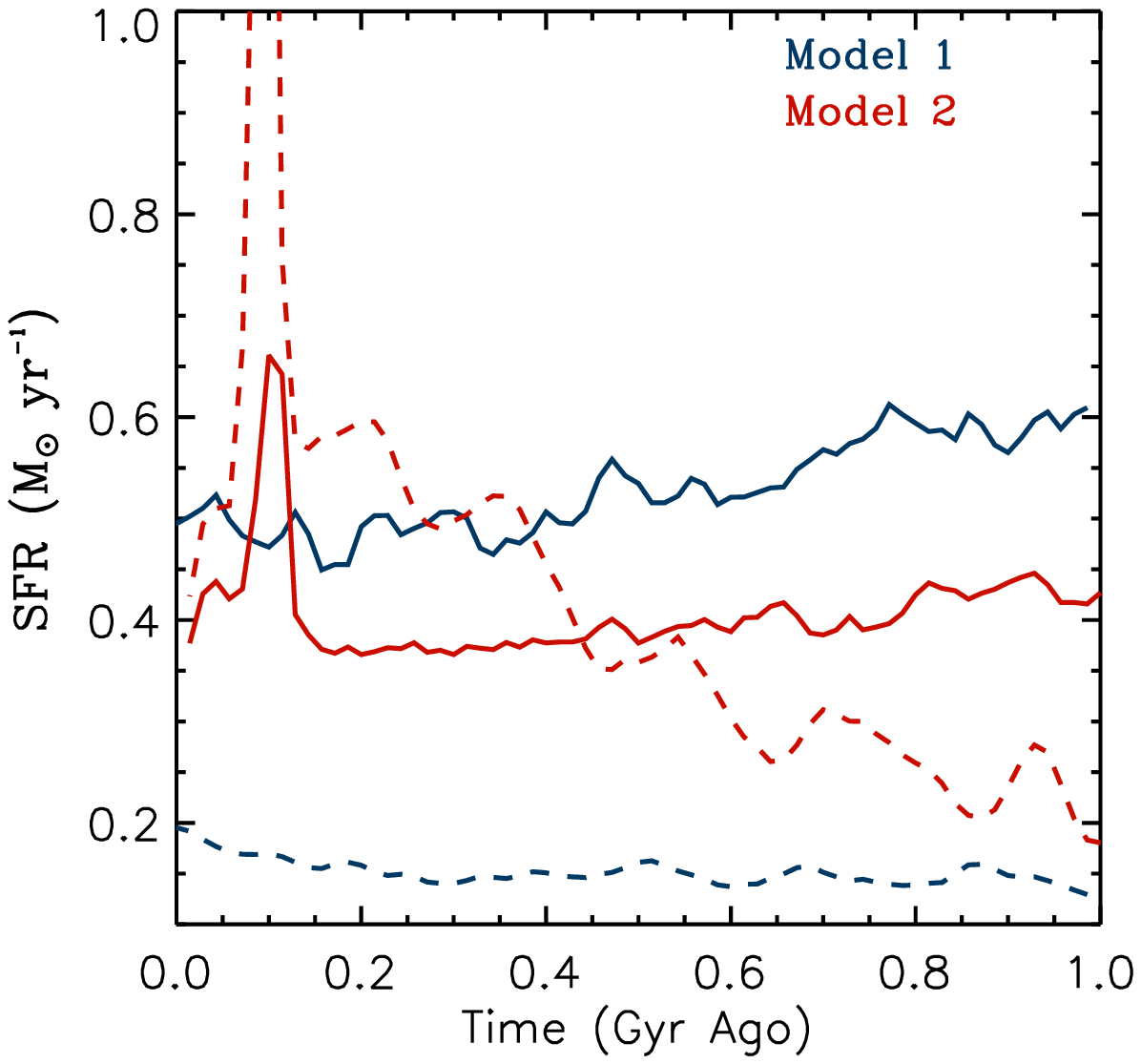}
 \caption{\label{ch5fig:SFRshort}  
  The simulated SFR is computed within a radius of 15 kpc for the LMC (solid lines) and 2 kpc for the SMC (dashed lines) and plotted as 
  a function of time since the MCs first crossed within R$_{200}$ of the MW for Model 1 (red) and Model 2 (blue).
 The corresponding orbital histories are plotted in Figure ~\ref{ch4fig:LSMWOrbit}. In Model 2, the MCs experience a close encounter
 at 0.1 Gyr ago (time 0 indicates today). }
 \end{figure*}

\begin{table*}
 \centering
 \begin{minipage}{200mm}
 \caption{Current Star Formation Rate \label{ch5tableSFR}}
 \begin{tabular}{@{}lcccc@{}}
 \hline
Galaxy & Model 1  & 	Model 2 	&  H$\alpha$ \& IR 	& Free-free Emission\\
			    &  ($\Msun$/yr)                & ($\Msun$/yr)   		&   ($\Msun$/yr)   			& ($\Msun$/yr) 	\\
\hline
\hline
LMC		&      $0.5 $				&  $0.4 $				&  0.25    					& 0.14 \\
SMC 	&      0.2 					&  0.4				&  0.05-0.08	 			& 0.015 \\
\hline
\end{tabular}
\tablecomments{
The SFR is computed within a radius of 15 kpc for the LMC and 2 kpc
for the SMC. The third column indicates the mean values determined from H$\alpha$ and MIPS emission by \citet{whitney2008} for the LMC 
and from IR and H$\alpha$ by \citet{wilke2004} and \citet{kennicutt1986}, respectively.
The fourth column presents a lower limit, determined from the free-free flux measured by \citet{mubdi2010} using WMAP.}   
\end{minipage}
\end{table*}

 A direct recent collision between the LMC and SMC would likely leave
notable marks in the star formation
histories (SFHs) of both of these galaxies. A correlated burst of star
formation during such a recent encounter
has been theorized in many previous numerical studies \citep[e.g., ][]{GN,
Bekki, bekki2007}.
Observationally, there is significant debate over the existence of
correlated bursts of
 star formation within both galaxies.  \citet{harris2009} claim that the
total SFR in the LMC was higher than average $\sim$100 and 500 Myr ago and
 in the SMC at $\sim$60 and 400 Myr ago.  Other authors claim that the LMC
shows global enhancements $\sim$ 125 and 800 Myr ago \citep{pietr2000b}
 and in the SMC at $\sim$100 Myr only \citep{pietr2000a}.  The general
consensus does appear to be that both galaxies show a steadily
increasing SFR
 over the past Gyr \citep{harris2009, noel2009, mccumber2005}.
Testing this scenario reliably depends on the accuracy of the timing of
the collision/interactions between the L/SMC and the adopted star
formation prescriptions.

 We can bracket the time range for this impact to be within 100-300 Myr.
The
upper limit on the collision timescale comes from the oldest detected
stellar populations in the Magellanic Bridge, which are believed to form
in-situ \citep{harris2007} and thus mark the formation of time of the
Bridge. As such, given the model parameter uncertainties, we cannot
use the models here to definitively predict the timing of the collision
and consequent star formation.

 The exact timing of the collision is strongly dependent on model
parameters. In the
presented Model 2, the most recent collision occurred $\sim$100 Myr ago. 
However, the MCs are too close together today in this model,
indicating that the true collision likely occurred earlier.

Given our adopted prescriptions, the modeled SFHs of both MCs are plotted as
a function of time since they
 first crossed within R$_{200}$ of the MW in Figure~\ref{ch5fig:SFRshort}.
 Note that the star formation rate is derived from the gas density above a set threshold 
 value.  
Contrary to Model 1, the SFH of the SMC over
 the past Gyr in Model 2 increases steadily, as observed  \citep{noel2009,
mccumber2005}. The Model 2 result occurs because the separation between
 the MCs is smaller than in Model 1, hence the relative importance of
tidal distortions from the LMC and consequent triggered star formation is
also stronger. 

  The LMC is observed to be unusually blue relative to analogs with similar R-band 
 magnitudes identified in SDSS \citep{tollerud2011}.
 This is likely a result of two factors: 1) if the MCs are on their first infall and just past their 
 first pericentric approach to our MW, they may be 
 experiencing triggered star formation induced by MW tides. 
2) Interactions between the MCs have likely kept the SFR in the LMC higher than 
 it would be if it did not have a companion. 
  The SFRs of galaxies are known to increase as a function of separation to a close companion 
  \citep[][and references therein]{larson1978, woods2010, patton2011}. 
 Since LMC-SMC pairs are rarely found around MW type hosts \citep{liu2010, BoylanBesla2011}, 
 it is natural that the LMC should have an anomalously high current SFR relative to the average 
 analog (which is more likely to be isolated). 
  This theory can be tested by comparing the LMC's color/SFR to a sample of 
 Magellanic Irregulars (LMC analogs) with known close companions.  

The recent collision between the MCs in Model 2 results in a dramatic increase in
the SFR in the SMC at that time. As discussed above, the exact timing  of
the true collision is quite uncertain. The magnitude of the simulated 
burst is inconsistent with observations of the SFH of the SMC
\citep{harris2006}. Furthermore, the modeled SFRs today are also higher in
both MCs than observed (see Table~\ref{ch5tableSFR}).  This likely points to
significant problems in our adopted star formation prescriptions.

The star formation prescription we have adopted in these simulations
follows a Kennicutt-Schmidt volume density law with a local volume
density cutoff for star formation of $n_H =0.13
cm^{-3}$. \cite{springel2003} showed that when combined with
appropriate star formation timescales and typical scale-heights this gives
a good match to the observed Kennicutt star formation surface density
relation for relatively massive galaxies.
However in the regime where galactic gas is
dominated by atomic hydrogen and where molecular, star forming, gas
constitutes only a fraction of the gaseous content, one has to properly account
for the formation of local density enhancements and molecular hydrogen
formation \citep{robertson2008, kuhlen2011, krumholz2009,
Hop2011}.
In low mass galaxies that tend to be metal poor, such as the SMC
\citep{fox2010}, molecular gas formation and therefore also the star
formation is very inefficient. Theoretical models show that not accounting
for the details of molecular gas formation and using global metallicity
independent, low density, threshold for the
ISM model can lead to serious overestimates of the star formation rates  of
metal-poor galaxies \citep{kuhlen2011}.
While larger improvement can be made by accounting for metallicity
dependent molecular gas formation in sub-resolution models, direct modeling
of processes that self-regulate formation of molecular clouds, their star
formation and related feedback requires more a complex ISM model and 
resolution that is beyond the simulations used in this work
\citep{Hop2011}.

Since the bulk of this study focuses on the morphological and kinematic
properties of the simulated MCs and the dynamics of the
Magellanic
System, the detailed star formation prescription does not alter any of the
main conclusions in this work.  However, it does limit the predictive
power of
the models concerning the chemical evolution and SFHs of the MCs.  Future
detailed studies of star formation prescriptions/feedback in a repeated
series of Model 2 collisions can be compared directly with the
multiples observational data sets for the SFHs of the MCs. Such a study
may be a powerful method of constraining the appropriate sub-resolution
physics for Magellanic
Irregulars and shed light on to the nature of bursts of star formation in
the histories of the MCs.

\subsection{Star Formation in the Bridge and Stream}

An additional check for our proposed Models is the predicted locations of ongoing star formation in the simulated Magellanic Bridge and Stream. 

\begin{figure*}
\begin{center}
\mbox{{\includegraphics[width=6.5in, clip=true, trim= 0 0.5in 0 0]{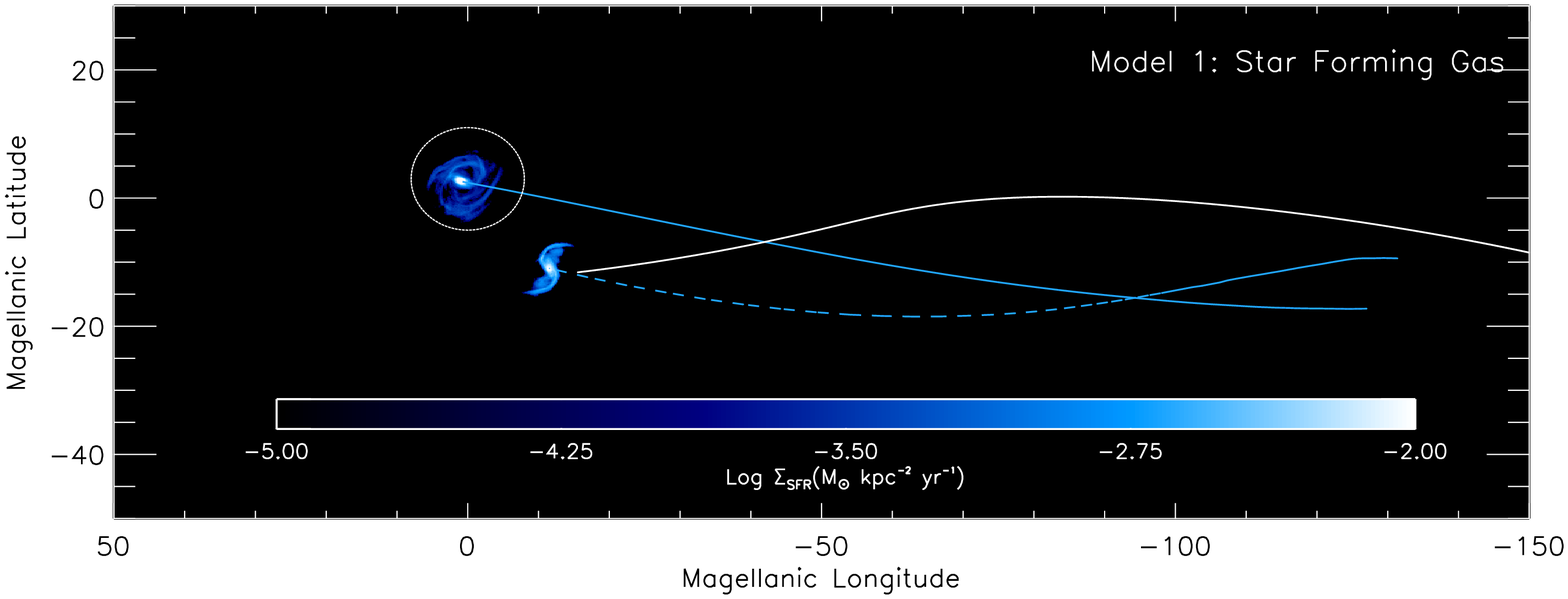}}}\\
 \mbox{ {\includegraphics[width=6.45in, clip=true, trim= 0.09in 0 0 0]{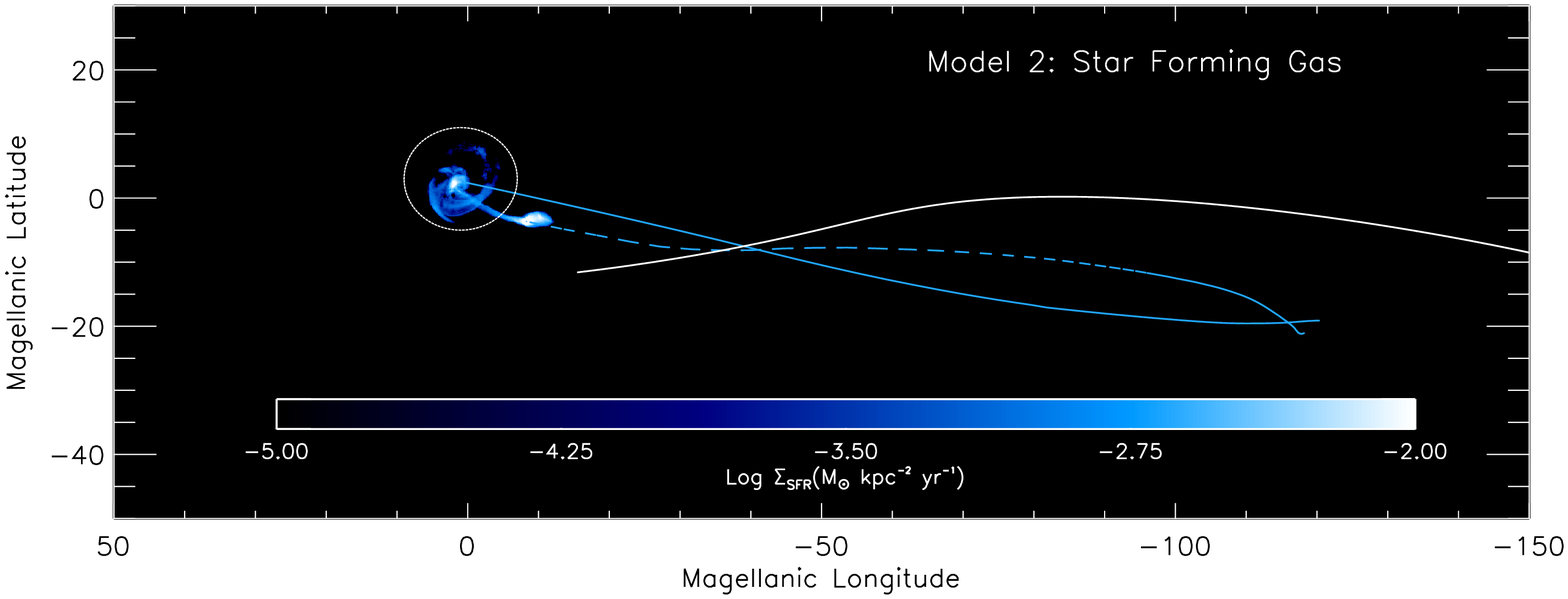}}}\\
 \end{center}
 \caption{\label{ch5fig:SFstream} The instantaneous SFR density for the simulated Magellanic System is plotted for Model 1 (top panel)
 and Model 2 (bottom) as a function of Magellanic Longitude. The SFR density is proportional to UV luminosity and identifies the location of gas 
 that is currently forming stars.  The white circle indicates the observed extent of the LMC's disk and the solid white line denotes the 
 observed location of the Stream. Blue solid(dashed) lines indicate the past orbits of center of mass of the LMC(SMC). }
 \end{figure*}

 Figure ~\ref{ch5fig:SFstream} shows the instantaneous SFR density in the Magellanic System for Models 1 and 2; the SFR density
 is derived from the gas density.
These plots indicate the location of gas with densities above the star formation threshold and should thus 
be forming stars.  While these plots may not be quantitatively accurate (the results clearly depend on the star formation prescription),
 they do highlight the location of high density gas that should be forming stars in any subgrid model. 

 In both cases the gas densities in the Magellanic Stream are too low to form stars. This is consistent with observations; molecular gas has not been 
detected in the densest cloudlets of the Stream, suggesting that star formation is not actively occurring there \citep{matthews2009}.

In Model 1 stars do not form in the Magellanic Bridge, whereas in Model 2 a well-defined bridge of star forming gas
is seen connecting both galaxies.   Stars as young as 10-40 Myr have been detected in the Bridge \citep{demers1998}, 
indicating that star formation is on-going there, as these stars could not have migrated from the SMC during their lifetimes. 
Furthermore, \citet{harris2007} was unable to locate stars older than 300 Myr in the Bridge, lending support to the idea that
the majority of young stars there were formed in-situ. 
 The different results between Model 1 and Model 2 indicate that 
a very close/direct encounter between the MCs may be required in order to trigger in-situ star formation in the Bridge, otherwise 
gas densities in the tidal bridge should not be significantly different from the Magellanic Stream (i.e. the tidal tail).
A close encounter has generally been invoked to explain the properties of the Bridge in previous studies \citep[e.g., ][]{Yoshizawa, Connors2006}.  
Clearly shock induced star formation is a relevant process in the direct collision scenario presented here and must be accounted for 
in order to characterize this environment accurately. 

Note that, although these results seem to favor Model 2, ram pressure compression of the Bridge region may also lead to triggered star formation 
\citep[e.g.,][]{mastro2009}. 
Although, in this case there likely should be stars forming in both the Leading Arm and the Magellanic Stream as well. 

The SFR along the Bridge in Model 2 increases steadily towards the SMC; this should be testable observationally. 
Indeed, \citet{harris2007} showed that the distribution of blue stars is more dense in the Bridge towards the 
SMC. Spitzer observations also confirm the presence of young stars  in the ``wing" of the SMC, which leads to the Bridge 
 \citep{gordon2009}.  Furthermore,  H$\alpha$ measurements of the Bridge by the Wisconsin H$\alpha$ Mapper 
(WHAM) survey indicate that the H$\alpha$ emission in the Bridge increases steadily towards the SMC (Barger et al. in prep). 

\begin{figure*}
\begin{center}
\mbox{ \includegraphics[width=2.5in, clip=true, trim=1.5in 3in 1.5in 3in]{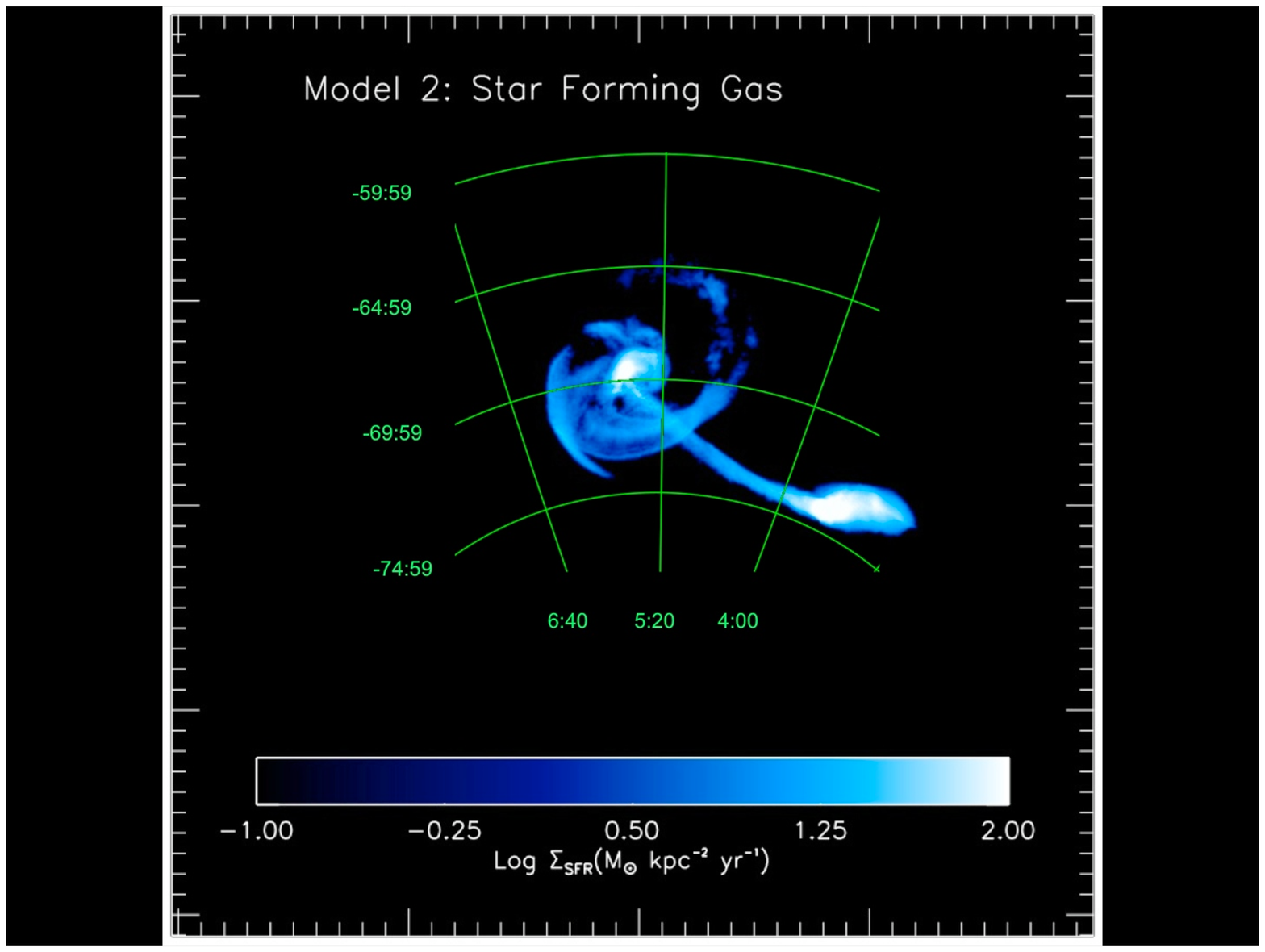}}
 \end{center}
 \caption{\label{ch5fig:SFstream2} A zoom-in on the instantaneous SFR density in the LMC and SMC in Model 2 from the bottom 
 panel of Figure~\ref{ch5fig:SFstream}. RA and DEC grids
 are overplotted in green (North is to the top and East is to the left). This image indicates where stars are expected to be forming
 in the LMC {\it today}:  star forming gas is concentrated in the South-East and a single spiral arm of star formation is seen in the North-West. 
 The SFR increases along the simulated bridge towards the SMC (South-East).   }
 \end{figure*}

In Figure~\ref{ch5fig:SFstream} the bar region of Model 1 is currently the dominant site of ongoing star formation in the LMC (contrary to observations). 
The distribution of star formation in the disk appears similar to an isolated galaxy with a number of spiral arms. 
On the other hand, in Model 2 the bar is not the most prominent site of star formation today. 
Figure~\ref{ch5fig:SFstream2} provides a zoomed in view of the Model 2 results in the line-of-sight frame for the LMC disk. 
 The center of mass of the stellar disk is located just to the right of an intense star-forming knot that 
 indicates the impact location of the SMC-LMC collision.
As discussed earlier, this occurs because the bar is warped with respect to the gas disk plane and is therefore inefficient 
at funneling gas along it. 
 
Without appropriate feedback prescriptions the appearance of the 
impact site today is difficult to assess.  The thickness of the gas disk near the impact site has increased relative to the rest of the disk
 (see where the SMC's trajectory crosses the LMC disk plane in Figure~\ref{ch6fig:LMCModel2}).  Intriguingly, this is also true of the active 30 Doradus star forming region \citep{padoan2001}, 
 which is located at roughly the same location. 
 However, the SFH of the true 30 Doradus region indicates that it has 
only been an active site of star formation for the past 12 Myr  \citep{harris2009}. 
Without accurately capturing star formation induced by shocks and including feedback, it is unclear whether the initial collision (100-300 Myr ago)
 is related to the triggering of star formation in 30 Doradus today or whether its remnant is related to any of the supergiant shells in the LMC disk
 \citep{kim1998, Book2008, Book2009}.

\subsection{Mass Breakdown}

The initial and final mass contained within characteristic radii of the simulated LMC and SMC and the resulting mass of the simulated
stream are listed in Table~\ref{ch4tableMass} and compared with 
observations.  Quoted gas mass estimates are for the {\it total} gas 
component; the neutral HI content will be lower than these values.  We do not include a UV ionization model in these
simulations and thus cannot accurately estimate the neutral gas fraction. 

In general, the mass estimates for the LMC and SMC agree with the 
observations within a factor of two. However, the final gas mass estimates for the MCs are lower 
than those observed and the simulated stream is about a factor of five lower. 

As discussed in $\S$~\ref{sec:SF}, given the adopted star formation prescriptions, the SFR in the MCs 
is being overestimated in the simulations. As a result, the gas consumption timescale is also overestimated.  The adoption of 
different prescriptions and/or changes in model parameters so that the SMC does not lose as much gas at early times (e.g. by reducing 
the scale length of the SMC's gas disk) may allow for the SMC to retain its gas for longer, allowing more material to be removed at 
later times.

The mass budget of the Stream is also likely underestimated 
 because ram pressure stripping has not been modeled here; 
 the bulk of the missing material is in the region closest to the MCs, rather than the tip of the tail. But this still means that more gas
 needs to be retained by the MCs themselves to provide a gas reservoir for this process to operate at late times.  This suggests that the larger 
 problem here is the adopted star formation prescriptions.
 
 Probably more significantly, the total initial gas budget of the MCs has been underestimated in these simulations; 
 the amount of gas initially modeled in both the 
 LMC and SMC cannot explain the total gas budget (including both neutral and {\it ionized} components) of the Magellanic System. 
Although the neutral HI gas content of the observed Magellanic System has been well quantified, the ionized gas fraction is poorly understood. 
\citet{fox2010} find that at the tip of the Stream as much as 95\% of the gas may be ionized. Along a similar line, \citet{lehner2008} find
 that gas in the Magellanic Bridge is 70-90 \% ionized.  This means that the total gas mass budget of the 
Stream (and consequently of its progenitor) is significantly underestimated.  Note that this statement implies that the LMC and SMC must have a
significant amount of dark matter in order to make their initial baryon fractions consistent with cosmological expectations; this is the main 
motivation for the large total infall masses adopted in this study.
 
 The fact that the stellar mass in the central regions of the SMC does not change substantially indicates that the bulk of the material
 being removed from the SMC via LMC tides is from the outskirts of the SMC. This explains why the stellar counterpart of the 
Stream is so faint.  We will comment on the observability of the this stellar stream counterpart in an upcoming paper. 

\begin{table*}
 \centering
 \begin{minipage}{160mm}
 \footnotesize
 \caption{Initial and Final Mass Distributions \label{ch4tableMass}}
  \begin{tabular}{@{}lccccc@{}}
\hline
Galaxy & Property	 & Initial               & Model 1     & Model 2 		&  Observed \\
			     &    				 &    ($10^9 \Msun$ )	          &    ($10^9 \Msun$)         & ($10^9 \Msun$)             & ($10^9 \Msun$)  \\ 
\hline
\hline
LMC     	            & Stars ($<$ 9 kpc)    	 &    $2.5   $  		&    $3.1 $ 			&  $3.1 $   		&    $2.7 $ (1) \\
                                &  Gas ($<$ 5 kpc)       	&    $0.87 $   	 	&    $0.17  $  			&    $0.26 $  		&  $0.441$  (2) \\  
                                &  Total ($<$9 kpc)     	&    $18 $    		& $21  $   		        		&  $21  $ 		&   $13 (\pm 3)$ (3) \\
                                &  Total ($<$4 kpc)       	 &    $6.9  $   		&  $7.3   $     			&  $7.1  $   		&  $5 $ (4) \\
\hline 
SMC			   &   Stars($<$ 3 kpc)       	&    $0.20   $  		&   $0.26  $   	 		& $0.20  $ 		&  $0.31  $ (5) \\
		  	   &    Gas ($<$ 3 kpc)       	&    $0.20   $ 		&   $ 0.16  $  			&  $0.18 $  		&  $0.42  $ (5) \\
			   &    Total ($<$ 3 kpc)    	&     $2.1   $  		&   $1.8  $    			& $1.1 $ 			&  $2.7-5.1 $ (6) \\
			   &                                      	 &                                  &                            		&                                   &  $2.4  $ (4) \\ 
			   &   Total($<$ 1.6 kpc)     	&    $0.84$    		&   $0.70$    			& $0.49 $  		& $1.4-1.9$ (6) \\
\hline
Stream\footnote{The Stream is defined as material at Magellanic Longitude less than -30.  We have also accounted for the distance 
in our simulated stream in order to properly compare to the observed mass, which is computed modulo the distance squared. }
	  & 	Gas 		&   				& $0.12$                           &     $0.10$                     &  $0.5$ (7) \\						 
 			  &    Stars 			&  				&  $0.6$                               &   $0.4$                       &                    \\
\hline
\tablecomments{(1) \citet{vanderMarel} : The outermost data point lies at 9 kpc; (2) \citet{Bruens2005};
 (3) \citet{vanderMarel2009} ; (4) \citet{kim1998}; (5) \citet{Stan2004}; (6) \citet{harris2006}; (7) Here we have accounted for the 
 average distance to the simulated streams. The measured value is $1.25\times10^{8} (d/55 \rm{kpc})^2$ \citep{Bruens2005} 
 + $2.0\times10^7(d/120)^2$ \citep{nidever2010}.
The quoted simulated gas mass refers to the total mass, not just that of the neutral HI, whereas the observed values refer to HI gas only.}
\end{tabular}
\end{minipage}
\end{table*}

\subsection{Tidally Captured SMC Stars}

In the presented models there is a continual transfer of material from the SMC to the LMC.  
In particular, there is expected to be a population of stars that are tidally stripped from the SMC and 
captured by the LMC in both models. 

Recently, \citet{olsen2011} discovered a population of metal poor RGB stars in the LMC field that have 
different kinematics from those of local stars in 
the LMC disk. \citet{graff2000} have also identified a possible kinematically distinct collection of carbon stars and suggest that this
 population lies outside of the LMC.  The discovery of such stars is a natural theoretical expectation from any tidal model for the 
LMC-SMC interaction.  To date, no stars have been detected in the Magellanic Stream \citep{NoStars} and only stars younger than 
300 Myr have been identified in the Bridge \citep{harris2007}.   Note that the 
\citet{harris2007} observations 
 focused on the leading ridgeline (location of the highest gas density) of the Magellanic Bridge.  This leading edge would
 currently be experiencing maximal ram pressure and so it is possible that the peak gas density in the bridge is displaced from 
 the tidal stellar population theorized to be there.
 \citet{harris2007} constrained the stellar density of a possible offset stellar population using 
 2MASS observations, but the 2MASS sensitivity limit of 20 Ks mag/arcsec$^2$ 
 is likely far too low to detect the expected faint stellar bridge ($>$30 Ks mag/arcsec$^2$; Besla et al. in prep).

The potential identification of tidally stripped SMC stars in orbit about the 
LMC may be a key discriminant between various model interpretations of the origin of the Magellanic Stream, as they should not be present
in a pure hydrodynamic model \citep{Mastro} or one that relies on stellar outflows \citep{nidever2008, olano2004}.

\begin{figure*}
\begin{center}
\mbox{\includegraphics[width=5in, trim=2.5in 0.57in 0. 0.05in, clip=true ]{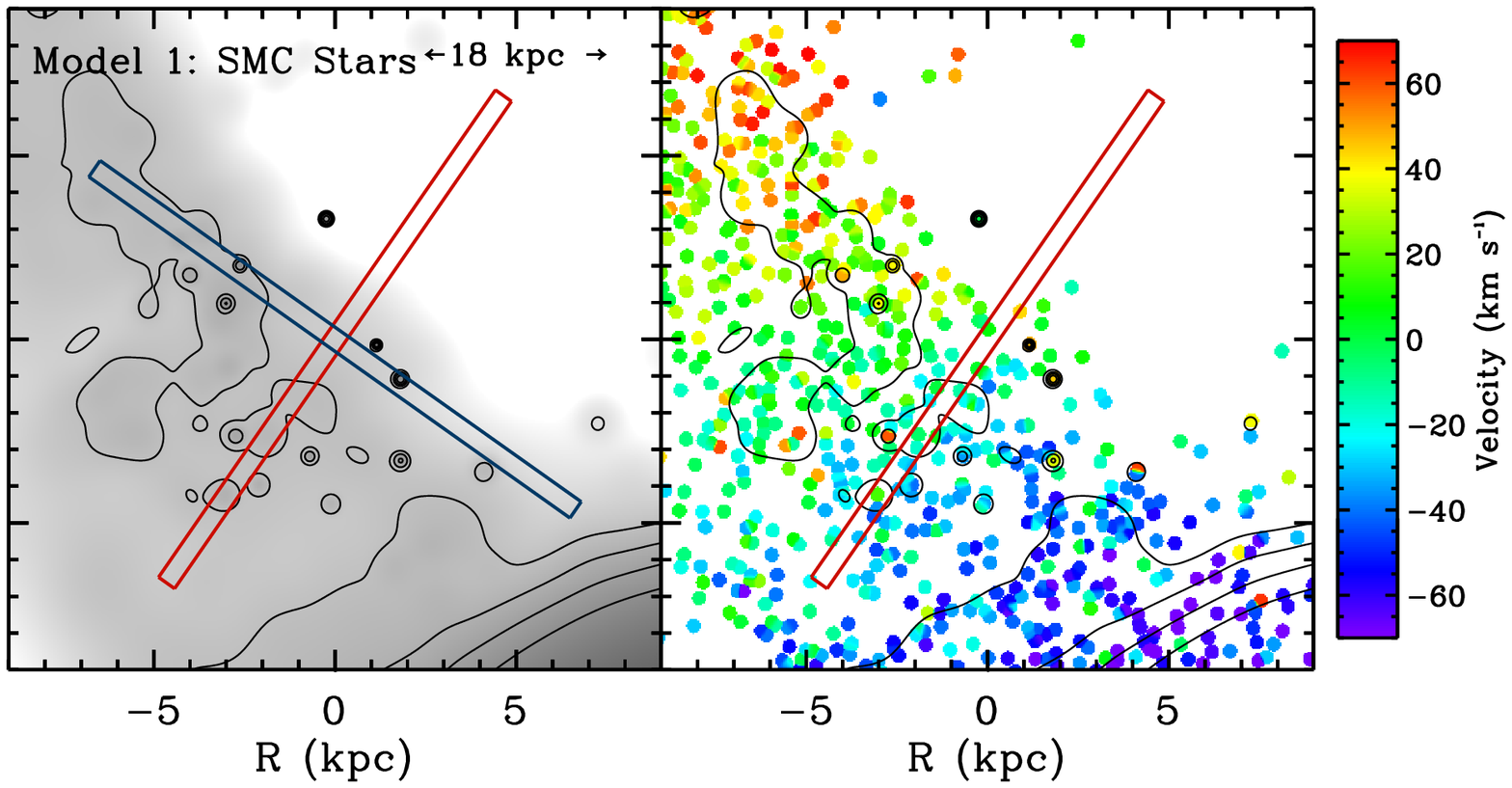}}\\
\mbox{ \includegraphics[width=5in, trim=2.5in 0.05in 0. 0.05in, clip=true ]{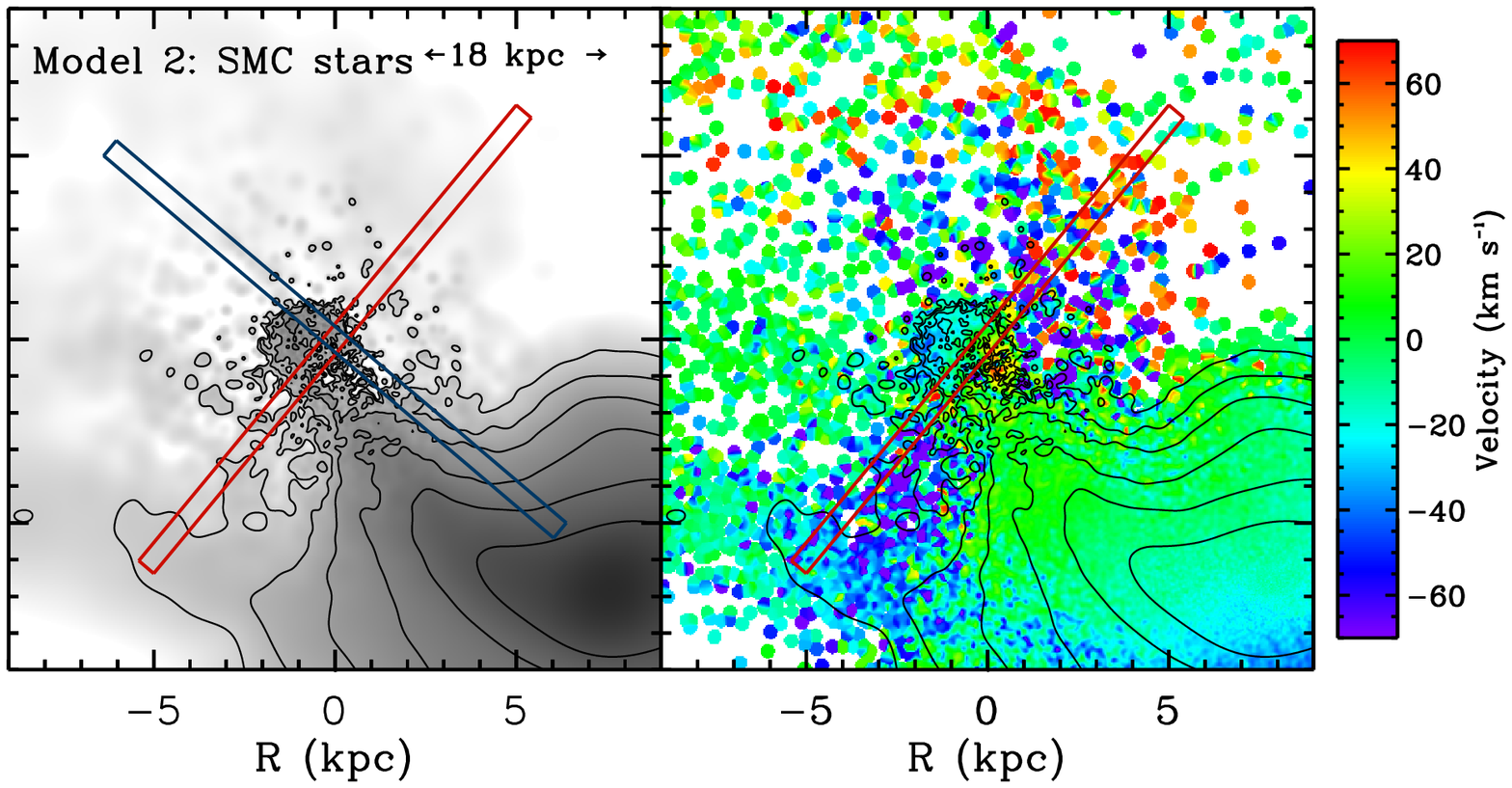} \hspace{0.01in}}\\
\caption{\label{ch6fig:TransferKin} The kinematics of SMC stars that are captured by the LMC in Model 1 (top panel) and Model 
 2 (bottom panel) are illustrated in the line-of-sight frame of the LMC's disk. Only particles initially belonging to the SMC are plotted. 
 The center-of-mass velocity of the LMC's stars has been subtracted from the 
 velocity field of the transferred material. 
  The left panels shows the stellar density map with the highest density contours overplotted. The size of each box is 11 kpc per side, the same
  as that in Figures ~\ref{ch6fig:LMCkinModel1} and ~\ref{ch6fig:LMCkinModel2} .  
  The slit along the major and minor kinematic axes of the LMC are overplotted for reference.
   The color gradient ranging from blue to red indicates material moving
 towards (blue) and away (red) from the observer. }
 \end{center}
 \end{figure*}

To test whether the simulation results for the tidal debris are consistent with the \citet{olsen2011} detections, we plot in
Figure ~\ref{ch6fig:TransferKin} the expected distribution and kinematics of the stars captured by the LMC from the SMC
for both Model 1 (top panel) and Model 2 (bottom panel). 
The box size and orientation is the same as in Figures ~\ref{ch6fig:LMCkinModel1} and ~\ref{ch6fig:LMCkinModel2}, that is, the field
of view is centered on the LMC.  
The stellar line-of-sight velocities have been corrected for the center of mass motion of the LMC.  The SMC is located towards the 
South-West in this viewing perspective. Note that the SMC is actually present in this field-of-view for the Model 2 results, as this simulation 
resulted in the MCs being closer together than observed.

In Model 1 these transferred stars form a well-defined arc that is in orbit about the LMC: these stars are located behind the LMC 
disk. Comparing the velocity field to the bottom-right panel of Figure~\ref{ch6fig:LMCkinModel1}, the kinematics of these stars 
appear to be offset by nearly 90 degrees relative to the velocity gradient in the LMC's disk. 
In Model 2, the stellar debris from the SMC exhibits a large range in velocities ($\pm$ 150 km/s).  We made a velocity floor and ceiling of 
$\pm$70 km/s to better compare to the LMC stellar disk kinematics. In the North-West there are stars that appear to be moving towards 
the observer and in the South-East there are stars moving away from the observer - this is opposite to the observed kinematics 
of the LMC stellar disk.  Thus in both models, stellar debris captured by the LMC from the SMC is expected to have kinematics that are 
distinct from those of the LMC disk stars - this is a generic prediction of the B10 model and is consistent with the \citet{olsen2011} observations. 

\citet{olsen2011} estimate the mass of the observed SMC debris population to be 5\% of that of the LMC's current disk mass (i.e. $\sim 1.4 \times 10^8 \Msun$).
In Model 1, only a modest amount of stars is transferred from the SMC to the LMC ($\sim 0.2\%$).  As such, Model 1 cannot account
for this population. On the other hand, in Model 2, the LMC accretes 1.5\% of its current disk mass from the SMC. 

The exact distribution of SMC debris is certainly dependent on a number of parameters, but overall we can conclude that the \citet{olsen2011} 
results support a model in which LMC tides have been actively distorting the SMC. 
 Furthermore, in both models the accreted stars are largely located behind the LMC disk and may provide a natural explanation for the origin 
of the observed MACHO microlensing events (Besla et al. in prep).

 \subsection{The Nature of Magellanic Irregulars }

Dwarf galaxies are broadly referred to as galaxies with luminosities $<$0.1-0.3 of L$_\ast$. This definition encompasses a wide range of 
objects of varying morphology, including both the LMC and SMC.  In this work we have introduced a mass model for the LMC with 
a total mass of $\sim10^{11} \Msun$; it is questionable as to whether such a massive galaxy should be included in the same 
category as dwarf Spheroidal or dwarf Irregulars galaxies.  In particular, detailed analysis of the geometry \citep{vanCioni2001} and 
kinematics \citep{vanderMarel} of the LMC prove that it is a disk galaxy. Furthermore, when looking at the distribution of intermediate-age and old stars out 
to large radii (i.e. ignoring the visible light in the bar region), the LMC does not look at all irregular, but clearly resembles a spiral 
galaxy with an asymmetric one-armed spiral \citep{vanderMarel2001}.   

\citet{deVaucouleurs} suggested that Magellanic Irregulars have more in common with spiral galaxies than dwarfs, referring to the 
LMC as an asymmetric, late-type barred spiral galaxy.  They argue that Magellanic Irregulars represent an extension of the Hubble spiral 
sequence (Sa, Sb, Sc, Sd, Sm, Im), where the subscript m denotes ``Magellanic" ~\citep[see also, section 4.1.1 of ][]{Binney1998}.  
Spiral structure ``decays" along the sequence, with Sc having irregular spiral patterns and Im none at all.  This is true also of the 
Barred Spiral sequence (SB), with asymmetry referring also to the appearance/location of the bar.    Magellanic Irregulars encompass 
the late stages of both barred and unbarred spirals (e.g. Sd-Im, SBd-SBm); the LMC is classified as SBm under this scheme and the SMC an Im.

This work postulates that many barred Magellanic Irregulars may be perturbed versions of symmetric low-mass, bulgeless, barred galaxies, 
such as SBc type galaxies, where the bar is typically well-centered.  We further illustrate a mode of inducing such perturbations,
namely interactions with lower mass companions.  In this picture, the LMC should not be thought of as a ``dwarf" galaxy except 
in the sense that it is less luminous than the MW.

Asymmetric bars are typically not seen in massive galaxies, making them a defining characteristic of Magellanic Irregulars. 
This might be explained by noting that Sc/SBc type galaxies and Magellanic Irregulars (Sd-Im) are low mass systems compared to MW type 
galaxies and, correspondingly, they sample very different environments.  Such galaxies do not have bulges, have shallower central potentials, 
are more dark matter dominated, have different halo concentration parameters,  higher gas fractions and different ratios between 
the ISM temperature and virial temperature of their halos: all of these differences will influence the response of the system to tidal perturbations.  
 As such, even if the physical scenario is similar (mass ratio and orbital configuration), the response of a MW type galaxy to a 1:10 mass ratio
 direct collision is expected to be different than an LMC-SMC ($\sim$1:10 mass ratio) encounter. 

It is possible, for example, that the presence of a bulge may aid in stabilizing the bar of a high mass galaxy, preventing comparable 
asymmetries in the bar from arising.  However, a detailed numerical study of such parameter space to test such conjectures is
 beyond the scope of the presented study.

Magellanic Irregulars are ubiquitous in our Local Volume. In light of the theory presented here, it must also be true that interactions between 
low-mass barred galaxies and smaller companions are a relatively frequent occurrence. 
From \citet{hopkins2010}, the expected galaxy major merger (mass ratio$ >$ 1/3) rate is relatively flat as a function of 
host galaxy mass at z=0, so it is not expected that such encounters would be more likely for low mass systems.  \citet{stewart2008} find that 
25-40\% of hosts with mass of order $\sim 10^{11} \Msun/h$ have accreted a 1:10 mass ratio subhalo within the past 6-8 Gyr, which is the 
timescale for our isolated LMC-SMC interaction. This scenario is thus not at odds with cosmological expectations.  

It has been pointed out by \citet{wilcots2004} that many Magellanic Irregulars do not currently have companions, despite earlier 
claims of a high frequency of pairs by ~\citet{odewahn1994}. However, the number of observed interacting dwarf systems is steadily increasing.  
Recently, \citet{martinez2011} have discovered a stellar stream about the Magellanic Irregular galaxy NGC 4449, which is an LMC analog in terms
of its absolute magnitude. 
Although the stellar mass ratio of the disrupted object and the host is 1:50, the inferred dynamical mass ratio is between 1:10 - 1:5, making this 
system an analog of the late stages of an LMC-SMC type tidal interaction.  NGC 4449 was long thought to be an isolated Magellanic Irregular 
until observations of associated HI streams indicated that it likely had an encounter with an unseen companion \citep{Hunter1998}. Moreover, 
an unusual globular cluster exists in this galaxy with properties consistent with the nucleus of a disrupted galaxy \citep{annibali2011a, annibali2011b};
 such observations indicate that the NGC 4449 may also have had more ancient accretion activity, which may partially explain the significant 
 amount of mass in the HI streams surrounding the system. 
  Signatures of earlier accretion events in dwarf galaxies has also been presented by \citet{geha2005}, who find evidence for a counterrotating core 
  in the elliptical dwarf galaxy NGC 770 that they attribute to a minor merger event. 

Such observations clearly illustrate that LMC mass objects do cannibalize smaller companions; however, the hallmarks of these encounters, such 
as faint tidal streams, are challenging to observe.  
The ~\citet{wilcots2004} conclusions may thus indicate that the perturbing companion has already been cannibalized, causing most Magellanic Irregulars to 
appear as isolated objects.  Moreover, the Mbaryon/Mtotal ratio is a steep function of mass for these low mass systems; many dwarfs in the local group 
have extremely large mass-to-light ratios.  As such, a 1:10 total mass ratio companion may have a very discrepant stellar mass ratio, making the identification 
of such a companion challenging observationally. 

 In the context of the work presented here, Magellanic Irregulars are therefore key targets for deep HI and optical follow up
 observations as they are expected to be associated with tidal HI and stellar streams. Particular attention should be paid to Magellanic Irregulars with high current star 
 formation rates, such as NGC 4449 and the LMC, which may point to ongoing tidal interactions with a low mass companion. 
   Furthermore, there is clearly need for future observational and 
 theoretical studies to better statistically quantify the frequency of interactions between LMC mass galaxies and smaller companions in order to assess the 
 ubiquity of the theory presented here for the nature of Magellanic Irregulars. 


\subsection{Assessment of the Models}
   
In this study we have explored the consequent large scale structure of the Magellanic System, the internal 
structure, kinematics of the LMC and the recent star formation histories 
of the MCs in a first infall scenario; i.e. without strong tidal torques from the MW. We focus on two different 
models for the interaction history of the LMC-SMC, one of which invokes a direct recent collision (Model 2), 
whereas in the other, the MCs never get closer than 20 kpc (Model 1). 
The ability of the presented models to reproduce key observed features of the Magellanic System are summarized in Table ~\ref{SUMMARY}. 

Both models are able to reproduce the global large scale structure of the Magellanic System.  Overall, however, Model 1 provides better
 agreement with the properties of the Magellanic Stream, whereas Model 2 provides 
significantly better agreement with the structure and kinematics of the LMC. 

\begin{table*}
 \centering
 \begin{minipage}{140mm}
\caption{Observed Properties of the Magellanic System and How the Models Fare \label{SUMMARY} }
\begin{tabular}{@{}lcccc@{}}
  \hline
Object & Property	            & Model 1     & Model 2    &  Alternative  \\ 
\hline
\hline
Large Scale Structure		&    A Leading Arm   			  &       Yes 			&     Yes    			&  \\
			&   Location of Leading Arm	& 		No 			&  No			&  Ram pressure  \\ 
			&    A  150 degree Stream		  &       Yes 			&  Yes 			& \\
			&   Stream location offset from orbit    & Yes 			&    roughly		&   I.C. : SMC disk inclination\\
			&     Stream velocity gradient     &        Yes				&   roughly  		&   I.C. : SMC disk inclination \\	
			& Stream bifurcation			&     roughly				&  No			& Hydro instabilities \\
			&   Stream column gradient       &        No			    &  No  				&  Ram Pressure \\ 
			& 						& 					& 				&  I.C. SMC gas mass \\
			&   Stream total mass 		&     No				& No 			&  Ram Pressure;   \\
 			& 						& 					& 				&  I.C. SMC gas mass\\
			&    A  Bridge				&      Yes				& Yes 			&  \\
			&    SF in Bridge 			&     No				&  Yes   			& \\
			&   No SF in Stream			&   Yes				& Yes   			& \\
\hline
LMC			&  Rotation Curve  			&   Yes				&   Yes   		 &\\
			&  Offset gas and stellar		&   No 				&  Yes 		& \\
			&  kinematics  				& 					&   			& \\
			&  Offset bar				&  No				&  Yes 		&\\
			&  Bar not seen in gas		&  No				&  Yes 		& \\
			&  Warped stellar disk		&   Yes				& Yes 		& \\
			&  Elliptical stellar disk		&   roughly				&  Yes 		& \\
			& Current SFR				& No					& No				& SF details, feedback \\ 	
\hline
			
SMC			&  Increasing SFR $<$ Gyr	&  No				&  Yes 			 &  \\
			&  Current SFR				& No					& No				& SF details, feedback \\			
\hline
\tablecomments{  SF stands for star formation. I.C. stand for initial conditions. The column marked Alternative indicates
other possible factors that may help fix discrepancies between the observations and the models. }
\end{tabular}
\end{minipage}
\end{table*}

While neither model reproduces every one of the features listed in Table ~\ref{SUMMARY} (and indeed the real answer is probably somewhere in between the 
two presented models) it is still rather remarkable that a single self-consistent model (namely Model 2) can simultaneously reproduce a large 
number of these features.  Generally where the models fail (e.g. the column density gradient or location of the Leading Arm) the likely 
missing ingredients are ram pressure stripping owing to the passage of the galaxies through the ambient hot gaseous halo of the MW or a detailed
search of model parameters (e.g. initial gas mass or orbital parameters). 
Inconsistencies with the current SFR and recent SFHs are almost certainly a result of the star formation prescriptions employed and the lack of 
stellar feedback. 

A number of the discrepancies with the orbital model (e.g the SMC's position and velocity) can likely be addressed by a complete parameter search of, 
e.g., plausible mass ratios between the two galaxies (we chose a fixed
mass ratio of 1:10) and different orientations of the SMC disk relative to the LMC-SMC binary orbital plane.  The inclination of the SMC's disk
  can dramatically alter the location and properties of the Stream, 
   e.g. a retrograde coplanar configuration would inhibit the formation of a stream entirely. It can also change the way LMC torques affect 
   the SMC's motion. 
    This is likely the explanation for why the Model 2 results do not reproduce the exact velocity field or location of the Stream.  

The Stream is also known to be both spatially and kinematically bifurcated, leading \citet{Putman2003} to describe it as a ``twisted helix''. 
 The Stream in our Model 1 is not of constant column density along its width; between Magellanic Longitudes of -50 and -100 the simulated 
 stream appears to split into two high column density filaments.
This bifurcation is largely a result of the rotation of the SMC's initial disk. Note also that because the Stream is seen in projection, the appearance of this 
bifurcation is highly dependent on the viewing orientation (which is model dependent). This is likely why similar structures are not 
seen as clearly in Model 2.  But it should be noted that in the Gadget simulations the ISM is smoothed by an effective equation of state, whereas 
in reality inhomogeneities should be present in the real ISM (e.g. molecular clouds).  Stripping from a clumpier SMC ISM would lead to differences
in the distribution of the stripped material. 
We further expect that hydrodynamic instabilities, which are not modeled here, would capitalize on any initial density inhomogeneities and 
augment this bifurcation. 
It is clear from the significant observed turbulence \citep{nigra2010} and the existence of head-tail structures in cloudlets within the Stream \citep{Putman2003, putman2011}, that
hydrodynamic instabilities are shaping its internal structure. 
Gas drag will operate differently on these clumps depending on their densities, naturally leading to a velocity bifurcation as well. 

Although strong perturbations in the LMC disk are expected from a close encounter with a 1:10 mass ratio companion, without a large parameter study 
of impact parameters for the collision and mass ratios, it is unclear how generically these results can be applied to Magellanic Irregulars. 
A future study of these parameters will assess the robustness and longevity of the presented asymmetric structures (i.e. offset bars and one-armed spirals).
 In particular, it remains unclear as to whether a direct collision is always required to produce such structures.  Regardless, the existence of Magellanic 
 Irregulars with close pairs connected by gaseous bridges is certainly suggestive of a link to a similar interaction history as 
 that of the MCs.

It is worth pointing out that 
while it has been speculated that many of the features listed in Table ~\ref{SUMMARY} were directly 
related to interactions with the SMC, most of these links have never been illustrated by numerical simulations which self-consistently 
reproduce the global large scale features of the system.  In this light, particular 
successes of the simulations presented in this work include the reproduction of a warped, off-centered bar that is neither detectable in the gaseous 
disk nor actively forming stars while simultaneously forming a Bridge, Leading Arm and trailing 150 degree long Stream. 
Furthermore, we have not yet discussed the remarkable agreement of the simulated SMC model with its observed
 structure and kinematics from this same Model 2. These results will be outlined in a subsequent paper.

\section{Conclusions}
\label{sec:Conc}

We have explored two models for the possible interaction history of the LMC and SMC in an effort to simultaneously reproduce both the 
large scale gaseous distribution of the Magellanic System and the internal structure and morphology of the LMC.  Here we summarize our 
findings for the Magellanic System and the implications for the study of Magellanic Irregular galaxies more generally.

\subsection{Conclusions for the Magellanic System}

The resulting kinematics and structure of the LMC strongly favor a scenario in which the MCs have recently (100-300 Myr) experienced a direct collision
(our Model 2). Orbital models where the SMC never gets closer than 20 kpc to the LMC (e.g. Model 1) are able to reproduce the 
large scale structure of the Magellanic System, but poorly match the LMC's internal properties.  In particular, without a direct collision with the 
SMC, the LMC would be better described as a normal, symmetric barred spiral galaxy.  An upcoming paper will also illustrate that the observed internal kinematics 
and morphology of the SMC are also better described by this same collision model (Besla et al., in prep). 

This study illustrates that, surprisingly, the LMC's disk can maintain a fairly smooth stellar velocity field 
despite a direct collision with the SMC and that such a scenario can explain a number of observed features of the LMC disk: 

\begin{enumerate}
\item The gas and stellar kinematic centers of the LMC disk are coincident; however they are offset from the photometric center \citep{cole2005}. 
\item The old stellar disk of the LMC is thick and warped. The warping at the edges may give it a flared appearance \citep{alves2000}. 
\item The bar is warped relative to the LMC disk plane by ten degrees and is off-center relative to the dynamical center of the gaseous disk. While an offset bar in the LMC
 has been suggested by numerous authors as being the result of collisions with the SMC \citep[e.g.,][]{sub2009b, sub2003, bekki2007} this is the first time it has been modeled self-consistently.
\item The bar is not seen in the gas distribution \citep{kim1998} or as a site of on-going star formation, likely because the bar is warped out of the plane, inhibiting efficient gas funneling.
\item Gaseous ``arms" similar to those seen in HI maps of the LMC \citep{nidever2010, nidever2008} are stripped out of the LMC by the SMC in the direction of 
the Magellanic Bridge. This ``arm" was formed during a violent collision and may have signatures in polarization maps of the LMC's magnetic field (Mao et al. in prep). 
 This result also implies that there should be a metallicity gradient along the Bridge, increasing towards the LMC owing to contamination by LMC gas. 
\item A one-armed spiral is induced in the LMC's disk and is a site of on-going star formation. 
\item Stellar debris from the SMC is expected in the same field of view as the LMC disk. These stars will have differing kinematics signatures from the local
LMC disk velocity field  \citep{olsen2011} and may be the source of microlensing events towards the LMC (Besla et al. in prep). Such tidally stripped stars 
are not expected in a pure ram pressure stripping model for the Stream. 
\item The gaseous Bridge that connects the two galaxies is the site of on-going star formation, where the SFR increases along its length towards the SMC.
\item The Leading Arm extends $>$70 degrees ahead of the MCs and the Magellanic Stream extends 150 degrees behind them, reproducing the full extent of the 
Magellanic System.  The Leading Arm is a younger structure than the Stream since it formed as a tidal tail during the SMC's most recent orbit about the LMC. 
\end{enumerate}

These listed properties are a direct consequence of interactions with the SMC, confirming the
 suspicions of \citet{deVaucouleurs} that the LMC's peculiar morphology does not owe to interactions with the MW.

\subsection{Implications for Magellanic Irregulars}

There is potentially much to learn about Magellanic Irregular galaxies as a class by studying the Magellanic System in detail.   
In this work we have shown that the off-center bar and one-armed spiral arm of the LMC may be a product of close encounters
with its smaller companion, the SMC.  We conclude that interactions between a massive dwarf and a smaller companion,  even one 
10 times smaller in mass, can significantly alter the morphology of an otherwise 
normal looking, low-mass spiral galaxy. This study thus indicates that dwarf-dwarf galaxy interactions can be important drivers of their morphological 
evolution, without relying on interactions with a massive host. 


Given that off-center bars and one-armed spirals are common characteristics of Magellanic Irregulars \citep{deVau1964}, it is possible 
that such galaxies currently have or once had small companions. Known examples of Magellanic Irregulars with a close companion
 are thus prime candidates for follow up HI surveys to map the extended HI distribution, 
as the bridges that connect them likely have tidal tail counterparts.  We argue here that if such systems were accreted by a massive spiral, they could
form an analogous Magellanic System, i.e. two interacting dwarfs surrounded by an extended HI complex.  A potential example could be the interacting pair of
MC analogs, NGC 4485/4490, which are surrounded by an extensive HI envelope but are not in close proximity to a massive host \citep{clemens1998}.
The recently identified stellar stream about the Magellanic Irregular galaxy NGC 4449 ~\citep{martinez2011} gives further credence to this theory.

Quantifying the efficiency with which such dwarf-dwarf tidal interactions can remove gaseous material is directly relevant to questions about how dwarf galaxies 
lose their gas and how gas is supplied to more massive galaxies.  The frequency of such dwarf-dwarf encounters has not yet been quantified observationally or
 theoretically. As such, the relative importance of such encounters remains to be determined.  
However, by reproducing the large scale structure of the Magellanic System in a LMC-SMC tidal scenario, we have
 shown here that at least 50\% of the original gas budget of a small dwarf can be easily removed by tidal interactions with a larger dwarf companion, making 
 such encounters a potentially important mode of baryon loss on these mass scales.

While the accretion of an LMC+SMC analog by a MW type host is a relatively rare event today \citep{BoylanBesla2011, liu2010},
 modern models for the cosmological assembly of galactic-scale structures suggest that the MW halo formed from the earlier accretion and
disruption of LMC-mass objects \citep{stewart2008}.  As such, simulations of isolated Magellanic Irregulars (LMC analogs) with low mass companions 
and the direct comparisons of such simulations to observed analogs, such as NGC 4027 \citep{phookun1992} and NGC 3664 \citep{wilcots2004}, 
 provides a logical testing ground for our understanding of the morphological, kinematic and chemical evolution of the fundamental building block of a MW-type galaxy.

\section*{Acknowledgments}

The simulations in this paper were run on the Odyssey cluster
supported by the FAS Science Division Research Computing
Group at Harvard University.
We thank Knut Olsen, Alar Toomre, David Nidever, 
 Crystal Martin, Patrik Jonsson, Mary Putman and Dennis Zaritsky 
 for useful discussions that have contributed to this paper.
GB acknowledges support from  NASA through Hubble Fellowship grant
 HST-HF-51284.01-A.

\appendix

\section{The Role of Ram Pressure Stripping in the Formation of the Stream}
\label{sec:Ram}

Evidence for the existence of a hot gaseous halo component of our MW
 comes from absorption lines towards distant AGN \citep{sembach2003}. 
The MCs are therefore known to be moving through some ambient medium, and so the effects of ram pressure stripping and gas drag likely
play some role in explaining the gas distribution of the Magellanic System. 
  Our contention is that the impact of ram pressure is to modify the properties (velocity, location, mass) 
  of the gas distribution, but does not in itself 
explain the existence of the Stream, Bridge and Leading Arm Feature.  

To model the impact of ram pressure from a physical ambient medium representing the MW's gaseous halo
 with either an SPH or Eulerian grid-based code would require extremely high numerical resolution, 
as the gaseous stream is very diffuse.  Instead, we have created a toy simulation, where the Gadget-3 code was modified to include
a ram pressure acceleration term that is applied uniformly to all gas particles and operates opposite to the direction of motion 
of those gas particles. Following the \citet{Gunn} prescription, the ram pressure experienced by a gas particle moving through the hot
gaseous halo of the MW (density, $\rho_{\rm hot}$), at a speed $v_{\rm gal}$ is expressed as follows:

\begin{equation}
P_{\rm ram} = \rho_{\rm hot} v_{\rm gal}^2.
\end{equation}

The acceleration experienced by a gas cloud of surface density $\Sigma_{\rm gas}$ owing to ram pressure is:

\begin{equation}
a_{\rm ram} = P_{\rm ram}/\Sigma_{\rm gas}.
\end{equation}

Following \citet{vollmer2001}, we can describe an HI gas cloud as having a characteristic column
density of $N_{\rm HI} = 7.5 \times 10^{20}$cm$^{-2}$ \citep{sanders1985}. Thus, $\Sigma_{\rm gas} \sim$ N$_{\rm HI} \times m_{\rm HI}$.
 For a given velocity component v$_{i = x,y,z}$, the corresponding acceleration applied in that direction per timestep is: 

\begin{equation}
{\bf a_i} =  - \frac{n_{\rm hot}}{  {\rm N}_{\rm HI}} \rm{v}^2  \frac{\bf{\rm{v}_{i}}}{\rm{v}},
\end{equation}

\noindent where the volume density ($\rho_{\rm hot}$) has been replaced by the number density of the MW's ambient halo medium ($n_{\rm hot}$)
multiplied by the mass of the average particle in the halo, which we have assumed is $\sim m_{\rm HI}$.
The quantity $n_{\rm hot}$ is largely unconstrained, although an upper limit of 
$5 \times 10^{-4}$  cm$^{-3}$ is estimated \citep{Rasmussen, maloney1999}.   
We consider three values for this parameter:  $5 \times 10^{-5}$, $10^{-4}$ and $5 \times 10^{-4}$ cm$^{-3}$. We further assume that this density is 
constant all the way to R$_{200}$ = 220 kpc of the MW.  The ram pressure acceleration is applied to all gas particles as soon as the MCs 
enter within R$_{200}$ of the MW. Note that this acceleration will not be constant as a function of time since the velocity of the
particles changes as the MCs fall towards the MW.

We stress that this is a crude model, as there is no actual ambient gas density present and the force is applied uniformly to all particles.  
In reality, gas in the inner parts of the disk should be shielded by those on the outside. 
Moreover, other hydrodynamical instabilities, such as Kelvin-Helmoltz and Rayleigh-Taylor, will not be modeled in this set-up and we will 
therefore underestimate the amount of gas loss the galaxies may incur. This toy model will, however, give us a rough idea of how the
 position of the simulated Leading Arm, Bridge and Stream will evolve over time.

\begin{figure*}
\begin{center}
\mbox{{\includegraphics[width=6in, trim=0 0.6in 0. 0, clip=true ]{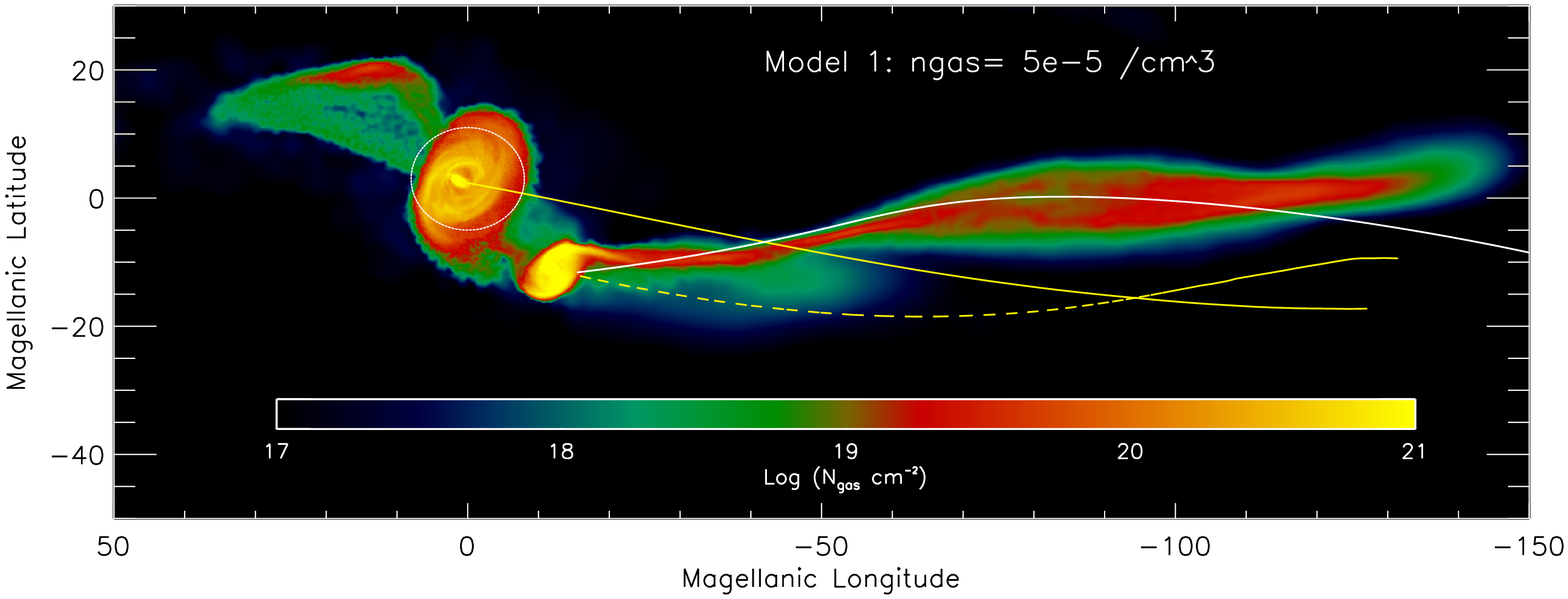}}}\\
\mbox{{\includegraphics[width=6in, trim=0 0.6in 0 0, clip=true]{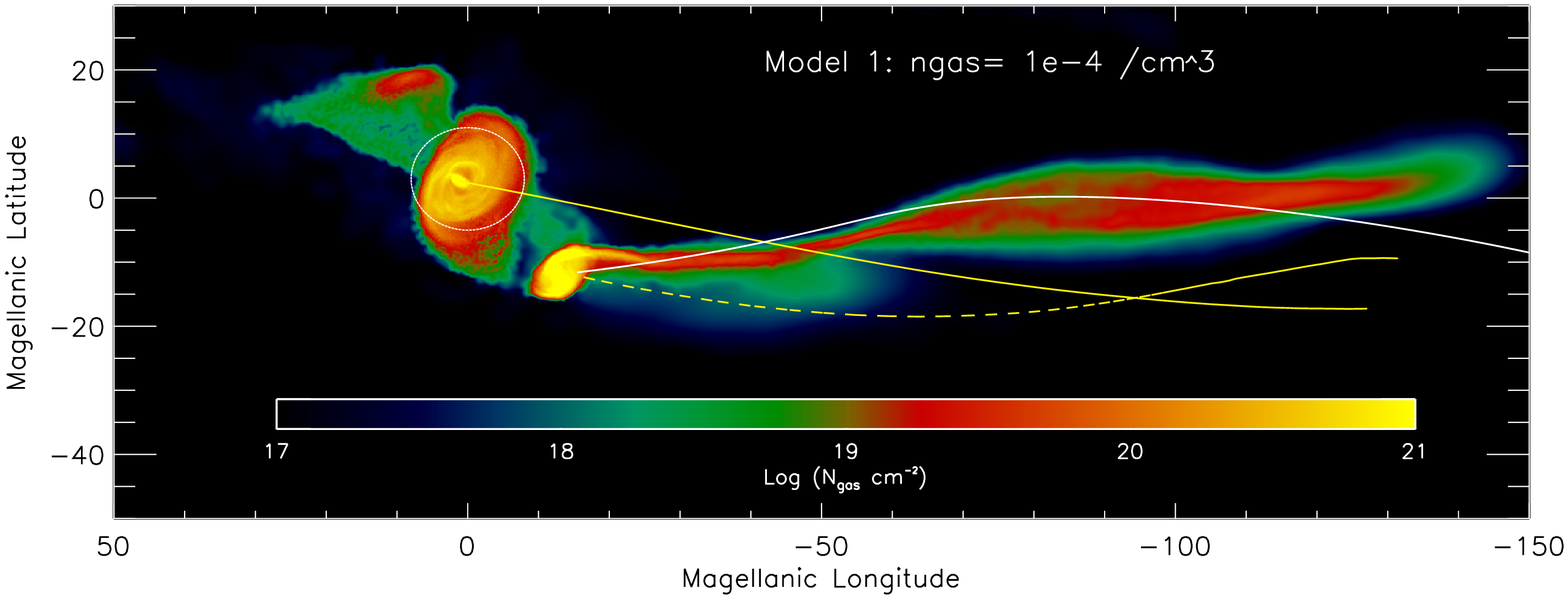}}}\\
\mbox{ {\includegraphics[width=5.95in, trim=0.1in 0 0 0, clip=true]{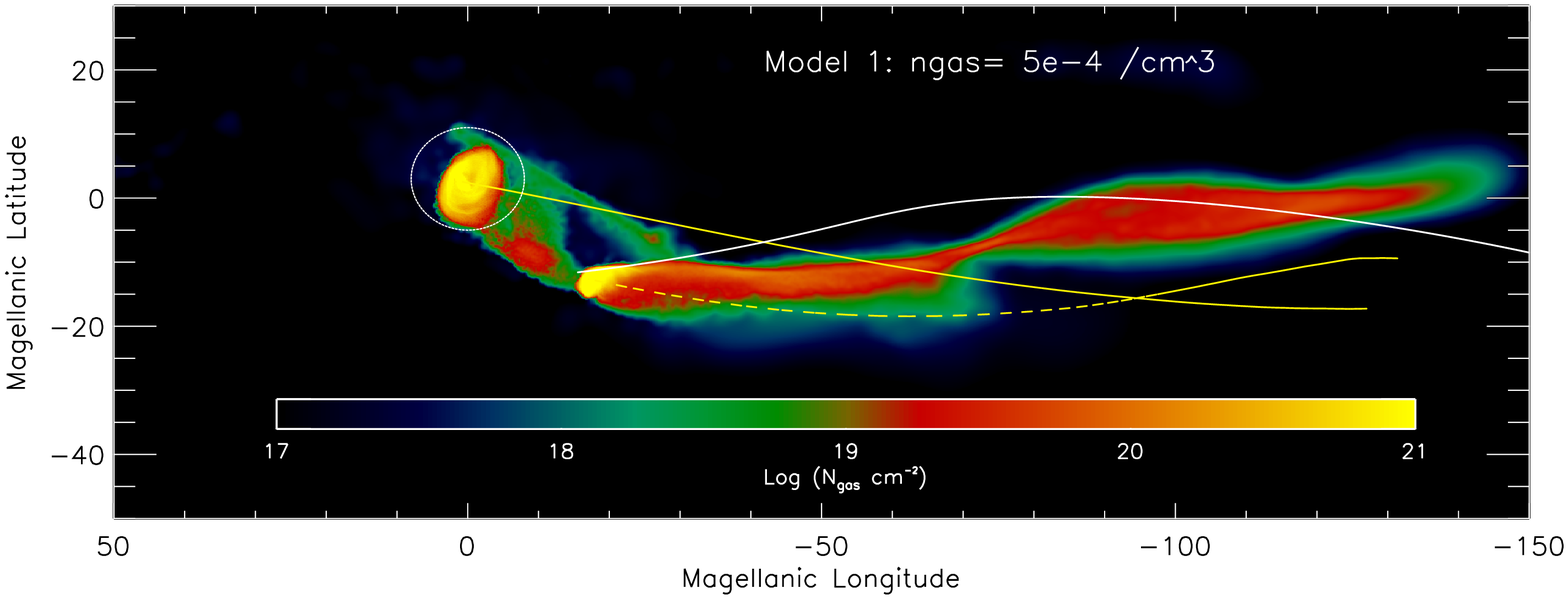}}}
 \end{center}
 \caption{\label{ch4fig:RamModel1} The gas column density in Magellanic Coordinates along the Magellanic System for Model 1 with a toy model ram pressure acceleration.  The ambient hot halo gas density increases 
 from top to bottom, as marked.  Ram pressure causes the simulated stream to trace the past orbits of the Clouds on the plane of the sky; this is opposite to the location of the true
 Stream (indicated by the white line).  The column density along the simulated stream also changes as the
 ram pressure becomes more important. }
 \end{figure*}


\begin{figure*}
\begin{center}
\mbox{{\includegraphics[width=6in, trim=0 0.6in 0. 0, clip=true ]{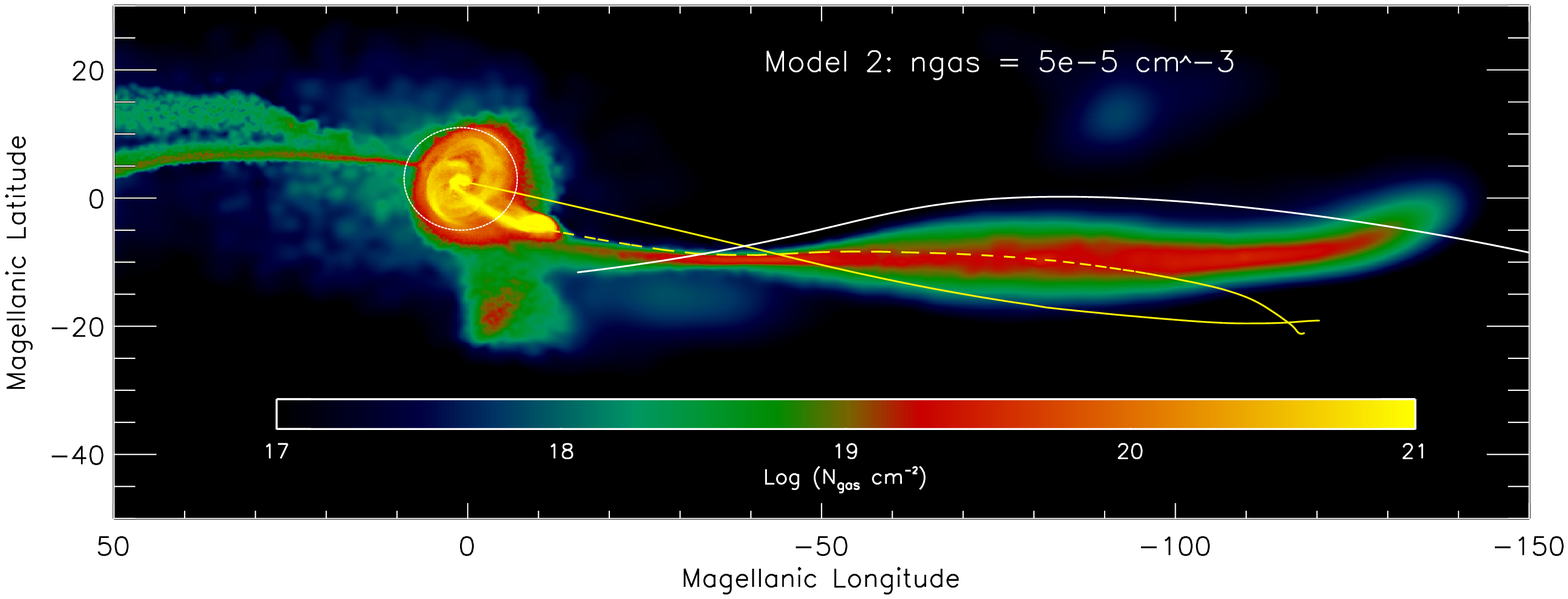}}}\\
\mbox{{\includegraphics[width=6in, trim=0 0.6in 0 0, clip=true]{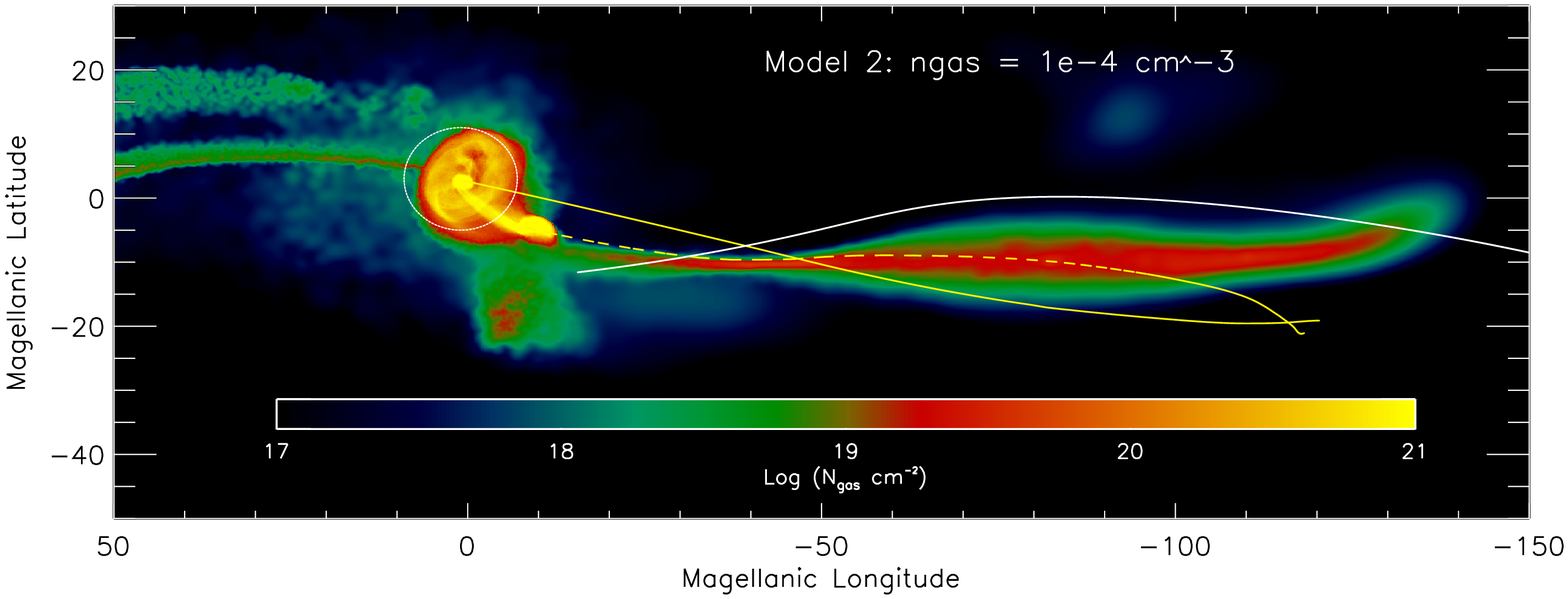}}}\\
\mbox{ {\includegraphics[width=5.95in, trim=0.1in 0 0 0, clip=true]{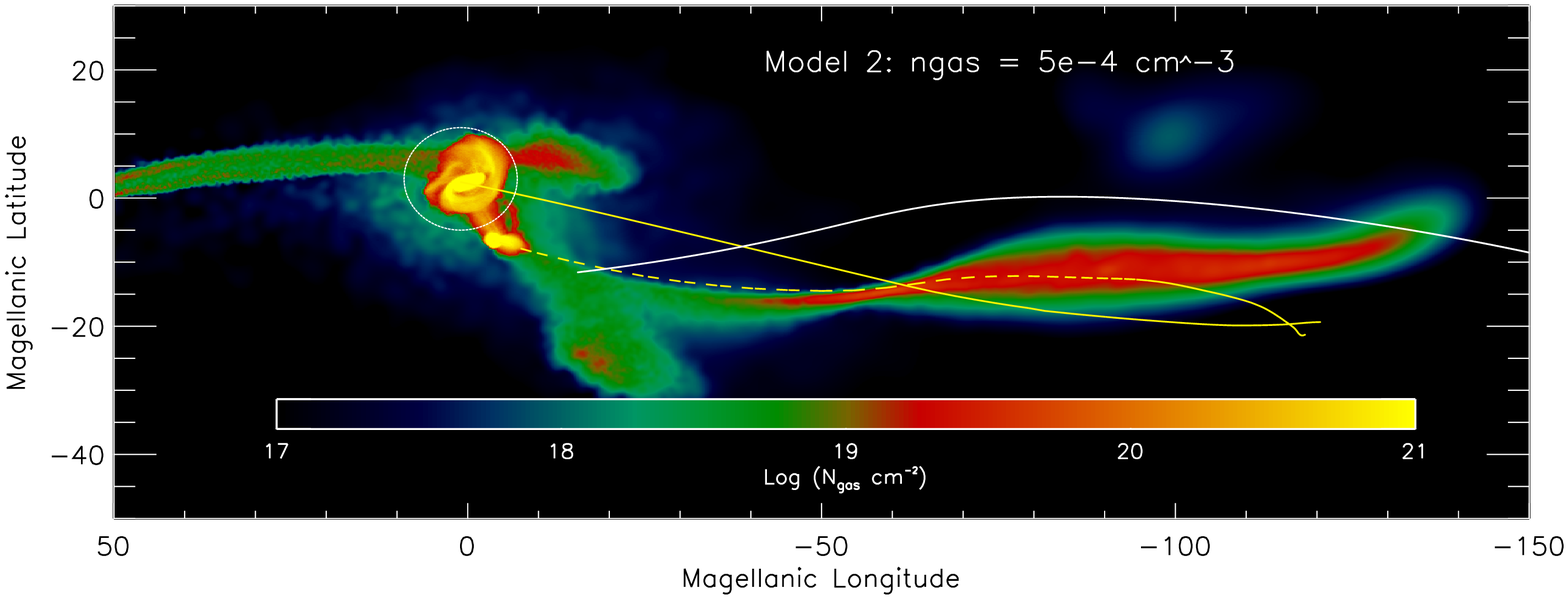}}}
 \end{center}
 \caption{\label{ch4fig:RamModel2} Same as Figure~\ref{ch4fig:RamModel1}, but for Model 2. }
 \end{figure*}

In Figure~\ref{ch4fig:RamModel1} and ~\ref{ch4fig:RamModel2} the gas column density of the Magellanic System 
is mapped in Magellanic coordinates for Model 1 and Model 2, respectively.  
As the background density is increased, the location of the Stream changes, and the gas begins to trace the orbit. 
In \citet{besla2007} it was shown that the location of the LMC past orbit deviates from the 
location of the Stream on the plane of the sky. This result was further shown to be insensitive to the MW 
model and robust within 3$\sigma$ of the measured proper motions.  The fact that ram pressure stripping works to {\it align} 
the stripped material and the past orbits suggests that it is not the main formation mechanism for the Magellanic Stream.

The Leading Arm also changes location and structure significantly in the toy model, particularly in Model 1. 
Facing gas densities larger than $10^{-4}$ cm$^{-3}$ it disappears entirely.   
Although the Bridge disappears in Model 1 at densities larger than $5 \times 10^{-5}$ cm$^{-3}$,
 it remains a strong feature in Model 2 at all densities. This is likely because, in Model 2,
the Bridge is a young feature that formed in the most recent collision between the MCs. 
It has thus not had enough time to experience significant ram pressure effects. 
In Model 2 the leading arm also gets closer to the MW, falling to a line-of-sight distance of 30 kpc.  
These models do not predict that components of the Leading 
Arm should be interacting with the gaseous disk of the MW, although there are claims that such a situation
 has been observed \citep{mcclure2008}.
 
 Note that as the ram pressure increases, the SMC's orbit changes; 
 the SMC is clearly not in the same location in the lower panel of Figures ~\ref{ch4fig:RamModel1} and ~\ref{ch4fig:RamModel2} as in 
 the respective upper panel.  
 This occurs because ram pressure decreases the velocity of the SMC. The LMC, on the other hand, is too massive for its motion to be affected. 
 A strong ram pressure headwind is thus likely at odds with the high relative velocity observed between the MCs \citep[$\sim$100 km/s, ][]{Nitya2}.

   
The structure of the LMC gas disk changes even with a mild ram pressure headwind.  The gas disk is rotating clockwise, and so the lower half
is rotating into the head wind and gets stalled. Gas therefore builds up in the lower left corner of the LMC's disk. This is in fact where the 30 Doradus
star-forming region is situated.  \citet{DeBoer} suggested that the HI overdensity seen in the South-East is a direct result of the interaction between 
the rotating LMC gaseous 
disk and a headwind from the ambient medium. Here, we can indeed see that this process occurs generically in both Models 1 and 2.  
The LMC's gaseous disk is observed to be truncated 
(i.e. it is not as extended as the stellar disk).  The simulated LMC gas disk is truncated if 
 the ambient gas density is at least of order $5 \times 10^{-5}$ cm$^{-3}$. 

Note that the column density along the Stream also changes as ram pressure become more important. The column density increases in the regions closest to the MCs,  
 indicating that ram pressure may be the solution for the mismatch between the observed maximal column density gradient along the Stream 
and the simulation results presented in Figure~\ref{ch4fig:Column}.


From this simple toy model we can conclude that ram pressure and other hydrodynamic instabilities 
will change the mass budget in the various components of the system and change their locations on the plane of the sky. In particular, 
ram pressure works to align the Stream with the past orbits, contrary to expectations from the proper motions \citep{besla2007}. 
 We estimate that if the background halo gas densities are in excess of 10$^{-4}$ /cm$^3$  the gas distribution of the 
simulated Magellanic System will be irreconcilable with observations. This estimate is in accord with observational upper limits \citep{Rasmussen}. 

A recent study by \citet{diaz2011ram} suggests that ram pressure at the ambient densities required by the \citet{Mastro} study  ($5 \times 10^{-5}$ /cm$^3$) 
would be too large for the survival of the Magellanic Stream.  In a first infall scenario, where the Magellanic System would be interacting with the 
ambient halo medium for a shorter period of time ($<$ 1 Gyr), we find that the Stream can indeed survive such densities and that ram pressure stripping 
likely plays a very important role in shaping the Magellanic Stream (particularly the Leading Arm) and increasing its mass budget.  It should also be noted 
that Faraday rotation has been detected in at least one high velocity cloud in the Leading Arm \citep{McClure2010}; magnetic fields
 may protect the cloudlets from evaporation and hydrodynamic instabilities, increasing the chance of survivability of the Stream in the face of a ram pressure 
 headwind.

\section{The Role of Stellar Feedback in the Formation of the Stream} 
\label{sec:Feedback}

There is evidence of stellar feedback in the vicinity of the MCs. 
\citet{lehner2009} have detected high velocity clouds (HVCs) between the MCs and the MW, moving at a velocity with respect to the 
local standard of rest ($v_{lsr}$) as high as 150 km/s. Using FUSE, they find on average that these HVCs have metallicities of
 [OI/HI] = -0.51 $^{+0.12}_{-0.16}$  and HI masses of (0.5-1) $\times 10^6 \Msun$, although these HVCs are
predominantly ionized. Such material is thus, on average, enriched relative to the LMC and SMC. 
Furthermore, \citet{lehner2007} have detected a highly ionized (OVI, CIV, SiIV, NV) corona of high velocity gas surrounding the LMC, 
suggestive of outflows. \citet{staveley2003} 
also report similar observations, and find some of this high velocity gas to be projected on HI voids in the LMC. 
Data from FUSE indicates that the SMC appears to also have an OVI corona \citep{hoopes2002}. 
These detected structures are in close proximity to the MCs, and so may be indicative of mass loss processes that are currently 
ongoing, rather than a long duration process that would be required to build the Magellanic Stream.  

Recently, \citet{nidever2008} found a coherent $v_{lsr}$ gradient from the LMC along one of the filaments in the Stream and in the Leading Arm. 
They also claim to detect sinusoidal velocity patterns in the Stream, which they interpret as signatures of the LMC's disk rotation. 
They thus conclude that as much as half of the mass within the Stream originates from the LMC and that the 
sinusoidal velocity pattern indicates that the bulk of this material is emanating from the south-east HI overdensity (SEHO) region 
of the LMC's HI gaseous disk as a stellar outflow over the past 1.7 Gyr.   The idea that the Magellanic Stream was formed from stellar outflows 
was also suggested by \citet{olano2004}. The LMC is pock-marked with giant superbubbles, indicating locations of strong stellar 
feedback/winds. If the column density were constant across the disk, then each superbubble would have originally contained roughly $10^7 \Msun$
 worth of material \citep{kim1998, kim1999}.   \citet{nidever2008} determine that 2-3 superbubbles losing $10^6 \Msun$ worth of material every 
 10 Myr over 1.7 Gyr would be sufficient to explain the mass budget of the Stream. 
 
 A stellar feedback model for the Stream is attractive in that it does not rely on strong 
MW tides to remove material from the LMC, making it consistent with a first infall scenario 
(although the leading component would still be an issue). 

However, it is unlikely that outflows generated by SNe feedback would be sufficiently
 energetic to be unbound from the LMC's gravitational potential. 
In our first infall LMC models ($M_{LMC} = 1.8 \times 10^{11} \Msun$), the escape speed is of order 250 km/s at 10 kpc.  \citet{martin2005} finds that 
terminal velocities expected from galaxies with maximum circular velocities of 100 km/s should also be of order 100 km/s, well below
the escape speed at 10 kpc. Moreover, galaxies with star formation rates as low as 0.1$\Msun$/yr \citep[i.e. the value in the 30 Doradus region today,][]{harris2009}
generally have outflow velocities less than 30 km/s (figure 6 of Martin 2005).

However, it is certainly possible that a ram pressure headwind can exploit the stellar feedback processes to aid in the removal of 
material from the deepest parts of the LMC's potential. 
This theory is testable by observations of the metallicity in the Stream:
 not only do the LMC and SMC have different metallicities, but any feedback scenario will result in the removal 
of enriched material \citep{maclow1999}.

Metallicities for various components of the Magellanic System are summarized in Table~\ref{metalTable}.
The current day LMC's metallicity is [O/H]$_{\rm LMC}$ = -0.34 $\pm$ 0.06 
\citep[][ updated to the latest solar abundances by Fox et al. 2010]{russell1992}. 
  But chemical enrichment models of \citet{pagel1998} suggest that 2 Gyr ago the LMC's metallicity could have been as low as 
   [O/H]$_{\rm LMC} \sim$ -0.5 (their Figure 3).  1-2 Gyr is the relevant timescale for the formation of the Stream 
  \citep{nidever2008, GN}; given the measured H$_\alpha$ emission along its length \citep{weiner1996} and the corresponding expected 
  ablation timescale for the neutral HI component of the Stream \citep{bland2007}, it is unlikely that the Stream could have survived for much 
  longer.

Recently, \citet{fox2010} have determined the oxygen abundance for a region near the tip of the Stream from absorption lines
towards the background quasar NGC 7469.  Observed [OI/HI] abundances are a close indication of the true oxygen abundance, [O/H], since 
oxygen is not strongly depleted onto interstellar dust \citep{jensen2005}.  They find: 

\begin{align}
\left [ \frac{ \rm O}{\rm H} \right] \sim \left [\frac{ \rm OI}{\rm HI} \right] & =   \rm{log} \left( \frac{N({\rm OI})}{N(\rm HI)} \right) - \rm{ log} \left(\frac{\rm O}{\rm H} \right)_\odot \\
\label{eqn:abund}
& =  \rm{log} \left( \frac{10^{14.32 \pm 0.04} }{10^{18.63 \pm  0.13}} \right )  - (-3.31)  \\
& = -1.00 \pm 0.05 (\rm {stat}) \pm 0.08 (\rm {syst}) 
\end{align}

In an outflow or ram pressure stripping scenario the 
metallicity of the Stream must be at least that of the original ISM. These measurements make it improbable that the Stream could have 
solely originated in the LMC, regardless of the formation mechanism. 
The SMC interstellar oxygen abundance is a better match to these observations (see Table ~\ref{metalTable}), especially since it 
was also less enriched 1-2 Gyr ago \citep{pagel1998}. 
 Interestingly, such low abundances are also measured
in the Bridge connecting the MCs \citep[Table~\ref{metalTable};][]{lehner2008}, suggestive of a common origin for both structures.

Closer to the MCs, the metallicity measurements derived from absorption lines towards background quasars
 are larger by a factor of 2-4 \citep{lu1998, gibson2000, sembach2001}.  But these other measurements
  also have much larger error bars, as they use tracers that are more sensitive to ionization corrections than the OI/HI ratio. 
     If we take an upper limit of a factor of 4 increase in the oxygen abundance in the Stream relative to the \cite{fox2010} values, 
 we can assess whether this material could have once ($\sim$2 Gyr ago) originated in a stellar wind from the LMC. 
 We follow the methodology of \citet{martin2002}, 
 who estimate the expected metallicity of outflows from the starbursting dwarf NGC 1569. 
 The total mass of the ejected wind $M_w$ is given by 
  
\begin{table*}
 \centering
 \begin{minipage}{200mm}
\caption{Oxygen Abundances \label{metalTable}}
\begin{tabular}{@{}lcc@{}}
\hline
Object 		& [O/H]\footnote{Log of the oxygen abundance relative to solar, as defined in equation~\ref{eqn:abund}.} 	& Reference \\
\hline
\hline
LMC (today)    	             	&  -0.34 $\pm$ 0.6 	 &  \citet{russell1992, fox2010}  \\
LMC (2 Gyr ago)		&  -0.5			&  \citet{pagel1998}  \\
SMC (today)			& - 0.66 $\pm$0.1	& \citet{russell1992, fox2010} \\ 
Stream 				&   -1.0 $\pm$0.13      & \citet{fox2010} \\ 
Stream (Upper Limit)\footnote{Upper limit is defined by increasing the observed oxygen column density of the Stream by a factor of four}	&   -0.4				& \citet{gibson2000, lu1998, sembach2001} \\ 
Bridge				&  -0.96$^{+0.13}_{-0.11}$ & \citet{lehner2008} \\
HVCs near LMC		&  -0.5$^{+0.12}_{-0.16}$  & \citet{lehner2009} \\ 
\hline
\end{tabular}
\end{minipage}
\end{table*}

 \begin{equation}
  M_w =  M_{ej} ( 1+ \chi ), 
 \label{eqMW}
 \end{equation}
 
\noindent where $M_{ej}$ is the mass in SNe ejecta
and $\chi$ is the mass loading of the wind.
We consider first the 
required mass loading from the ISM of the LMC needed to dilute the metallicity so as to not violate the upper limits for the Stream. 
The oxygen abundance of the wind can be expressed as, 
\begin{equation}
\rm{Z}_{\rm O, w} = \frac{ M_{ej} (\rm{Z}_{\rm O, SN} + \chi \rm{Z}_{\rm O, ISM} )}{M_w}.  
\label{eqZO}
\end{equation}

\noindent Written in terms of the solar values,  the abundance of the ISM 2 Gyr ago is Z$_{\rm O, ISM}  = 10^{-0.5}$ = 0.32. 
 Following \citet{martin2002}, the IMF-averaged metallicity of SNe ejecta is Z$_{\rm O, SN} \sim$ 8 times solar for oxygen. We 
 choose three values for Z$_{\rm O, SN} =$ 4, 8 and 12.  
 We can rewrite equation ~\ref{eqZO} to solve for $\chi$ in a mass independent way using equation~\ref{eqMW}:
 
\begin{equation}
\chi = \frac{ \rm{Z}_{\rm O, SN} -  \rm{Z}_{\rm O, w}}{\rm{Z}_{\rm O,w} - \rm{Z}_{\rm O, ISM} }.
\label{eqChi}
\end{equation}

Given that we know the current upper limit for the metallicity of the wind, Z$_{\rm O,w} = 10^{-0.4} = 0.4$, then
the required mass loading is $\chi$(Z$_{\rm O, SN} = 4,8,12) = (45, 95, 145)$.  

Mass loading in excess of factors of 10 are not observed. The wind from M82 is expected to be mass loaded by a 
factor of 3-6 \citep{suchkov1996} and  \citet{martin2002} find a value of $\chi = 9$ to be favored for NGC 1569. 
Furthermore, numerical simulations by \citet{maclow1999} find it difficult to accelerate any ambient cool gas 
that is swept up with the expanding outflow; instead, much of that material remains bound to the galaxy.

Ram pressure stripping might be the key missing ingredient that could help entrain ambient gas and explain such high mass loading factors. 
Assuming the addition of ram pressure stripping results in mass loading in excess of a factor of 45, then the resulting total wind mass (equation ~\ref{eqMW})
over a timescale of 1.7 Gyr (the lifetime of the Stream according to the Nidever et al. 2008 model) 
 would be of order $M_{w} \sim 10^{9} \Msun$\footnote{$M_{ej} = \Gamma \dot{\rm M_\ast} T_{MS}$ is determined assuming a star formation rate 
of $\dot{\rm M_\ast} = 0.1 \Msun$/yr (the current 
rate in 30 Doradus), a SNe formation rate of $\Gamma = $0.12 determined from Starburst99 \citep{leitherer1999} with a Kroupa IMF, 
over a timescale of $T_{MS} = $1.7 Gyr}. 
While this appears to be more than enough material to explain the total amount 
of neutral HI observed in the Stream ($5\times10^{8}\Msun$; Table~\ref{ch4tableMass}),  it would imply that the entire Stream 
should be metal enriched, contrary to observations. 
 This scenario thus 
over-predicts the mass budget of the Stream and requires mass loading factors that are more extreme than observed.

\bibliography{ms}

\label{lastpage}
\end{document}